\documentclass[12pt]{article}
\usepackage{jheppub}
\usepackage{rotating}
\usepackage{amsmath,amssymb,amsfonts,pstricks,mathtools,epsfig}
\usepackage{float}
\usepackage{subfig}
\usepackage{scalefnt}
\usepackage{wrapfig}
\usepackage{fancybox}
\usepackage{mathrsfs}

\newlength{\arrow}
\newcommand{\de}{\mathrm{d}}

\DeclareMathOperator{\Tr}{Tr}
\DeclareMathOperator{\Str}{Str}
\DeclareMathOperator{\diag}{diag}
\newcommand{\Gam}[1]{\ensuremath{\Gamma\!\left({#1}\right)}}
\renewcommand{\Re}{\ensuremath{\operatorname{Re}}}
\renewcommand{\Im}{\ensuremath{\operatorname{Im}}}

\newcommand{\ceil}[1]{\ensuremath{\left\lceil{#1}\right\rceil}}

\newcommand{\noeq}{\ensuremath{\mathrel{\phantom{=}}}}

{\setlength{\fboxsep}{15pt}
\setlength{\mylength}{\linewidth}%
\addtolength{\mylength}{-2\fboxsep}%
\addtolength{\mylength}{-2\fboxrule}%
\Sbox
\minipage{\mylength}%
\setlength{\abovedisplayskip}{0pt}%
\setlength{\belowdisplayskip}{0pt}%
\equation}%
{\endequation\endminipage\endSbox
\[\fbox{\TheSbox}\]}

\newcommand\cR{{\mathcal R}}

\DeclareMathOperator\im{{\mathrm{Im}}}
\newcommand\half{\frac12}

\newcommand{\gs}{g_{\mathrm s}}
\newcommand{\ls}{\ell_{\mathrm s}}

\usepackage{tikz}
\usetikzlibrary{shapes.geometric}
\usepackage{tensor}
\usepackage{braket}
\usetikzlibrary{arrows}
\usetikzlibrary{decorations.markings}
\tikzset{
  big arrow/.style={
    decoration={markings,mark=at position 1 with {\arrow[scale=2,#1]{>}}},
    postaction={decorate},
    shorten >=0.4pt},
  big arrow/.default=black}

\begin{document}

\title{Negative Branes, Supergroups and the Signature of Spacetime}

\author[\dagger]{Robbert Dijkgraaf,}
\author[\ast]{Ben Heidenreich,}
\author[\ast]{Patrick Jefferson}
\author[\ast] {and Cumrun Vafa}
\affiliation[\dagger]{Institute for Advanced Study, Princeton, NJ 08540, USA}
\affiliation[\ast]{Jefferson Physical Laboratory, Harvard University, Cambridge, MA 02138, USA}

\abstract{We study the realization of supergroup gauge theories using negative branes in string theory. We show that negative branes are intimately connected with the possibility of timelike compactification and exotic spacetime signatures previously studied by Hull. Isolated negative branes dynamically generate a change in spacetime signature near their worldvolumes, and are related by string dualities to a smooth M-theory geometry with closed timelike curves.
Using negative D3 branes, we show that $SU(0|N)$ supergroup theories are holographically dual to an exotic variant of type IIB string theory on dS$_{3,2}\times \bar{\text{S}}^5$, for which the emergent dimensions are timelike. Using branes, mirror symmetry and Nekrasov's instanton calculus, all of which agree, we derive the Seiberg-Witten curve for $\mathcal{N}=2$ $SU(N|M)$ gauge theories. Together with our exploration of holography and string dualities for negative branes, this suggests that supergroup gauge theories may be non-perturbatively well-defined objects, though several puzzles remain.
}

\maketitle

\section{Introduction}

Lie supergroups are a natural extension of Lie groups by fermionic generators.
A prominent example of their use in physics is to describe the symmetries of supersymmetric quantum
field theories.
 In this way they are part of the
global symmetries of unitary supersymmetric theories.  On the other hand, Lie groups have been
used in the context of quantum field theories (QFTs) as gauge symmetries, leading to a successful construction
of the Standard Model. It is thus natural to ask what happens if one uses a Lie supergroup as the gauge symmetry group of a QFT. If the supergroup gauge symmetries lead to consistent QFTs, they will necessarily be \emph{non-unitary}, as the gauge fields corresponding to the fermionic generators violate the spin-statistics theorem.

A seemingly unrelated topic is the signature of spacetime.  Ordinary physics happens in Lorentzian signature, but the reason for this restriction is not wholly self-evident. Some analog of physics may be possible in Euclidean signature, or in theories with multiple times, but these possibilities are not often studied, and raise many unanswered questions (for a review of two-time physics, see~\cite{Bars:2000qm}.)
 What would physics with more than one time mean?
What determines the signature of spacetime? Can the signature of spacetime change dynamically?
A closely related question in the context of holography is how to generate time from
a purely Euclidean theory.

Supersymmetry has been a powerful organizing principle in the context of string theories, so it is natural to ask if supersymmetry fixes the signature of spacetime.  It was shown in~\cite{Hull:1998ym}
that this is not the case and a diverse range of signatures are consistent with supersymmetry and
supergravity.  This raises the question of whether or not non-Lorentzian signatures should be incorporated into string theory.
It turns out that the string duality web can accommodate a number of additional theories with unusual spacetime signature, suggesting that the full collection of these theories may form a
mathematically consistent structure, regardless of their applicability to our universe.

In this paper we show that supergroup gauge theories and dynamical change of signature
of spacetime are intimately connected.  Our study was motivated by the observation in~\cite{Vafa:2014iua} that ${\cal N}=4$ $SU(N|M)$ gauge theories, should they exist, must be holographically dual to $\text{AdS}_5\times \text{S}^5$, since they are indistinguishable from $SU(N-M)$ gauge theories to all orders in $1/(N-M)$ (assuming $N>M$). This observation raises the question
of uniqueness of non-perturbative completions of gravity if we include non-unitary gauge theories in our considerations.

Supergroup gauge symmetries can be realized in string theory by introducing \emph{negative} branes~\cite{Vafa:2001qf,Okuda:2006fb}. Originally introduced as `topological anti-branes' and `ghost'  branes, negative branes are defined to be the extended objects that completely cancel the effects of ordinary branes. As a consequence, the Chan-Paton
factors associated to string endpoints sitting on negative branes have extra minus signs. Thus, $SU(N|M)$ gauge symmetry can be realized by a stack of $N$ ordinary D-branes and $M$ negative D-branes. Negative branes exhibit many unusual properties. The most unsettling is their negative tension, and one might be particularly interested in exploring the gravitational backreaction of negative branes on the geometry of spacetime.
A preliminary study of this was done in~\cite{Okuda:2006fb} where it was found that negative brane solutions in supergravity produce naked curvature singularities in spacetime.

One aim of the present paper is to study these backreactions more systematically and uncover
their physical meaning.  What we find is that negative branes are surrounded by a bubble of spacetime where the metric signature has changed.  In other words we find that {\it negative branes induce a dynamical
change of space-time signature}!  In particular, we learn that the directions transverse (or parallel, depending on convention) to the brane worldvolume flip signature inside the bubble surrounding the brane. Using this fact, we
obtain all the supergravities with diverse signatures anticipated in~\cite{Hull:1998ym}.  For example
in M-theory negative M2 branes flip the signature of the eight transverse directions leading to a theory in signature $(2,9)$ (or $(9,2)$).\footnote{We denote signature by the pair $(s,t)$, with $s$ being the number of spacelike dimensions and $t$ being the number of timelike dimensions.}
Similarly negative M5 branes flipping the signature of five transverse directions lead to a theory in signature $(5,6)$ (or $(6,5)$).
These are precisely the signatures anticipated in~\cite{Hull:1998ym} for M-theory, in addition to the usual $(10,1)$ (or
$(1,10)$) (see also~\cite{Hull:1998fh,Hull:1999mt,Blencowe:1988sk}). 

We next use string theory to study strong coupling
aspects of supergroup gauge theories. We find that the large $N$ dual
to ${\cal N}=4$ SYM for $SU(N|M)$ when $N<M$ is still $\text{AdS}_5\times \text{S}^5$ but with signature
$(7,3)$ rather than $(9,1)$.  In particular $SU(0|M)$ is of this type and is dual to supergravity
with this unconventional signature.   While not a proof of the non-perturbative
existence of ${\cal N}=4$ supergroup gauge theories, this observation does show---at least to all orders in the $1/N$
expansion guaranteed to exist for these theories---that the holographic dual should agree with the supergravity theory in signature $(3,7)$ defined by string perturbation theory.

We find a consistent picture of dualities involving negative branes which fits very well with the
structure found in \cite{Hull:1998ym}, suggesting that these theories actually exist beyond string perturbation theory. To check these statements we consider a sample of ${\cal N}=2$ supersymmetric
gauge theories with $SU(N|M)$ gauge supergroup and find their exact non-perturbative vacuum geometry.
We find the corresponding Seiberg-Witten curve using three different methods:  brane constructions,
geometric engineering of the theory, and direct Nekrasov calculus.  We find that all three methods agree
with one another and yield the same result. These checks lend further support to the the claim that these
theories exist non-perturbatively.

The organization of this paper is as follows.  In~\S\ref{sec:supergroup}, we review selected aspects of negative branes and supergroups.
In~\S\ref{sec:signaturechange}, we consider the gravitational backreaction of negative branes and argue that they dynamically change the signature of the spacetime surrounding them.  In~\S\ref{sec:diffsigs}, we review
the results in~\cite{Hull:1998ym} involving string theories and their respective low energy limits with diverse signatures.  In~\S\ref{sec:AdSCFT},
we study the near-horizon limit of negative brane geometries and conjecture holographic
duals for these theories.  In~\S\ref{sec:curvature} we discuss the role of $\mathcal{R}^4$ curvature corrections,
which in some cases make the action complex.  In~\S\ref{sec:worldsheet} we discuss aspects of the worldsheet description of the various string theories.
In~\S\ref{sec:seibergwitten}, we discuss some non-perturbative aspects of these theories and in particular show how the non-perturbative vacuum geometry of ${\cal N}=2$ supersymmetric
theories based on supergroups can be solved in three different ways, thus giving further evidence for the existence
of these theories.  In~\S\ref{sec:problems}, we end with discussion of some issues that need to be resolved in future work.

\section{Negative Branes and Supergroups} \label{sec:supergroup}

The connection between D-branes and gauge groups is one of the most important features of string theory. On the worldsheet, D-branes appear as boundaries. Adding an extra ``Chan-Paton'' label to each boundary---specifying on which brane the string ends---is all that is required to introduce multiple D-branes. Surprisingly, although the Chan-Paton label has no worldsheet dynamics, introducing $N$ labels automatically generates a $U(N)$ gauge theory in the target space, so that gauge theories emerge naturally from string theory. The reverse is true as well: drawing Feynman diagrams in 't Hooft's double line notation, the propagators and vertices form a worldsheet whose boundaries lie on D-branes in the corresponding string theory description.

When closed strings scatter off a stack of $N$ D-branes, each worldsheet boundary contributes a factor of $\Tr 1 = N$ to the amplitude. While from the perspective of closed strings in perturbation theory $N$ could be any number, non-pertubatively the Dirac quantization condition for D-branes requires $N$ to be an integer. This does not, however, exclude the possibility that $N$ is negative. Explicitly, negative $N$ arises when we associate an extra minus sign to each boundary that carries one of a designated subset of the Chan-Paton labels. These labels correspond to what we will call ``negative branes,'' where boundaries without any vertex operators contribute $N_+ - N_-$ to the amplitude for $N_+$ positive D-branes and $N_-$ negative D-branes.

What are the consequences of negative branes for gauge theory? The extra signs imply that a string stretched between a positive brane and a negative brane has the opposite of the usual statistics, picking up an extra minus sign in the corresponding annulus diagram. This means that when positive and negative branes are brought together anticommuting vectors (``$W$ fermions'') become light, enhancing the $U(N_+) \times U(N_-)$ gauge group to the supergroup $U(N_+ | N_-)$~\cite{Vafa:2001qf, Okuda:2006fb}, i.e.\ the group of unitary supermatrices
\begin{equation}
U = \begin{pmatrix} A & B \\ C & D\end{pmatrix} \,,
\end{equation}
where $A$ and $B$ are $N_+ \times N_+$ and $N_- \times N_-$ matrices with c-number entries and $B$ and $C$ are $N_+ \times N_-$ and $N_- \times N_+$ matrices with Grassmann number entries.
The $U(N_+ | N_-)$ invariant trace is $\Str U \equiv \Tr A - \Tr B$, so that each hole in the 't Hooft diagram contributes $\Str 1 = N_+ - N_-$, as in the worldsheet picture.

Some features of negative branes at first appear parallel to anti-branes.\footnote{In topological string theory, negative branes and anti-branes are equivalent, cf.~\cite{Vafa:2001qf}.} For instance, closed strings see only the difference $N_+ - N_-$, similar to the result of placing $N$ branes atop $\bar{N}$ anti-branes and allowing them to annihilate. Indeed, negative branes carry the same Ramond-Ramond (RR) charge as anti-branes, but unlike anti-branes negative branes have \emph{negative} tension: both the R-R and NS-NS components of the negative D-brane CFT boundary states differ from those of ordinary D-brane boundary states by a minus sign~\cite{Okuda:2006fb,Parkhomenko:2008dt}. As a consequence, while branes and anti-branes are not mutually BPS, allowing them to annihilate, positive and negative branes preserve all the same supersymmetries.

Further differences appear upon closer inspection. Since the $W$ fermions violate the spin-statistics theorem, it follows that supergroup gauge theories are \emph{non-unitary}, as is string theory in a background containing negative branes. Moreover, these theories contain negative energy states, most prominently the negative D-brane itself. We show later that the true properties of negative branes are still more bizarre: each negative brane is surrounded by a bubble where the signature of spacetime changes. Depending on the brane, the space inside the bubble may have no time direction, or multiple time directions, and consequently familiar physical concepts such as unitarity cease to have any meaning.

Thus, a seemingly innocuous change to the worldsheet theory has profound consequences for the target space. Given the strange and unexpected results, it is natural to ask whether string theories with negative branes exist non-perturbatively. A simpler question---though closely related by the AdS/CFT correspondence---is whether supergroup gauge theories themselves exist non-perturbatively. We consider the balance of evidence briefly before returning to our discussion of negative branes.

\subsection{Supergroup gauge theories:  Do they exist?} \label{subsec:SGexist}

Consider ${\cal N}=4$ Super Yang-Mills with gauge supergroup $U(N_+|N_-)$ realized
on the worldvolume of $N_+$ ordinary D3-branes and $N_-$ negative D3-branes.
The Lagrangian density of this theory contains the terms
\begin{equation}
{\frac{1}{\gs}}\big[\Str F^2 +\sum_{i=1}^6 \Str (D\Phi^i)^2+\cdots\big]={\frac{1}{\gs}}\big[ \Tr F_+^2-\Tr F_-^2+\sum_{i=1}^6(\Tr (D\Phi^i_+)^2- \Tr (D\Phi^i_-)^2)+\cdots\big]
\end{equation}
where the $\pm{}$ subscripts label the $U(N_{\pm})$ blocks.
This theory exhibits an unbounded energy spectrum and so is non-unitary, and there is no obviously convergent expression for the path integral.  However, it is possible to define the above theory to all orders in $\gs=g_{\text{YM}}^2/4 \pi$,
consistent with the fact that in string theory one can compute all amplitudes to any order in $\gs$.
  Moreover, the usual arguments
for finiteness of ${\cal N}=4$ $U(N)$ SYM theory still apply, suggesting that at least perturbatively,
this particular example of a supergroup gauge theory is finite. The fact that the 't Hooft diagrams of the $U(N_+|N_-)$ theory can be obtained from those of the usual $U(N)$ theory by replacing $N$ with $N_+ - N_-$ further supports this conclusion. Nevertheless, this argument does not prove that such a theory has a consistent non-perturbative completion.\footnote{The analogous supermatrix models can be shown to be non-trivial and consistent after gauge fixing the supergroup~\cite{supermatrix}, which suggests that a path integral definition could exist if the supergroup gauge symmetry is gauge-fixed in an appropriate manner.}

We pause here to remark that non-unitary conformal field theories are known to exist in two dimensions~\cite{Cardy:1985yy,Belavin:1984vu,Fisher:1978pf,Quella:2007hr} and there is evidence suggesting their existence in higher dimensions~\cite{Fei:2014yja, Mati:2016wjn}.  It is possible that some of these conformal field theories
are IR fixed points of supergroup gauge theories; in this sense, ${\cal N}=4$ supersymmetry
may be quite useful in attempting to prove their existence.

Although we do not attempt an existence proof in this paper (in fact, the existence of non-trivial unitary
theories in more than two dimensions remains unproven), we nevertheless provide evidence that these supergroup gauge theories pass the same checks as their usual unitary counterparts. We have already argued perturbative consistency.  Later, we argue that (at least for some classes) consistent non-perturbative corrections
involving amplitudes that preserve some supersymmetry also exist.
Specifically, we obtain Seiberg-Witten curves for the ${\cal N}=2$ version of these theories.  We also argue that when their unitary counterparts have holographic duals, the non-unitary theories also have holographic duals that
can be used to compute amplitudes to all orders in $1/N$. We will not attempt to check the non-perturbative existence of amplitudes that completely break supersymmetry. While this has not even been done even for unitary theories, in the unitary case lattice regularization techniques
in Euclidean space provide a working definition beyond perturbation theory, whereas
this remains to be shown for supersymmetric supergroup gauge theories.

In summary, the above arguments suggest that supersymmetric $U(N_+|N_-)$ gauge theories are perturbatively (and perhaps even nonperturbatively) well-defined. The fact that these theories are closely connected to a consistent web of dualities in string theory reinforces the possibility of their non-perturbative existence.

 We should also mention here that supergroup gauge theories have proven to be a convenient framework to discuss exact renormalization group techniques preserving manifest gauge symmetry \cite{Arnone:2000qd}. In this setup the original  $SU(N)$ Yang-Mills theory is replaced by a theory with supergroup $SU(N|N)$. The model is then studied in the phase where the supergroup is broken as
$$
SU(N|N) \to SU(N) \times SU(N).
$$
The makes all the fermion gauge modes massive. The second factor essentially acts as a Pauli-Villars regulator, without the need of gauge fixing. Consequently, this method preserves the original $SU(N)$ gauge symmetry. While it is true that this setup embeds a unitary theory in a non-unitary theory, if the symmetry breaking scale is high enough, the unphysical modes will not be excited.

\section{Backreaction and Dynamical Signature Change} \label{sec:signaturechange}

We return to our discussion of negative branes. The backreaction of isolated negative branes is problematic. Since the brane tension is negative, the gravitational backreaction for $N_- - N_+ \gg 1$ generates a naked singularity. As this configuration preserves 16 supercharges, the possible curvature corrections are strongly constrained --- if any are allowed at all, cf.~\cite{Metsaev:1998it,Kallosh:1998qs} --- and it is not immediately clear how to resolve this singularity.

To see the problem, we first review the black-brane geometry sourced by a large number of coincident D-branes, subsequently generalizing the result to negative D-branes. In string frame, the black D-brane geometry is of the form:\footnote{Our conventions are chosen to agree with the supergravity effective action, cf.~(\ref{eqn:sugraaction0}--\ref{eqn:sugraactionCS}) with $\alpha=\beta=1$. Textbook treatments often omit the factor of $g_s^{-1}$ in $F_{p+2}$, presumably at the expense of explicit factors of $g_s$ in the supergravity action.}
\begin{equation} \label{eqn:negbranebkg}
\begin{split}
\de s^2 &= H^{-\half} \de s^2_{p+1}+ H^{\half} \de s^2_{9-p} \,,\\
e^{-2 \Phi} &= \gs^{-2} H^{\frac{p-3}{2}} \,,\\
F_{p+2} &= \gs^{-1} \de H^{-1} \wedge \Omega_{p+1} \,,
\end{split}
\end{equation}
for $p\ne 3$, where $\Omega_{p+1}$ is the volume-form along the branes, $\gs$ is the asymptotic value of the string coupling, and
\begin{equation}
H(r) = 1 +\frac{(2 \sqrt{\pi} \ls)^{7-p}\, \Gamma\!\left(\frac{7-p}{2}\right)}{4 \pi} \sum_i \frac{\gs N_i}{|r - r_i|^{7-p}}
\end{equation}
is a harmonic function in the transverse directions\footnote{For $p=7$, we have $H \sim \log |r-r_i|$ instead.} describing stacks of $N_i$ D$p$ branes at positions $r_i$. For $p=3$, the geometry is the same, but with a self-dual flux
\begin{equation}
F_5 =\gs^{-1}  (1+\star) (\de H^{-1} \wedge \Omega_{4})
\end{equation}

\begin{figure}
\begin{center}
\includegraphics[height=2in]{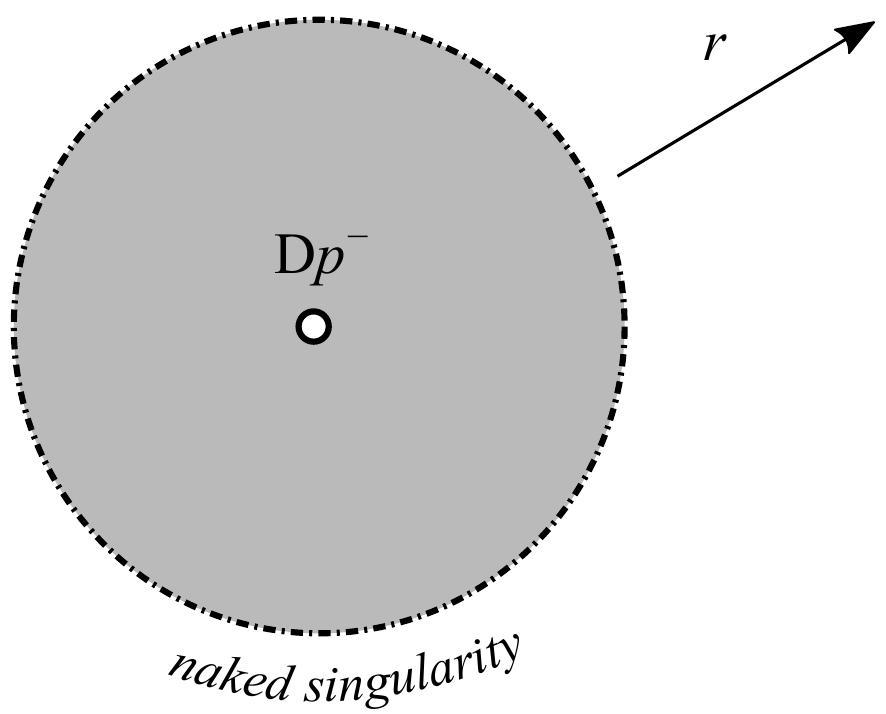}
\caption{Negative branes are surrounded by a naked singularity at a finite distance from the brane, forming a ``bubble'' around the brane.}
\label{fig:nakedsing}
\end{center}
\end{figure}

To generalize to negative branes, we replace $N_i \to N_i^+ - N_i^-$. If $N_i^- > N_i^+$, $H\to - \infty$ near the branes, hence there is an interface at finite distance where $H=0$ and the curvature is singular. For instance, the Ricci scalar is
\begin{equation}
R = - \frac{(p+1)(p-3)}{4 H^{5/2}} (\nabla H)^2\,,
\end{equation}
which diverges as $H \to 0$, except when $p=3$. In the latter case, other curvature invariants diverge. For instance,
\begin{equation}
R^{m n} R_{m n} = \frac{5}{8 H^5} (\nabla H)^4
\end{equation}
for $p=3$, which again diverges as $H \to 0$. This situation is illustrated in Figure~\ref{fig:nakedsing}.

To determine the nature of this singularity, we probe it with BPS branes. Recall that the DBI action for a D$p$ brane takes the form
\begin{equation}
S_{\rm DBI} = - \frac{1}{(2\pi)^p \ls^{p+1}} \int_{\Sigma} \de^{p+1}x \,e^{-\Phi} \sqrt{-\det(\Sigma^{\ast}(g_{m n} +B_{m n})+2 \pi \ls^2 F_{m n})}
\end{equation}
where $F = \de A$ is the field-strength of the world-volume gauge theory and $\Sigma^{\ast}$ is the pullback map associated to the cycle $\Sigma$. A probe D-brane is mutually supersymmetric with the negative D$p$-brane background if the number of world-volume directions of the probe brane parallel ($n_{\parallel}$) and perpendicular ($n_{\perp}$) to the negative brane satisfy:
\begin{equation} \label{eqn:Dp_calb}
n_{\perp}+ (p+1)- n_{\parallel} = 4k\,, \qquad (k \in \mathbb{Z}_{\ge 0})
\end{equation}
where $4 k \ge 0$ is the number of directions along which one, but not both, of the two branes extend. Using the calibration condition~(\ref{eqn:Dp_calb}), we find the dependence of the DBI action-density on the warp factor to be
\begin{equation}
\mathcal{L}_{\rm DBI} \propto H^{k-1} \,.
\end{equation}
If $k=0$ then the probe-brane is parallel to the interface, and generically does not intersect it. Thus, for a brane crossing the interface we have $k>0$, hence the action density is finite.

Another example of this is an F-string connecting a positive brane to a negative D3 brane beyond the interface, representing a half-BPS ``$W$ fermion'' on the Coulomb branch of the supergroup. Since $n_{\parallel} = n_{\perp} = 1$, the Nambu-Goto action $S_{\rm NG}=-\frac{1}{2 \pi \ls^2} \int_{\Sigma} \de^2 \xi \sqrt{-\det (\Sigma^{\ast} g)}$ does not depend on the warp factor $H$, and the $W$ mass is finite and uncorrected.

This suggests that supersymmetric probes can pass through the $H=0$ interface, and raises the question of what lies beyond it. To answer this question, we first consider the negative D0 brane solution, obtained by setting $p=0$ in~(\ref{eqn:negbranebkg}). The type IIA background~(\ref{eqn:negbranebkg}) lifts to an eleven-dimensional metric
\begin{equation}
\de s_{11}^2 = e^{-\frac{2}{3} \Phi} \de s_{10}^2 + e^{\frac{4}{3} \Phi} (\de y+A_1)^2
\end{equation}
where $y \cong y+2 \pi \ell_{\text{P}}$ with $\ell_\text{P}$ the eleven-dimensional Planck-length. Using $A_1 = \gs^{-1} H^{-1} \de t$, we obtain:
\begin{equation} \label{eqn:D0ppwave}
\de s_{11}^2 = \de s_9^2 + 2 \de t \de y + H \de y^2
\end{equation}
after rescaling $y \to \gs^{-2/3} y$, $t \to \gs^{1/3} t$, $x^i \to \gs^{1/3} x^i$, where $x^i$ are the coordinates transverse to the D0 branes with $\de s_9^2 = \sum_i (\de x^i)^2$ and now $y \cong y+ 2 \pi \gs^{2/3} \ell_{\text{P}}$.\footnote{As in~(\ref{eqn:negbranebkg}), our conventions are self-consistent, but differ slightly from the usual textbook treatment in the placement of factors of $g_s$.} 

\begin{figure}
\begin{center}
\includegraphics[height=2.75in]{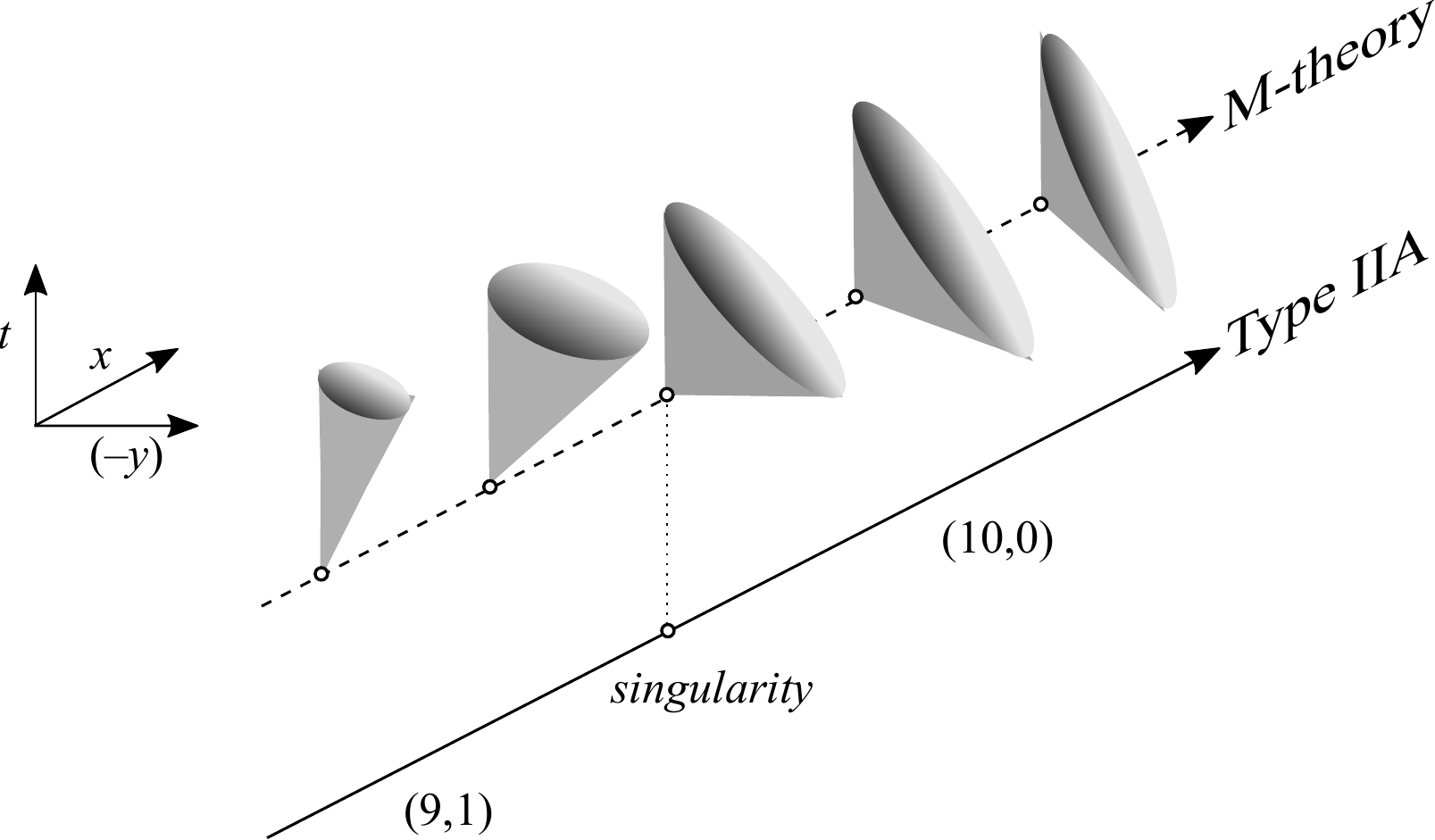} 
\caption{The forward light cone in the $y$--$t$ plane of the M-theory pp-wave background $\de s_{11}^2 = \de s_9^2 + 2 \de t \de y + H \de y^2$ as a function of some transverse coordinate $x$. When $\partial_y$ becomes null, the signature of the compact circle changes from spacelike to timelike, resulting in a naked singularity and a change of spacetime signature in the type IIA description.}
\label{fig:signaturechange}
\end{center}
\end{figure}

This is a pp-wave background representing momentum around the M-theory circle. As a pp-wave, all curvature invariants vanish identically. More importantly, the metric is smooth at the $H=0$ interface, allowing us to pass beyond it. For $H<0$, (\ref{eqn:D0ppwave}) describes M-theory compactified on a \emph{time-like} circle, resulting in a ten-dimensional limit with spacetime signature $(10,0)$, which we identify as the far side of the singular interface discussed above, see Figure~\ref{fig:signaturechange}. The physics inside this ``bubble'' surrounding the negative D0 branes is quite different! For instance, fundamental strings arise from M2 branes---with worldvolume signature $(2,1)$---wrapping a timelike circle, hence they are \emph{Euclidean} strings, with worldvolume signature $(2,0)$. On the other hand, Euclidean D2 branes are absent because they would lift to M2 branes with worldvolume signature $(3,0)$, which do not occur in M-theory.

Thus, negative D0 branes are intimately connected to timelike compactifications of M-theory.\footnote{As F-strings are T-dual to a pp-wave background, negative F-strings induce a change of signature in much the same way as negative D0 branes.} Since D0 branes are related to all other branes in type II string theory and M-theory via string dualities, a similar conclusion applies to any negative brane: the existence of negative branes is directly tied to the consistency of timelike compactification in string theory and M-theory. For each type of negative brane, the nature of the bubble surrounding the brane will be different. In order to derive these differences, we will use Hull's results on timelike compactification of string theory~\cite{Hull:1998ym}, reviewed in~\S\ref{sec:diffsigs}.

\subsection{Singularity crossing} \label{subsec:crossing}

Before diving into this analysis, we preview the results via a useful heuristic argument.
Neglecting any possible curvature corrections, let us take the background~(\ref{eqn:negbranebkg}) near the $H=0$ interface seriously for the time being. To continue past the singularity, we analytically continue the background as a function of $H$, avoiding the singularity at $H=0$.\footnote{see~\cite{Bars:2011aa,Araya:2015fva} for a discussion of singularity-crossing in cosmology} We obtain:
\begin{equation} \label{eqn:negbranecontinued}
\begin{split}
\de s^2 &= \omega^{-1} \bar{H}^{-\half} \de s^2_{p+1}+ \omega \bar{H}^{\half} \de s^2_{9-p} \,,\\
e^{-2 \Phi} &= \omega^{p-3} \gs^{-2} \bar{H}^{\frac{p-3}{2}} \,,\\
F_{p+2} &= -\gs^{-1} \de \bar{H}^{-1} \wedge \Omega_{p+1} \,,
\end{split}
\end{equation}
where $\bar{H} \equiv - H$ is positive in the region beyond the interface and $\omega = \pm i$, depending on which way we encircle $H=0$ in the complex plane. While the metric at first appears imaginary, we can remove an overall factor using a Weyl transformation, leaving the real metric
\begin{equation}
\label{eqn:redefmetric}
\de s^2 = -\bar{H}^{-\half} \de s^2_{p+1} + \bar{H}^{\half} \de s^2_{9-p} \,,
\end{equation}
up to an arbitrary overall sign. A similar field redefinition can be used to remove the complex factor in front of the dilaton profile in~(\ref{eqn:negbranecontinued}). The resulting background is real, but in spacetime signature $(10-p,p)$ instead of $(9,1)$. For the special case $p=0$, this is precisely the background obtained by dimensionally reducing the metric~(\ref{eqn:D0ppwave}) on the far side of the $H=0$ interface, where the overall sign in~(\ref{eqn:redefmetric}) was chosen to agree with~(\ref{eqn:D0ppwave}). Moreover, the field redefinitions change some of the signs in the supergravity action so that it matches a timelike dimensional reduction of M-theory. 

This heuristic reasoning can be extended to BPS probe branes crossing the singular interface. For instance, the $W$ fermion considered above arises from a string with $(1,1)$ worldsheet signature. After crossing the interface into signature $(10-p,p)$ the timelike worldsheet coordinate becomes spacelike, resulting in a worldsheet signature of $(2,0)$, in agreement with the observation about Euclidean strings given previously.

More generally, this kind of ``singularity crossing'' argument suggests that negative D$p$ branes live in a string theory of spacetime signature $(10-p,p)$, within which their worldvolume signature is $(1,p)$. After reviewing Hull's work, we will show that this is indeed the case.

\section{String Theory and M-Theory in Different Signatures} \label{sec:diffsigs}

The starting point of Hull's analysis~\cite{Hull:1998ym} is to T-dualize type IIA or type IIB string theory on a timelike circle~\cite{Moore:1993zc,Moore:1993qe,Hull:1998vg}. As this introduces a closed timelike curve (CTC), one should expect the results to be exotic. Indeed, the T-dual of type IIA (IIB) string theory on a timelike circle is not ordinary type IIB (IIA) string theory, but rather a variant with a different spectrum of branes.

Recall that D$p$ branes in ordinary string theory have worldvolume signature $(p,1)$. ``Euclidean branes,'' with worldvolume signature $(p+1,0)$ are not part of the theory.\footnote{Euclidean branes can appear as instantons after Wick-rotation, but are not present as stable physical objects in the $(9,1)$ Lorentzian spacetime. However, it was suggested in~\cite{Gutperle:2002ai} that Euclidean branes (``S-branes'') can appears as unstable solutions of appropriate worldvolume theories.} For instance, extremal black-brane solutions to the low-energy supergravity action have Lorentzian worldvolume signature; extremal solutions with a Euclidean world-volume require a different sign for the $F_{p+2}$ kinetic term.\footnote{To see why the sign of $|F_{p+1}|^2$ affects the existence of extremal solutions, consider the ordinary Reissner-Nordstr\"om black hole in four dimensions.
The outer and inner horizons are located at $r_\pm = M \pm \sqrt{M^2 - Q^2}$ in appropriate units. Changing the sign of $|F_2|^2$ takes $Q^2 \to - Q^2$, hence $r_-<0$ and there is no inner horizon and no extremal solutions.} 

However, T-dualizing type II string theory along a timelike circle removes the timelike direction from the worldvolume of every D-brane. Thus, the T-dual contains only Euclidean D-branes! To determine the available NS branes, recall that the T-dual of a long string with neither momentum nor winding around the compact circle is another long string, whereas the T-dual of an NS5 brane wrapping the compact circle is another NS5 brane wrapping the compact circle. Consequently, these exotic string theories appear to have Euclidean D-branes but Lorentzian F-strings and NS5 branes.

To keep track of the brane spectrum, we invent a new notation (different from~\cite{Hull:1998ym}). As usual, we denote type II theories with D0, \ldots, D8 branes as type IIA variants and those with D1, \ldots, D7 branes as type IIB variants. To denote the available brane signatures, we write IIA$^{++}$, IIA$^{+-}$, IIA$^{-+}$, etc., where the first sign indicates whether fundamental strings are Lorentzian ($+$) or Euclidean ($-$) and the second indicates whether D1 / D2 branes are Lorentzian or Euclidean. Similarly, we write M$^+$ (M$^-$) for the M-theory variant with Lorentzian (Euclidean) M2 branes. When we wish to be explicit, we denote the spacetime signature with a subscript, e.g.\ IIA$^{+-}_{9,1}$ for a IIA variant in signature $(9,1)$ with Lorentzian strings and Euclidean D2 branes. As we will see, this notation --- summarized in Table~\ref{tab:exoticstringsnotation} --- is sufficient to describe all the theories found by Hull. In particular, the signatures of the remaining branes and the low energy effective action are fixed once these low-dimensional branes are specified.

\begin{table}
\newcommand{\Lor}{Lor}
\newcommand{\Euc}{Euc}
\begin{center}
\begin{tabular}{ccc}
	\begin{tabular}{c|cc} 
		 & F1 & D2 \\\hline
		IIA$_{s,t}^{++}$ &  \Lor &  \Lor \\
		IIA$_{s,t}^{+-}$ & \Lor & \Euc  \\
		IIA$_{s,t}^{-+}$ & \Euc &  \Lor \\
		IIA$_{s,t}^{--}$ & \Euc  & \Euc 
	\end{tabular} &
	\begin{tabular}{c|cc} 
		 & F1 & D1 \\\hline
		IIB$_{s,t}^{++}$ & \Lor &  \Lor \\
		IIB$_{s,t}^{+-}$ & \Lor &  \Euc \\
		IIB$_{s,t}^{-+}$ & \Euc &  \Lor \\
		IIB$_{s,t}^{--}$ & \Euc &  \Euc \\
	\end{tabular} &
	\begin{tabular}{c|c} 
		 & M2 \\\hline
		M$_{s,t}^+$ & \Lor \\
		M$_{s,t}^-$ & \Euc
	\end{tabular}
\end{tabular}
\end{center}
\caption{Notation for exotic string theories and M-theories, where $(s,t)$ denotes the spacetime signature. The superscript signs $+$ and $-$ denote whether certain branes are Lorentzian (\Lor) or Euclidean (\Euc), respectively. For spacetime signatures other than $(D-1,1)$, ``Euclidean'' (``Lorentzian'') indicates an even (odd) number of worldvolume times. IIA$^{++}_{9,1}$, IIB$^{++}_{9,1}$, and M$^+_{10,1}$ denote the usual string and M-theories.}
\label{tab:exoticstringsnotation}
\end{table}

To keep score in the following discussion, we refer the reader to Figures~\ref{fig:dualityweb1}, \ref{fig:dualityweb2} and Table~\ref{tab:branetable}, depicting the results of Hull's analysis in our notation. In the following discussion, we refer to IIA$_{9,1}^{++}$ and IIB$_{9,1}^{++}$ (M$_{10,1}^+$) as ``ordinary'' string theories (M-theory). We call the new theories introduced by Hull ``exotic'' string theories (M-theories) when we wish to distinguish them from their ordinary counterparts.

It should be emphasized that there may be more than one string theory with the spectrum of branes indicated by, e.g., IIB$^{+-}_{9,1}$.  Different sequences of dualities which lead to the same brane spectrum can in principle lead to distinct string theories, and there may be other data which is important for the non-perturbative definition of the theory, such as discrete theta angles in gauge theory~\cite{Aharony:2013hda} or string theory~\cite{Sethi:2013hra}. Thus, conservatively our theory labels indicate classes of string theories, which may contain more than one member. Checking whether this occurs is an interesting direction for future research.

\subsection{The duality web} \label{subsec:dualityweb}

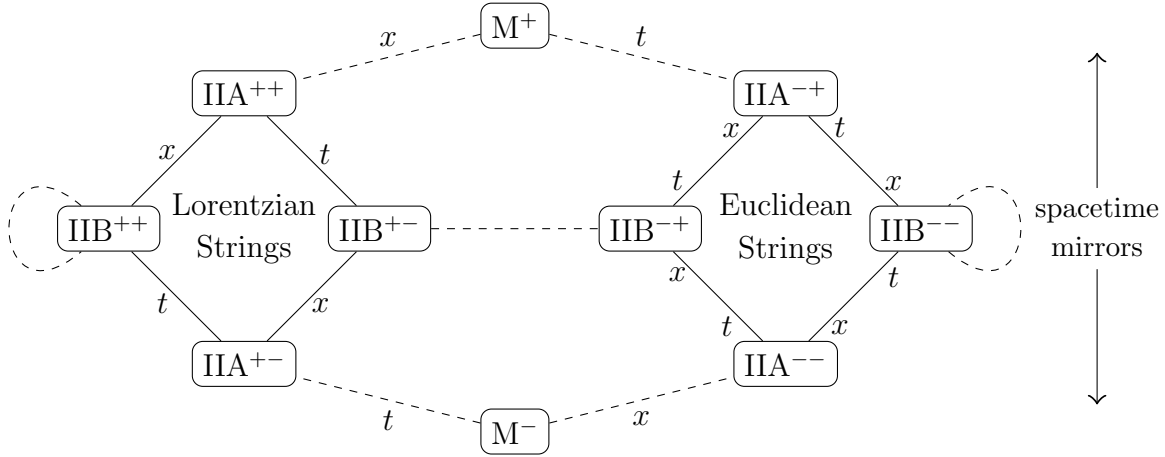
\begin{figure}
	\begin{center}
		\begin{tikzpicture}[scale=1.8]
			\node[rectangle,rounded corners,draw](M+) at (0,-1.5) {M$^+$};
			\node[rectangle,rounded corners,draw](IIA++) at (-2,-2) {IIA$^{++}$};
			\node[rectangle,rounded corners,draw](IIA-+) at (2,-2) {IIA$^{-+}$};
			\node[rectangle,rounded corners,draw](IIB++) at (-3,-3) {IIB$^{++}$};
			\node[rectangle,rounded corners,draw](IIB+-) at (-1,-3) {IIB$^{+-}$};
			\node[rectangle,rounded corners,draw](IIB-+) at (1,-3) {IIB$^{-+}$};
			\node[rectangle,rounded corners,draw](IIB--) at (3,-3) {IIB$^{--}$};
			\node[rectangle,rounded corners,draw](IIA+-) at (-2,-4) {IIA$^{+-}$};
			\node[rectangle,rounded corners,draw](IIA--) at (2,-4) {IIA$^{--}$};
			\node[rectangle,rounded corners,draw](M-) at (0,-4.5) {M$^-$};
			\node[align=center](LS) at (-2,-3) {Lorentzian\\Strings};
			\node[align=center](ES) at (2,-3) {Euclidean\\Strings};
			\draw[dashed] (M+) to node[above,midway]{$x$}(IIA++);
			\draw[dashed] (M+) to node[above,midway]{$t$} (IIA-+);
			\draw[dashed] (M-) to node[below,midway]{$t$} (IIA+-);
			\draw[dashed] (M-) to node[below,midway]{$x$} (IIA--);
			\draw[dashed] (IIB+-) to (IIB-+);
			\draw[] (IIA++) to node[above,pos=0.6]{$x$} (IIB++) to node[,below,pos=.35] {$t$}(IIA+-) to node[,below,pos=.6]{$x$}(IIB+-) to node[,above,pos=.35]{$t$}(IIA++);
			\draw[] (IIA-+) to node[above,pos=.35]{$x$} node[above,pos=0.95]{$t$} (IIB-+) to node[below,pos=.05]{$x$} node[below,pos=.6]{$t$}(IIA--) to node[below,,pos=.35]{$x$} node[below,pos=.95]{$t$}(IIB--) to node[above,pos=.05]{$x$} node[above,,pos=.65]{$t$} (IIA-+);
			\draw[dashed,out=140,in=220,looseness=7] (IIB++) to (IIB++);
			\draw[dashed,out=140-180,in=220-180,looseness=7] (IIB--) to (IIB--);				
			\draw[big arrow] (4.3,-2.7) to (4.3,-1.7);
			\draw[big arrow] (4.3,-3.3) to (4.3,-4.3);
			\node at (4.3,-3) {\small $\begin{array}{c} \text{spacetime} \\\text{mirrors} \end{array}$};
		\end{tikzpicture}
		\caption{T-dualities (solid lines) and S-dualities (dashed lines) relating type II string theories and M-theory. In this diagram we suppress the spacetime signature for simplicity (cf.\ Figure~\ref{fig:dualityweb2}). The label $x$ ($t$) indicates dualities arising from compactification on a spatial (timelike) circle. The left (right) diamond consists of theories with Lorentzian (Euclidean) F-strings. Theories above and below the center line are related by exchanging space and time directions, see (\ref{eqn:exchangepairs}).}
		\label{fig:dualityweb1}
	\end{center}
\end{figure}

In our notation, the timelike T-dualities discussed above are
\begin{equation}
\mathrm{IIA}^{++}_{9,1} \longleftrightarrow \mathrm{IIB}^{+-}_{9,1} \qquad,\qquad \mathrm{IIB}^{++}_{9,1} \longleftrightarrow \mathrm{IIA}^{+-}_{9,1} \,,
\end{equation}
where $\mathrm{IIA}^{++}_{9,1}$ and $\mathrm{IIB}^{++}_{9,1}$ denote the usual type IIA and type IIB string theories, whereas $\mathrm{IIA}^{+-}_{9,1}$ and $\mathrm{IIB}^{+-}_{9,1}$ denote the newly-discovered theories with all Euclidean D-branes. If we further T-dualize one of the latter on a \emph{spacelike} circle then the D-branes are still Euclidean, so it appears that $\mathrm{IIA}^{+-}_{9,1}$ and $\mathrm{IIB}^{+-}_{9,1}$ are related by ordinary (spacelike) T-duality. The four type II string theories related by spacelike and timelike T-duality are depicted in the lefthand diamond of Figure~\ref{fig:dualityweb1}.

Next, consider the $\gs \to \infty$ limit of $\mathrm{IIB}^{+-}_{9,1}$.\footnote{Here we assume that the RR axion $C_0 = 0$ for simplicity.} As usual, S-duality will exchange the fundamental string with the D1-brane, and the D5-brane with the NS5-brane, leaving the D3-brane invariant. Since the F-string and NS5-brane are Lorentzian in $\mathrm{IIB}^{+-}_{9,1}$, whereas the D1, D3 and D5 are Euclidean, the new theory has Euclidean strings and NS5-branes, Lorentzian D1 and D5-branes, and Euclidean D3-branes. Thus, it is $\mathrm{IIB}^{-+}_{9,1}$ in our nomenclature. Tracking D7-branes through the S-duality requires an F-theory description; instead, we will determine their signature once the role of further T-dualities is understood.

Consider the T-dual of $\mathrm{IIB}^{-+}_{9,1}$ compactified on a spacelike circle. If we assume that this is dual to some theory ``X'' compactified on another spacelike circle, then by wrapping a D5-brane on the compact circle we infer that X has Lorentzian D4-branes. However, if we instead consider a D3-brane transverse to the compact circle, we would conclude that X has Euclidean D4-branes! Both cannot be true, because reversing the chain of T- and S-dualities would then incorrectly imply that ordinary string theory has both Euclidean and Lorentzian D-branes.

The resolution is that $\mathrm{IIB}^{-+}_{9,1}$ compactified on a spacelike circle must be T-dual to X compactified on a \emph{timelike} circle. In this case, both arguments imply that X has Lorentzian D4-branes. Wrapping a D3-brane on the compact circle, we conclude that X has Euclidean D2-branes, hence it is IIA$^{--}_{8,2}$ in our nomenclature.

By the same argument, consistency with T-duality will require that in $\mathrm{IIB}^{-+}_{9,1}$, IIA$^{--}_{8,2}$ and any further T-duals, the D$p$-branes alternate between Euclidean and Lorentzian as $p \to p+2$. This implies, for instance, that $\mathrm{IIB}^{-+}_{9,1}$ has Euclidean D7-branes. Moreover, all T-duals must share the property that the T-dual of some theory X on a spatial circle is another theory Y on a timelike circle and vice versa, so that the spacetime signature changes with each T-duality. This was derived in~\cite{Hull:1998ym} from the worldsheet theory, and is a general property of string theories with Euclidean strings.

As we venture beyond signature $(9,1)$, worldvolume signatures beyond Euclidean --- $(p+1,0)$ --- and Lorentzian --- $(p,1)$ --- become available. For instance, T-dualizing on a spatial circle transverse to a D1-brane in $\mathrm{IIB}^{-+}_{9,1}$, we learn that IIA$^{--}_{8,2}$ has D2-branes with signature $(1,2)$ as well as the $(3,0)$ Euclidean ones. More generally, we observe that whenever a D-brane with signature $(p,q)$ is available, so is one with $(p-2,q+2)$ and vice versa, so long as it will ``fit'' in the spacetime signature. This is obviously true in the $(9,1)$ theories, whereas it is readily seen to be preserved under T-duality and S-duality. Because of this, we will sometimes abuse terminology and call branes with an even-number of worldvolume times ``Euclidean'' and those with an odd number ``Lorentzian.'' Consequently, the $(1,2)$ D2-brane is ``Euclidean.''

\begin{figure}
	\begin{center}
		\begin{tikzpicture}[scale=1.9]
			\node(IIA++91) at (-2.2,-1.7) {\scriptsize $(9,1)$};
			\node(IIA++55) at (-2.2,-2.1) {\scriptsize $(5,5)$};
			\node(IIA++19) at (-2.2,-2.5) {\scriptsize $(1,9)$};
			\node(IIA+-91) at (-2.2,-4.3) {\scriptsize $(9,1)$};
			\node(IIA+-55) at (-2.2,-3.9) {\scriptsize $(5,5)$};
			\node(IIA+-19) at (-2.2,-3.5) {\scriptsize $(1,9)$};
			\node(IIB++91) at (-3.6,-3) {\scriptsize $(9,1)$};
			\node(IIB++55) at (-3.15,-3) {\scriptsize $(5,5)$};
			\node(IIB++19) at (-2.7,-3) {\scriptsize $(1,9)$};
			\node(IIB+-91) at (-0.8,-3) {\scriptsize $(9,1)$};
			\node(IIB+-55) at (-1.25,-3) {\scriptsize $(5,5)$};
			\node(IIB+-19) at (-1.7,-3) {\scriptsize $(1,9)$};
			\draw[] (IIA++91) to (IIB++91) to (IIA+-91) to (IIB+-91) to (IIA++91);
			\draw[] (IIA++55) to (IIB++55) to (IIA+-55) to (IIB+-55) to (IIA++55);
			\draw[] (IIA++19) to (IIB++19) to (IIA+-19) to (IIB+-19) to (IIA++19);
			\node(IIA-+100) at (2.25625,-1.6) {\scriptsize $(10,0)$};
			\node(IIB-+91) at (0.7375,-2.95) {\scriptsize $(9,1)$};
			\node(IIA--82) at (2.14375,-4.2) {\scriptsize $(8,2)$};
			\node(IIB--73) at (3.4375,-3.05) {\scriptsize $(7,3)$};
			\node(IIA-+64) at (2.25625,-2.0) {\scriptsize $(6,4)$};
			\node(IIB-+55) at (1.1875,-2.95) {\scriptsize $(5,5)$};
			\node(IIA--46) at (2.14375,-3.8) {\scriptsize $(4,6)$};
			\node(IIB--37) at (2.9875,-3.05) {\scriptsize $(3,7)$};
			\node(IIA-+28) at (2.25625,-2.4) {\scriptsize $(2,8)$};
			\node(IIB-+19) at (1.6375,-2.95) {\scriptsize $(1,9)$};
			\node(IIA--010) at (2.14375,-3.4) {\scriptsize $(0,10)$};
			\draw[] (IIA-+100) to (IIB-+91) to (IIA--82) to (IIB--73) to (IIA-+64) to (IIB-+55) to (IIA--46) to (IIB--37) to (IIA-+28) to (IIB-+19) to (IIA--010);
			\draw[dashed] (IIB+-91) to[out=-13,in=197] (IIB-+91);
			\draw[dashed] (IIB+-55) to[out=-18,in=202] (IIB-+55);
			\draw[dashed] (IIB+-19) to[out=-23,in=207] (IIB-+19);
			\draw[dashed] (IIB++19) to[out=120,in=240,looseness=20] (IIB++19);
			\draw[dashed] (IIB++55) to[out=120,in=240,looseness=14] (IIB++55);
			\draw[dashed] (IIB++91) to[out=120,in=240,looseness=6] (IIB++91);
			\draw[dashed] (IIB--37) to[out=-60,in=60,looseness=14] (IIB--37);
			\draw[dashed] (IIB--73) to[out=-60,in=60,looseness=6] (IIB--73);
			\node(M+101) at (0,-1.3) {\scriptsize $(10,1)$};
			\node(M+65) at (0,-1.7) {\scriptsize $(6,5)$};
			\node(M+29) at (0,-2.1) {\scriptsize $(2,9)$};
			\draw[dashed] (IIA++91) to (M+101) to (IIA-+100);
			\draw[dashed] (IIA++55) to (M+65) to (IIA-+64);
			\draw[dashed] (IIA++19) to (M+29) to (IIA-+28);
			\node(M-110) at (0,-3.9) {\scriptsize $(1,10)$};
			\node(M-56) at (0,-4.3) {\scriptsize $(5,6)$};
			\node(M-92) at (0,-4.7) {\scriptsize $(9,2)$};
			\draw[dashed] (IIA+-91) to (M-92) to (IIA--82);
			\draw[dashed] (IIA+-55) to (M-56) to (IIA--46);
			\draw[dashed] (IIA+-19) to (M-110) to (IIA--010);
			\node[rectangle](M+) at (-0.4,-1.1) {M$^+$};
			\node[rectangle](M-) at (0.5,-4.8) {M$^-$};
			\node[rectangle,rounded corners](IIA++) at (-2.8,-1.5) {IIA$^{++}$};
			\node[rectangle,rounded corners](IIA-+) at (2.9,-1.4) {IIA$^{-+}$};
			\node[rectangle,rounded corners](IIA+-) at (-2.8,-4.5) {IIA$^{+-}$};
			\node[rectangle,rounded corners](IIA--) at (2.7,-4.4) {IIA$^{--}$};
			\node[rectangle,rounded corners](IIB+-) at (-.75,-2.6) {IIB$^{+-}$};
			\node[rectangle,rounded corners](IIB-+) at (.75,-2.6) {IIB$^{-+}$};
			\node[rectangle,rounded corners](IIB++) at (-3.7,-2.3) {IIB$^{++}$};
			\node[rectangle,rounded corners](IIB--) at (3.5,-2.3) {IIB$^{--}$};			
		\end{tikzpicture}
		\caption{T-dualities (solid lines) and S-dualities (dashed lines) relating type II strings theories and M-theory, keeping track of spacetime signature. See Figure~\ref{fig:dualityweb1} for labels distinguishing timelike and spacelike T- and S-dualities.\label{fig:dualityweb2}}
	\end{center}
\end{figure}
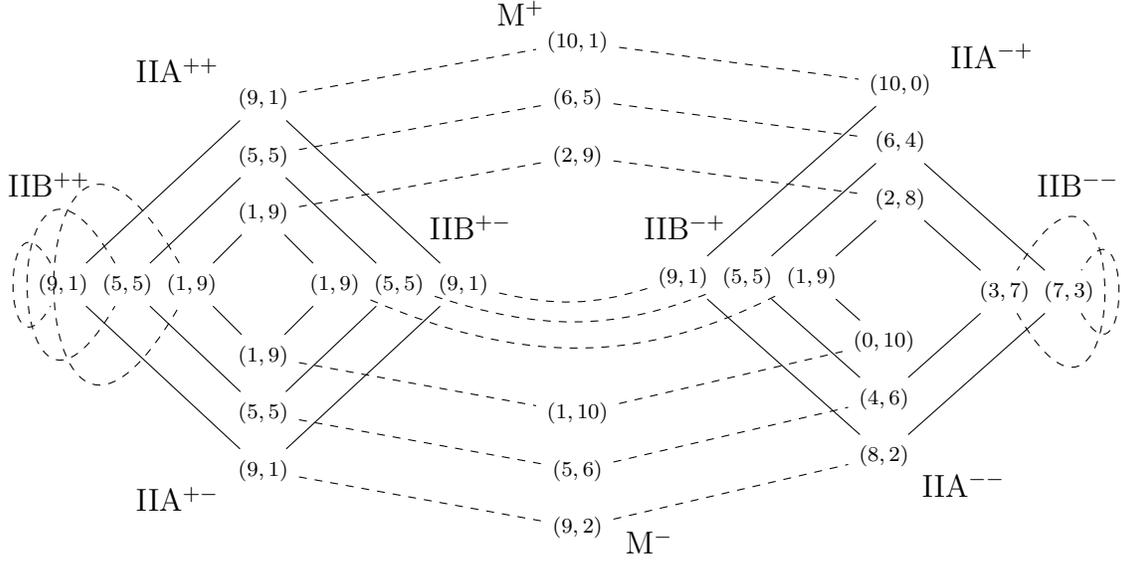

Having understood T-duality for Euclidean string theories, we can now fill out the remaining string theories found by Hull. T-dualizing $\mathrm{IIB}^{-+}_{9,1}$ on a \emph{timelike} circle, we obtain a theory with signature $(10,0)$ and Euclidean D4-branes. This is $\mathrm{IIA}^{-+}_{10,0}$ in our nomenclature.\footnote{This is also the theory which describes the inside of the bubble surrounding negative D0-branes.} Note that this theory has no D2-branes at all, but is labeled as a theory with Lorentzian D2-branes, since there are no Euclidean D2-branes despite fitting into the spacetime signature, whereas Lorentzian D2-branes do not fit, explaining their absence. Starting with this theory and repeatedly T-dualizing along spatial circles, we obtain the chain of theories
\begin{equation} \label{eqn:TdualSigChange}
\begin{split}
\mathrm{IIA}^{-+}_{10,0} &\longrightarrow \mathrm{IIB}^{-+}_{9,1} \longrightarrow \mathrm{IIA}^{--}_{8,2}\longrightarrow \mathrm{IIB}^{--}_{7,3} \longrightarrow \mathrm{IIA}^{-+}_{6,4}\longrightarrow \mathrm{IIB}^{-+}_{5,5} \\
&\longrightarrow \mathrm{IIA}^{--}_{4,6} \longrightarrow \mathrm{IIB}^{--}_{3,7} \longrightarrow \mathrm{IIA}^{-+}_{2,8} \longrightarrow \mathrm{IIB}^{-+}_{1,9} \longrightarrow \mathrm{IIA}^{--}_{0,10} \,,
\end{split}
\end{equation}
as shown on the righthand side of Figure~\ref{fig:dualityweb2}.
We can condense this sequence by suppressing the spacetime signature, so that it becomes formally periodic
\begin{equation} \label{eqn:TdualNoSig}
\mathrm{IIA}^{-+} \longrightarrow \mathrm{IIB}^{-+} \longrightarrow  \mathrm{IIA}^{--} \longrightarrow \mathrm{IIB}^{--} \longrightarrow \mathrm{IIA}^{-+} \longrightarrow \ldots \,,
\end{equation}
which is the righthand diamond of Figure~\ref{fig:dualityweb1}.
In fact, the allowed brane signatures --- summarized in Table~\ref{tab:branetable} --- \emph{are} periodic up to the question of which branes fit inside the spacetime signature. The periodic sequence~(\ref{eqn:TdualNoSig}) has length four, hence --- as can be seen in (\ref{eqn:TdualSigChange}) --- whenever one of these string theories exists in spacetime signature $(p,q)$, it also exists in signature $(p+4k,q-4k)$. For instance, IIA$^{-+}$ occurs in signatures $(10,0)$, $(6,4)$ and $(2,8)$.

\begin{table}
\begin{center}
$\begin{array}{c}
	\begin{array}{c|c|cc} 
		 & \text{Spacetime} & \text{M2} & \text{M5} \\\hline
		\text{M}^+ & (10,1), (6,5), (2,9) & (+,-) & (-,-) \\
		\text{M}^- & (9,2), (5,6), (1,10) & (-,+) &(-,-) 
	\end{array}\\ \\
	\begin{array}{c|c|ccccccc}
		& \text{Spacetime} &\text{D0} &  \text{D2} &  \text{D4} &  \text{D6} &  \text{D8} &  \text{F1} &  \text{NS5}   \\\hline
		\text{IIA}^{++} & (9,1), (5,5), (1,9) & (+,-) & (+,-) & (+,-) & (+,-) &(+,-) &(-,-) &( -,-)  \\
		\text{IIA}^{+-} & (9,1), (5,5), (1,9) & (-,+)&(-,+)& (-,+)&(-,+) &(-,+) & (-,-) & (-,-) \\
		\text{IIA}^{-+} & (10,0), (6,4), (2,8) & (-,+) & (+,-) & (- ,+)& (+,-) & (-,+) &(+,+) &(-,-) \\
		\text{IIA}^{--} & (8,2), (4,6), (0,10)  & (+,-) & (-,+) & (+,-) &(-,+) &(+,-)& (+,+)&(-,-)
	\end{array}\\ \\
	\begin{array}{c|c|cccccccc} 
		& \text{Spacetime} & \text{D}(-1) & \text{D1} & \text{D3} &\text{D5} &\text{D7} &\text{D9} & \text{F1} &\text{NS5} \\\hline
		\text{IIB}^{++} & (9,1), (5,5), (1,9)  &\text{---}& (-,-) & (-,-) & (-,-)& (-,-)& \checkmark &(-,-)& (-,-) \\
		\text{IIB}^{+-} & (9,1), (5,5), (1,9) &\checkmark& (+,+) & (+,+) & (+,+) &(+,+) &  \text{---} &(-,-) & (-,-) \\
		\text{IIB}^{-+} & (9,1), (5,5), (1,9)  &\checkmark& (-,-) &(+,+ ) &(-,-) &(+,+) & \checkmark &(+,+ ) &(+,+)  \\
		\text{IIB}^{--} & (7,3), (3,7) & \text{---} & (+,+) &(-,-) &(+,+)&(-,-) & \text{---} &(+,+) &(+,+) 
	\end{array}
\end{array}$
\end{center}
\caption{The available worldvolume and spacetime signatures for various type IIA, type IIB and M-theory variants. Here $(+,-)$ indicates an even ($+$) number of worldvolume spatial directions and an odd ($-$) number of worldvolume timelike directions, etc. For D$(-1)$ and D9-branes we need only indicate whether the brane is present in the theory, since the former has no worldvolume and the latter fills spacetime.}
\label{tab:branetable}
\end{table}

We now consider the S-duals of these theories, beginning with the S-duals of the type IIB variants. Note that since $\mathrm{IIB}^{-+}_{9,1}$ has Euclidean NS5-branes, T-duality implies that all $\mathrm{IIB}^{-+}$ and $\mathrm{IIB}^{--}$ theories have Euclidean NS5-branes, whereas all $\mathrm{IIA}^{-+}$ and $\mathrm{IIA}^{--}$ theories have Lorentzian NS5-branes. Consider $\mathrm{IIB}^{--}_{7,3}$ and $\mathrm{IIB}^{--}_{3,7}$. Since the F-string, NS5-brane, D1-brane and D5-brane are all Euclidean, S-duality maps the spectrum of branes to itself, and we infer that these theories are self-dual. Conversely, we already saw that $\mathrm{IIB}^{-+}_{9,1}$ is S-dual to $\mathrm{IIB}^{+-}_{9,1}$. Similarly $\mathrm{IIB}^{-+}_{5,5}$ and $\mathrm{IIB}^{-+}_{1,9}$ are S-dual the new theories $\mathrm{IIB}^{+-}_{5,5}$ and $\mathrm{IIB}^{+-}_{1,9}$, respectively. T-dualizing these theories, we obtain $\mathrm{IIA}^{++}$, $\mathrm{IIA}^{+-}$ and $\mathrm{IIB}^{++}$ in signatures $(5,5)$ and $(1,9)$. Since $\mathrm{IIB}^{++}_{5,5}$ and $\mathrm{IIB}^{++}_{1,9}$ have only Lorentzian branes, S-duality maps the spectrum of branes to itself, and we infer that these theories are self-dual.

All that remains to be considered are the S-duals of the IIA variants, which will be variants of M-theory. As usual, the S-dual of IIA$^{++}_{9,1}$ is an eleven-dimensional theory with Lorentzian M2 and M5-branes, where the D0-brane in IIA corresponds to the KK mode on the M-theory circle and the D6-brane is the KK monopole in eleven dimensions (the Taub-NUT geometry). In our nomenclature, this is M$^+_{10,1}$. Consider instead the S-dual of IIA$^{-+}_{10,0}$. This theory has Euclidean F-strings, D0-branes and D4-branes and no D2, D6, or NS5-branes, which is consistent with the dimensional reduction of M$^+_{10,1}$ on a timelike circle.\footnote{Whereas the KK-mode on a spacelike circle is a particle with a timelike worldline, the KK-mode on a timelike circle is instead a particle with a \emph{spacelike} worldline, which explains why the D0-brane is Euclidean (spacelike) in this example.} Because IIA$^{++}$ (IIA$^{-+}$) also exists in signatures $(5,5)$ and $(1,9)$ (signatures $(6,4)$ and $(2,8)$), we conclude that the M-theory variants M$^+_{6,5}$ and M$^+_{2,9}$ should also exist as their S-duals, where now M2-branes can have signatures $(2,1)$ or $(0,3)$ and M5-branes $(5,1)$, $(3,3)$ or $(1,5)$, depending on what fits within the spacetime signature.

Finally, consider the S-dual of IIA$^{+-}_{9,1}$. This theory has Lorentzian F-strings and NS5-branes but Euclidean D0, D2 and D4-branes. This is consistent with the dimensional reduction of an eleven-dimensional theory in signature $(9,2)$ on a timelike circle, now with M2-branes of signature $(3,0)$ and $(1,2)$ and M5-branes of signature $(5,1)$. In our nomenclature this is M$^-_{9,2}$, where reduction on a spatial circle leads to IIA$^{--}_{8,2}$, and variants M$^-_{5,6}$ and M$^-_{1,10}$ are also possible by the same reasoning as above.

The complete set of S- and T-dualities are shown in Figure~\ref{fig:dualityweb1}, with the explicit spacetime signatures indicated in Figure~\ref{fig:dualityweb2}.

\medskip

The above classification has a two-fold redundancy: for each string theory or M-theory variant described above, there is a related theory obtained by exchanging the roles of space and time, flipping the spacetime and brane signatures from $(p,q)$ to $(q,p)$.\footnote{However, as in the rest of our discussion a specific theory of the type $\mathrm{M}^+_{s,t}$ may or may not be equivalent to a specific theory of the type $\mathrm{M}^-_{t,s}$, simply because each class of theories could contain more than one member. See~\cite{Duff:2006iy, Duff:2006ix} for a discussion of invariance under signature reversal $g \rightarrow - g$.} In particular, the various classes of theories are mapped to each other as follows (as illustrated in Figure~\ref{fig:dualityweb1}):
\begin{equation} \label{eqn:exchangepairs}
\mathrm{M}^+_{s,t} \longleftrightarrow \mathrm{M}^-_{t,s} \;\;,\;\; \mathrm{IIA}^{\alpha \beta}_{s,t} \longleftrightarrow \mathrm{IIA}^{\alpha (-\beta)}_{t,s} \;\;,\;\; \mathrm{IIB}^{\alpha \beta}_{s,t} \longleftrightarrow \mathrm{IIB}^{\alpha \beta}_{t,s} \,,
\end{equation}
preserving the brane spectrum shown in Table~\ref{tab:branetable}. Following Hull, we refer to the pairs of theories in~(\ref{eqn:exchangepairs}) as ``spacetime mirrors'', not to be confused with mirror symmetry of Calabi-Yau manifolds (an unrelated phenomenon). 

\subsection{The low-energy limit} \label{subsec:SGaction}

The type II string theory and M-theory variants considered above each correspond to a distinct low-energy effective supergravity. The supergravity effective action for the new theories can be derived from the known type IIA, type IIB, and eleven-dimensional supergravity actions by considering the relevant T- and S-dualities in the low-energy theory. This is done in Appendix~\ref{app:supergravity}. For the eleven-dimensional theories we find the low-energy effective action
\begin{equation}
S[\mathrm{M}^{\pm}] = \frac{1}{2 \kappa_{11}^2} \int \de^{11} x \sqrt{|\det g|} \left[\cR \mp \frac{1}{2} |F_4|^2\right] - \frac{1}{6} \int C_3 \wedge F_4 \wedge F_4 \,,
\end{equation}
where $F_4 = \de C_3$,  $| F_p |^2 \equiv \frac{1}{p!} F^{{\mu}_1 \ldots {\mu}_p}
  F_{{\mu}_1 \ldots {\mu}_p}$, and we omit the fermions for simplicity.

Likewise, for the ten dimensional theories we find the bosonic action
\begin{equation} \label{eqn:sugraaction0}
S = S_{\rm NS} + S_{\rm R} + S_{\rm CS}
\end{equation}
with
\begin{equation}
\label{eqn:sugraaction}
\begin{split}
S_{\rm NS}[\mathrm{IIA/B}^{\alpha \beta}] &= \frac{1}{2 \kappa_{10}^2} \int \de^{10} x \sqrt{|\det g|} \ e^{-2 \Phi}\! \left[\cR + 4(\nabla \Phi)^2 - \frac{\alpha}{2} |H_3|^2 \right] \,, \\
S_{\rm R}[\mathrm{IIA}^{\alpha \beta}] &= -\frac{1}{2 \kappa_{10}^2} \int \de^{10} x \sqrt{|\det g|} \left[\frac{\alpha \beta}{2} |F_2|^2 + \frac{\beta}{2} |\tilde{F}_4|^2 \right] \,, \\
S_{\rm R}[\mathrm{IIB}^{\alpha \beta}] &= -\frac{1}{2 \kappa_{10}^2} \int \de^{10} x \sqrt{|\det g|} \left[\frac{\alpha \beta}{2} |F_1|^2 + \frac{\beta}{2} |\tilde{F}_3|^2 + \frac{\alpha \beta}{4} |\tilde{F}_5|^2 \right] \,, \\
\end{split}
\end{equation}
where $\alpha,\beta = \pm$, $H_3 = d B_2$, $F_p = d C_{p-1}$, and $\tilde{F}_p = F_p - H_3 \wedge C_{p-3}$.
The Chern-Simons term can be taken to be independent of $\alpha, \beta$
\begin{equation} \label{eqn:sugraactionCS}
S_{\rm CS}[\mathrm{IIA}] = -\frac{1}{4 \kappa_{10}^2} \int B_2\wedge F_4 \wedge F_4 \;\;,\;\;
S_{\rm CS}[\mathrm{IIB}] =  -\frac{1}{4 \kappa_{10}^2} \int B_2 \wedge F_3 \wedge F_5 \,,
\end{equation}
since the overall signs of the Chern-Simons terms (as well as the relative signs in the definitions of $\tilde{F}_p$) are arbitrary up to redefinitions of the $p$-form potentials and spacetime parity reflection. For simplicity, we have omitted the mass term in the type IIA variants. The action for the type IIB variants is a pseudo-action, requiring the constraint $\tilde{F}_5 = \alpha \beta \star \tilde{F}_5$ in addition to the equations of motion.\footnote{The extra sign $\alpha \beta$ in the chirality constraint is required to match the $\tilde{F}_5$ Bianchi identity with the equations of motion.} In all cases, the low-energy effective action takes the same form in all allowed signatures once the class of theories (e.g.\ IIA$^{+-}$ or M$^-$) is specified.\footnote{This should hold for the terms involving fermions as well, consistent with the fact that the permissible reality and chirality conditions on $\mathrm{Spin}(p,q)$ spinors only depend on $(p-q) \bmod 8$, which is fixed for theories in the same class.}

We now comment on the relation between the spacetime mirrors of (\ref{eqn:exchangepairs}). It is straightforward to check that
\begin{equation} \label{eqn:mirrormatch}
S[M^-_{s,t}] = - S[M^+_{t,s}] (g_{\mu \nu} \to - g_{\mu \nu}) \,,
\end{equation}
so the actions are classically equivalent, but differ by an overall sign. At first, this sign difference appears to have quantum mechanical significance, suggesting that the loop expansions differ between the mirror theories. However, we show in~\S\ref{subsec:quantmirrors} that this sign difference is in fact required to compensate for a difference in the $\epsilon$ prescription induced by the signature change, $g_{\mu \nu} \to - g_{\mu \nu}$, hence the loop expansions match between the mirror theories. Analogous results apply to the IIA and IIB spacetime mirrors, except that in the latter case a spacetime parity flip is required alongside the signature flip to preserve the chirality of $C_4$.

\subsection{Consequences for negative branes} \label{subsec:negbraneresults}

We now use what we have learned about string dualities with timelike compactification to relate all possible negative branes to dynamic signature changes, confirming the heuristic arguments of~\S\ref{sec:signaturechange}. The reasoning is very simple. As we have already established, the negative D0-brane is related to a smooth pp-wave geometry in M-theory, in which the signature of the compact circle changes dynamically from spacelike to timelike. By applying further T- and S-dualities to both sides of the interface, we can generate the dynamical signature changes associated to other types of negative branes (which transform in the same way as the associated positive branes and T- and S-dualities).

\begin{figure}
\begin{center}
\includegraphics[width=\textwidth]{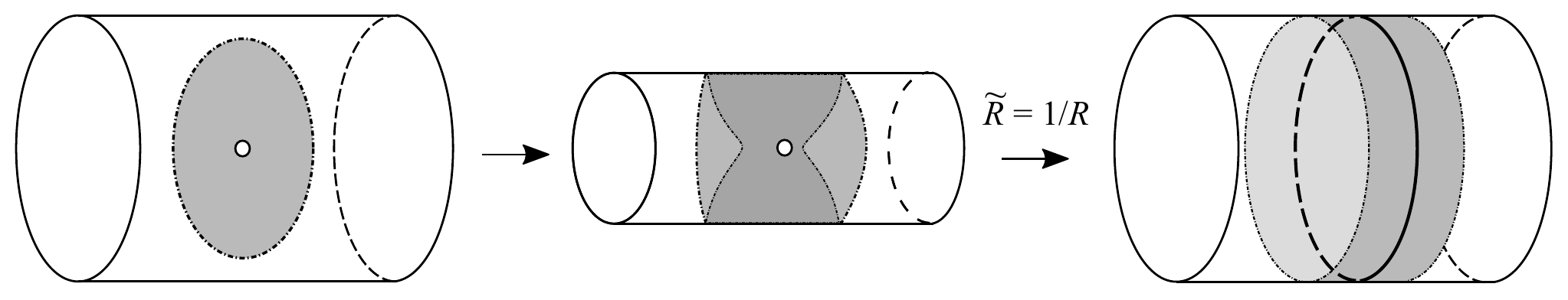}
\caption{T-duality relates negative D-branes of different dimensions. As the radius of the compact circle shrinks the inside of the bubble is compactified, and its T-dual can be found from Figures~\ref{fig:dualityweb1}, \ref{fig:dualityweb2}. This allows us to fix the exotic string theory inside the bubble by relating it to the negative D0-brane, whose strong coupling limit is a smooth geometry in M-theory.}
\label{fig:T-duality}
\end{center}
\end{figure}

For instance, compactifying one of the spatial directions transverse to the negative D0-brane and T-dualizing, we obtain a negative D1-brane. Applying the same transformation to the backreacted geometry, we find that the T-duality transforms the inside of the bubble from IIA$^{-+}_{10,0}$ to IIB$^{-+}_{9,1}$, as illustrated in Figure~\ref{fig:T-duality}. Proceeding in the same fashion, we can derive the dynamical signature changes associated to all types of negative D$p$-branes. For instance, we find that a negative D5-brane induces a dynamical signature change to IIB$^{-+}_{5,5}$. Applying S-duality, we conclude that a negative B-type NS5-brane induces a change to IIB$^{+-}_{5,5}$. T-dualizing one of the spatial directions on the NS5 worldvolume, we find that an A-type NS5-brane induces a change to IIA$^{++}_{5,5}$.\footnote{Because the worldvolume directions flip signature upon crossing the interface, this is a timelike T-duality inside the bubble.} Similar arguments allow us to fix the dynamical signature change associated to all possible types of negative branes, as summarized in Table~\ref{tab:bubbletable}.

\begin{table}
\begin{center}
\renewcommand*{\arraystretch}{1.2}
\setlength{\arraycolsep}{4pt}
$\begin{array}{c|cccccccc} 
	& \text{D0/D1} & \text{D2/D3} & \text{D4/D5} & \text{D6/D7} & \text{D8} & \text{F1/M2} & \text{NS5/M5} & \text{TN} \\\hline
	\text{IIA}   & \text{IIA}^{-+}_{10,0} & \text{IIA}^{--}_{8,2} & \text{IIA}^{-+}_{6,4} &  \text{IIA}^{--}_{4,6} & \text{IIA}^{-+}_{2,8} & \text{IIA}^{+-}_{9,1} & \text{IIA}^{++}_{5,5} &  \text{IIA}^{+-}_{5,5} \\
	\text{IIB}   & \text{IIB}^{-+}_{9,1} & \text{IIB}^{--}_{7,3} & \text{IIB}^{-+}_{5,5} &  \text{IIB}^{--}_{3,7} & \text{---} & \text{IIB}^{+-}_{9,1} & \text{IIB}^{+-}_{5,5} & \text{IIB}^{++}_{5,5} \\
	\text{M} & \text{---} & \text{---} & \text{---} & \text{---} & \text{---} & \text{M}^-_{9,2} & \text{M}^+_{6,5} & \text{M}^-_{5,6}
\end{array}$
\end{center}
\caption{The exotic string theory inside the bubble surrounding negative branes of various types in string and M-theory. Here D0/D1 denotes a negative D0 (D1)-brane for type IIA (IIB) string theory, etc., and ``TN'' denotes a Taub-NUT geometry with negative charge (reviewed in~\S\ref{subsec:geomeng}), which is related by T- and S-dualities to negative NS5 and D6-branes in string and M-theory, respectively.}
\label{tab:bubbletable}
\end{table}

These results are consistent with the heuristic ``singularity crossing'' argument introduced in~\S\ref{sec:signaturechange}. For instance, notice that the inside of the bubble has signature $(D-p,p)$ for a negative $p$-brane,\footnote{A TN-brane has codimension four, and counts as a fivebrane in type II string theory and a sixbrane in M-theory.} arising from reversing the signature of the worldvolume directions while preserving the signature of the remaining directions. This agrees with the singularity crossing argument because the backreacted metric generically takes the form $H^{-1} \de s_{p,1}^2 + \de s_{D-p-1,0}^2$ up to a Weyl transformation. To identify the string theory living inside the bubble, we place BPS probe branes across the interface and read off their signatures on the far side. For instance, consider a negative D$p$-brane and a probe D$q$-brane sharing $r+1$ spacetime directions with the negative brane and perpendicular to it in the others. The probe brane is BPS if $p+q-2r\equiv 0 \pmod 4$, whereas the signature of the brane within the bubble is $(q-r+1,r)$. Applying this formula to D1 and D2-branes, we conclude that the string theory inside the bubble is either\footnote{Assume the equivalence of spacetime mirror theories, we can apply~(\ref{eqn:exchangepairs}) to re-express the resulting string theory in the flipped signature $(p,10-p)$, so that the directions transverse (rather than parallel) to the brane flip signature at the interface. The two descriptions are physically equivalent.}
\begin{equation} \label{eqn:negDpbubble}
\text{IIA}^{-(-)^{\frac{p}{2}}}_{10-p,p} \qquad \text{or} \qquad \text{IIB}_{10-p,p}^{-(-)^{\frac{p-1}{2}}}\,,
\end{equation}
depending on whether $p$ is even or odd, where we use the fact that strings inside the bubble are Euclidean, as observed in~\S\ref{sec:signaturechange}. This is consistent with Table~\ref{tab:bubbletable}, where the signatures of the remaining BPS D-branes can be determined analogously.

Another way to identify the correct string theory is to apply the field redefinitions mandated by the singularity crossing to the low energy effective action. For a negative D$p$-brane (cf.\ (\ref{eqn:negbranebkg})), the metric and dilaton pick up factors
\begin{align} \label{eqn:crossing}
g_{\mu \nu} &\to \omega g_{\mu \nu}\,, & \det e_\mu^a &\to \omega^{4-p} \det e_\mu^a\,, & e^{-2 \Phi} &\to \omega^{p-3} e^{-2 \Phi} \,,
\end{align}
on crossing the singularity, where $\omega = \pm i$ depending on which way we go around the branch cut. Here we replace $\sqrt{|\det g|} = \det e_\mu^a$ to avoid branch cuts in the action, where $e_\mu^a$ is the vielbein and $\det e_\mu^a$ picks up a factor of $\omega$ ($\omega^{-1}$) for each direction which is transverse (parallel) to the brane. Applying this to the effective action for type II string theory and comparing with the results of~\S\ref{subsec:SGaction} we reproduce~(\ref{eqn:negDpbubble}), irrespective of the choice of branch cut. Similar reasoning can be used to reproduce the rest of Table~\ref{tab:bubbletable} using singularity crossing arguments. We leave this as an exercise for the interested reader.

Thus, accounting for backreaction, we conclude that negative branes are surrounded by a bubble containing an exotic string theory whose spacetime signature is in general not Lorentzian. Remarkably, although we began with a negative tension object, within this bubble the ``negative brane'' appears to have positive tension. To see this, consider the negative D0-brane, for which the warp factor is
\begin{equation} \label{eqn:negD0warp}
\bar{H} = -1 +\frac{(2 \sqrt{\pi} \ls)^{7}\, \Gamma\!\left(\frac{7}{2}\right)}{4 \pi} \sum_i \frac{\gs N_i^-}{|r - r_i|^{7}} \,,
\end{equation}
inside the IIA$^{-+}_{10,0}$ bubble. Near the negative brane, we have approximately
\begin{equation} \label{eqn:negD0warp2}
\bar{H} \simeq \frac{(2 \sqrt{\pi} \ls)^{7}\, \Gamma\!\left(\frac{7}{2}\right)}{4 \pi} \sum_i \frac{\gs N_i^-}{|r - r_i|^{7}}\,,
\end{equation}
but this is the same near-brane behavior that we expect for a D0-brane in IIA$^{-+}_{10,0}$, for which $H = 1 + (\ldots)/|r-r_i|^7$. To see that they are the same object, recall that the negative D0-brane descends from a pp-wave in M-theory
\begin{equation}
\de s_{11}^2 = \de s_9^2 + 2 \de t \de y - \bar{H} \de y^2 \,,
\end{equation}
but there cannot be more than one object corresponding to the same geometry. Since~(\ref{eqn:negD0warp}) and~(\ref{eqn:negD0warp2}) only differ far away from the brane (near the interface with IIA$^{++}_{9,1}$), we conclude that both correspond to the same object placed in different backgrounds, IIA$^{++}_{9,1}$ and IIA$^{-+}_{10,0}$ respectively.

All that remains to be established is that a D0-brane in IIA$^{-+}_{10,0}$ has positive tension, in some appropriate sense. We take the ADM definition of tension, which is positive because the second summand in~(\ref{eqn:negD0warp2}) is positive.

\begin{figure}
\begin{center}
\includegraphics[height=2.5in]{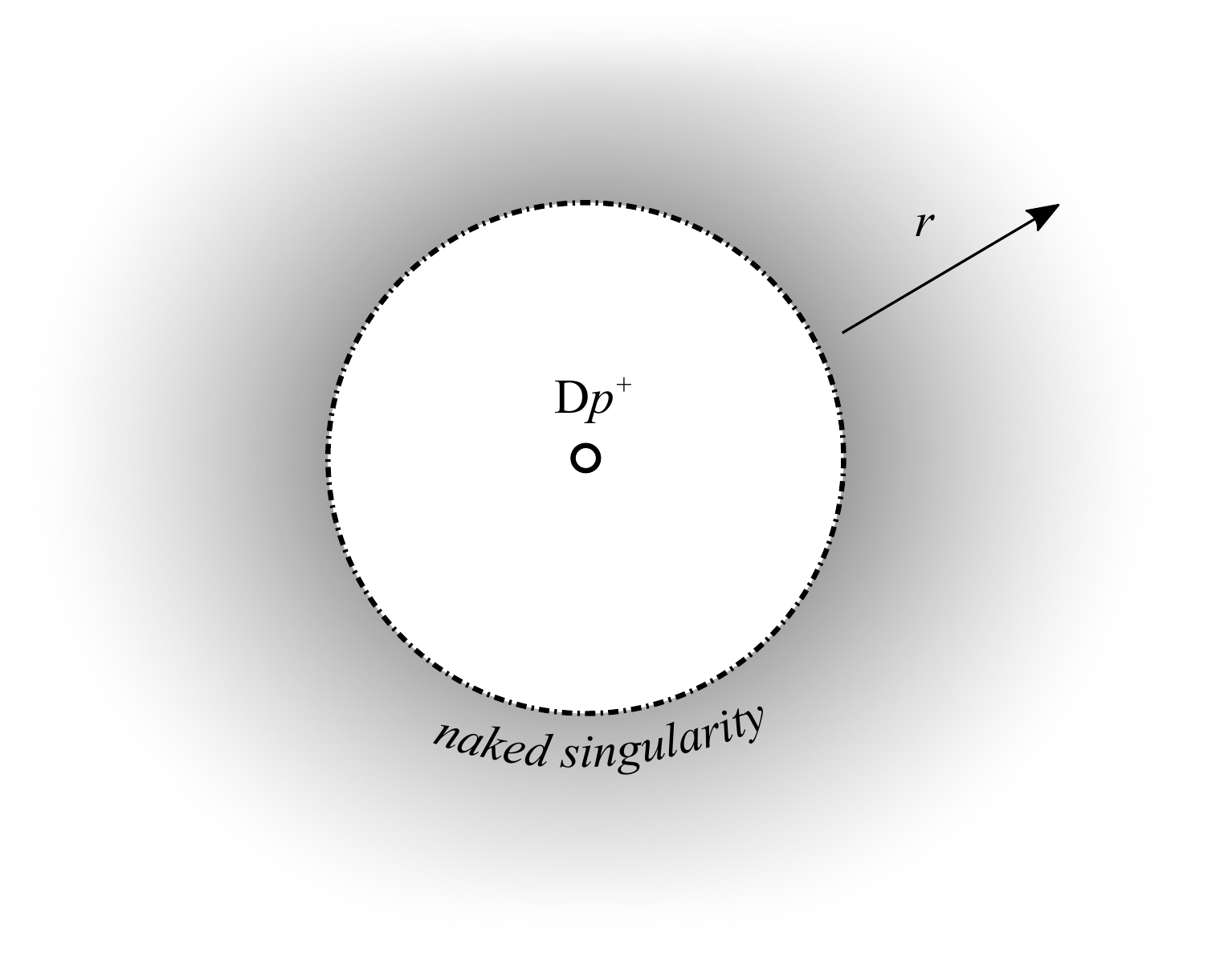}
\caption{Negative branes from the perspective of the exotic string theory living within the bubble. From this viewpoint, they are positive tension objects.}
\label{fig:insidebubble}
\end{center}
\end{figure}

This is suggests a possible resolution to the mysterious role of negative branes in ordinary string theories. Although they appear as exotic, negative tension objects in these theories, they are more naturally viewed as positive tension branes in an exotic string theory (see Figure~\ref{fig:insidebubble}), where the presence of the negative brane in ordinary string theory dynamically induces a change of signature to the exotic string theory.

\subsection{Signature-changing domain walls}

This dynamic change of signature can occur even without the presence of negative branes. Consider the eleven-dimensional metric
\begin{equation} \label{eqn:smoothwall}
\de s_{11}^2 = 2 \de t \de y +\frac{x}{L_{11}} \de y^2 + \de x^2+  \de s_8^2 \,,
\end{equation}
where $y \cong y + 2 \pi R_{11}$, and $L_{11}$ is some arbitrary length scale. This metric---a pp-wave that is everywhere smooth---is the same as the metric near the $H=0$ interface in~(\ref{eqn:D0ppwave}). Reducing to type IIA on the $y$ circle, we obtain the IIA$^{++}_{9,1}$ background
\begin{equation} \label{eqn:F2wall}
\begin{aligned}
\de s^2 &= -\left(\frac{x}{L_{10}}\right)^{-\half} \de t^2+ \left(\frac{x}{L_{10}}\right)^{\half} (\de x^2 + \de s^2_8) \,, &
e^{2 \Phi} &= \gs^{2} \left(\frac{x}{L_{10}}\right)^{\frac{3}{2}} \,,\\
F_{2} &= \frac{L_{10}}{\gs}\, \de t \wedge \frac{\de x}{x^2}  \,,
\end{aligned}
\end{equation}
for $x>0$, whereas for $x<0$ 
we obtain the IIA$^{-+}_{10,0}$ background
\begin{align} \label{eqn:F2wall2}
\de s^2 &= \left(-\frac{x}{L_{10}}\right)^{-\half} \de t^2+ \left(-\frac{x}{L_{10}}\right)^{\half} (\de x^2 + \de s^2_8) \,, &
e^{2 \Phi} &= \gs^{2} \left(-\frac{x}{L_{10}}\right)^{\frac{3}{2}} \,,
\end{align}
with $F_2$ the same as~(\ref{eqn:F2wall}). We conclude that the singular interface $x=0$ is a signature-changing domain wall. Since it is described by a smooth metric in M-theory, this domain wall is a sensible object in type IIA string theory, provided that the timelike compactification itself makes sense.\footnote{This geometry and its connection to negative tension D8-branes in type IIA string theory were studied in~\cite{Cornalba:2003kd}.}

We remark in passing that the metric~(\ref{eqn:smoothwall}) is actually flat, unlike the pp-wave metric~(\ref{eqn:D0ppwave}) describing the full backreaction of the negative D0 brane in M-theory. It can be written in light-cone form $\de s^2 = 2 \de T \de Y+\de X^2 + \de s_8^2$ in terms of the coordinates
\begin{align}
X &= x - \frac{y^2}{4 L_{11}}\,, & T &= t + \frac{x y}{2 L_{11}} - \frac{y^3}{24 L_{11}^2}\,, & Y &= y\,,
\end{align}
where the periodic identification $y\cong y+2 \pi R_{11}$ takes the form of a Poincare transformation
\begin{equation} \label{eqn:Poincarequot}
(T, X, Y) \cong \left( T + \beta X - \frac{\beta^2}{2} Y, X - \beta Y, Y
   \right) + 2 \pi R_{11} \left( - \frac{\beta^2}{6}, - \frac{\beta}{2}, 1 \right) \,,
\end{equation}
for $\beta \equiv \frac{\pi R_{11}}{L_{11}}$; note that a similar construction (without closed timelike curves) was considered in~\cite{Liu:2002ft}. It would be interesting to consider other Poincare quotients $x \cong \Lambda x + a$ and their dimensional reductions.\footnote{A necessary and sufficient condition to avoid fixed points is that $\Lambda$ has a unit eigenvector $\lambda \Lambda = \lambda$ such that $\lambda\cdot a \ne 0$. Using this observation, one can show that~(\ref{eqn:Poincarequot}) is one of only three families of orientation preserving smooth Poincare quotients of $\mathbb{R}_{2,1}$, with $\lambda \propto \de y$ null. The other two have spacelike (timelike) $\lambda$ and can be written as a boost (rotation) followed by a translation in the orthogonal direction. There are no non-trivial ($\Lambda \ne 1$) examples for $D<3$.}

Applying T- and S-dualities to both sides of the interface, we obtain a wide variety of signature-changing domain walls in both string theory and M-theory. As in the above example, each domain wall is characterized by the flux supporting it, of the form $F_{p+1} \propto \de x \wedge \Omega_p$, where the directions spanned by $\Omega_p$ reverse signature at the singular interface and the remaining directions are unaffected. The Taub-NUT geometry occurs as a special case, where
\begin{equation} \label{eqn:negTaubNUT}
\de s^2 = \de s_{D-4}^2+V \de s_3^2 + \frac{1}{V} (\de \theta+A)^2 \,.
\end{equation}
The one-form connection $A$ on the $U(1)$ bundle satisfies $\de V=\star_3 \de A$, so that $\Omega_{D-4} \wedge \de V$ plays the role of magnetic flux, where $V=x/L$ for a signature-changing domain wall.

\begin{figure}
\begin{center}
\includegraphics[height=2in]{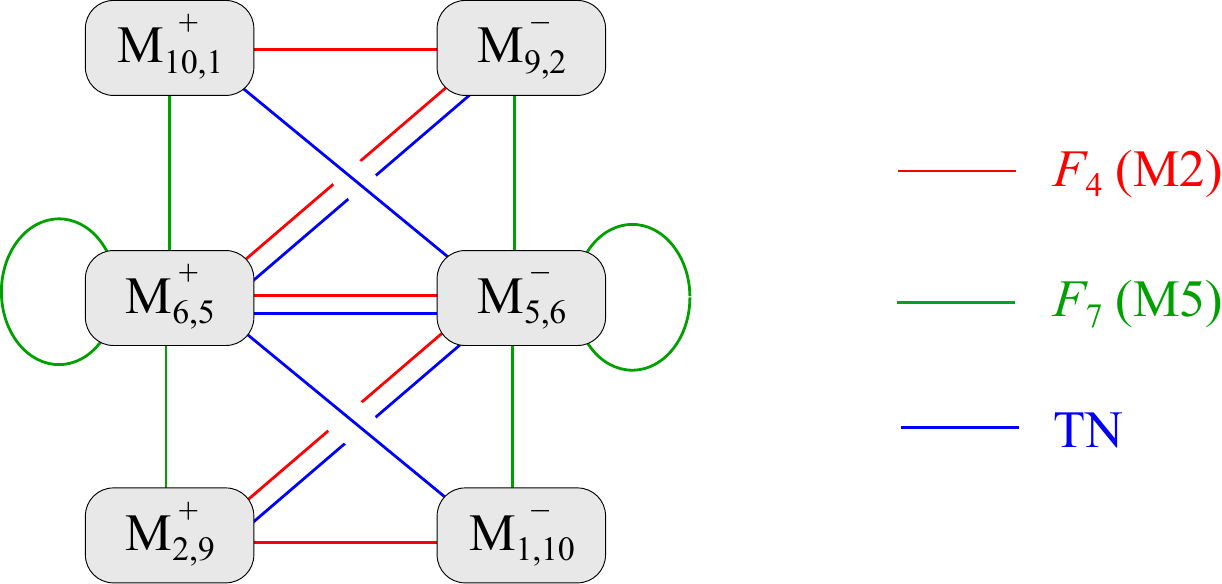}
\caption{The possible signature-changing domain walls in M-theory, classified by the theories they connect and the flux supporting them, where $F_7 = \star F_4$ indicates that the magnetic flux $F_7$ has a leg perpendicular to the interface, rather than the electric flux $F_4$. For each flux, we indicate the type of negative brane which sources it. TN denotes the Taub-NUT geometry with linear potential $V \propto x$.
}
\label{fig:Mthywalls}
\end{center}
\end{figure}

\begin{figure}
\begin{center}
\includegraphics[height=2in]{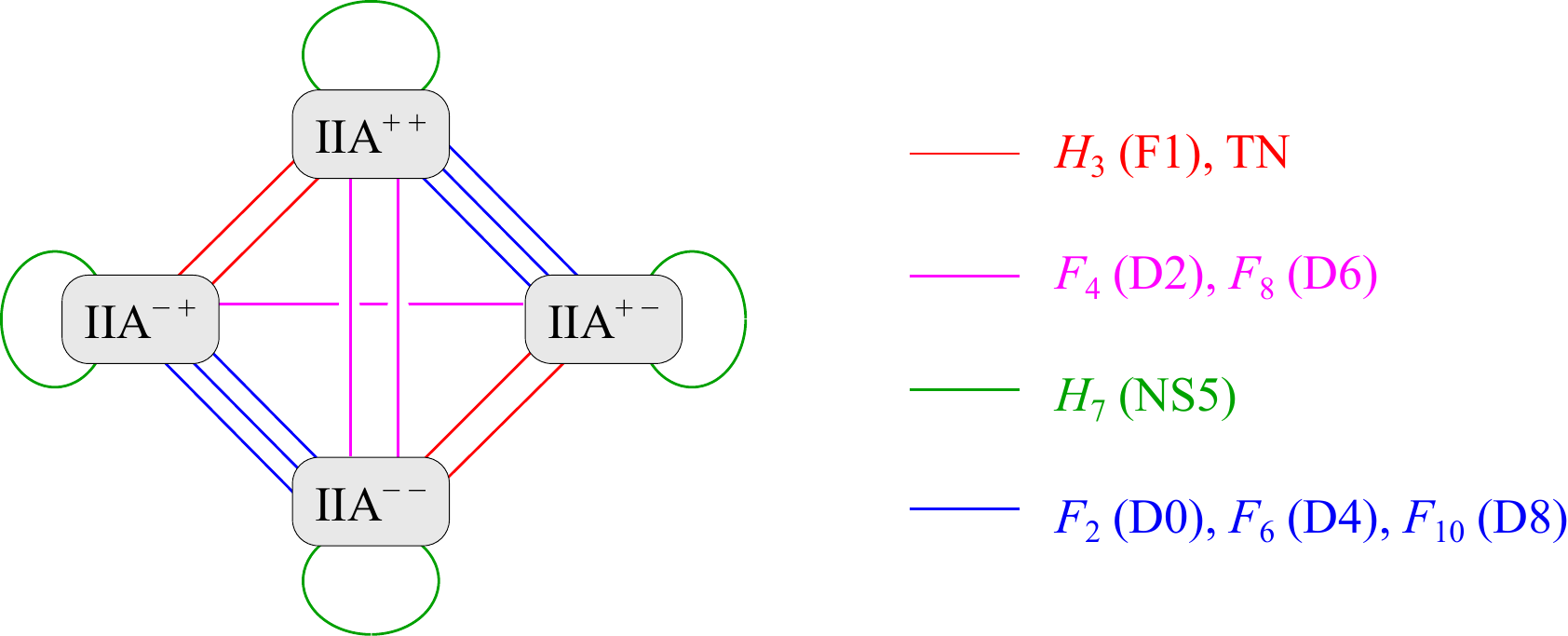}
\caption{The possible signature changing domain walls in type IIA string theory, with the spacetime signatures suppressed for simplicity.}
\label{fig:IIAwalls}
\end{center}
\end{figure}

\begin{figure}
\begin{center}
\includegraphics[height=2in]{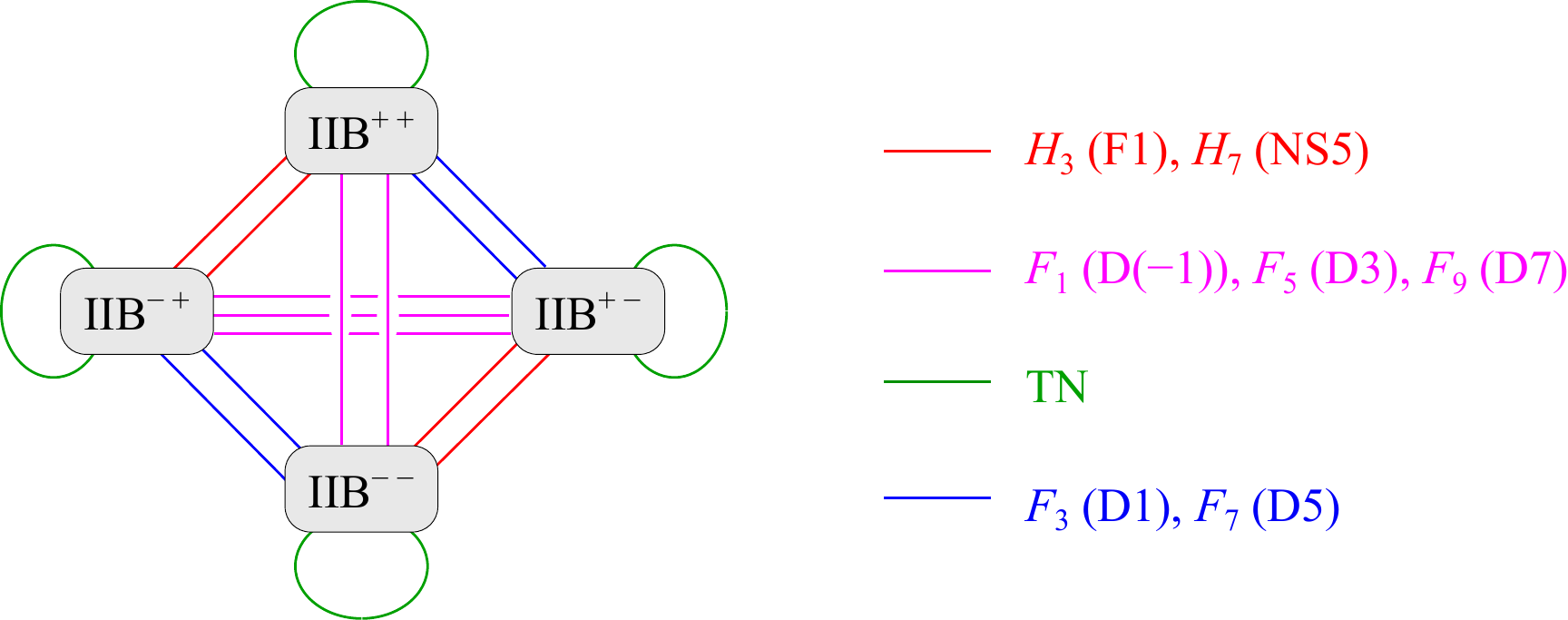}
\caption{The possible signature changing domain walls in type IIB string theory, with the spacetime signatures suppressed for simplicity.}
\label{fig:IIBwalls}
\end{center}
\end{figure}

All the possible domain walls connecting two M-theories are pictured in Figure~\ref{fig:Mthywalls}, classified by the pair of theories they connect and the type of flux supporting them. There are numerous examples connecting two string theories, as summarized in Figures~\ref{fig:IIAwalls}, \ref{fig:IIBwalls}. In some cases, these correspond to the signature changes induced by negative branes in ordinary string theory, cf.\ Table~\ref{tab:bubbletable}. In other cases, the domain walls connect two exotic string theories.

\section{AdS/CFT For Negative Branes} \label{sec:AdSCFT}

In the next few sections, we discuss various consistency checks of our proposed link between negative branes and signature change, as well as consistency checks of Hull's exotic string theories themselves.

Our first topic is the AdS/CFT correspondence for negative branes. On the gauge theory side, the large $N$ planar loop expansion is convergent, rather than asymptotic, implying that the large $N$ behavior of supergroup gauge theories can be understood without needing to address the non-perturbative subtleties discussed in~\S\ref{subsec:SGexist}. In particular, the $\mathcal{N}=4$ theories with gauge groups $U(N|0)$ and $U(0|N)$ are related by $\lambda \to - \lambda$ for $N$ fixed, with $\lambda \equiv g^2_{\rm YM} N$. Since planar quantities are analytic near $\lambda = 0$, the supergravity limit $\lambda \gg 1$ can be reached by analytic continuation, allowing for non-trivial comparisons with the AdS dual. The only difference is that (relative to the normal case) we are interested in $\lambda \to -\infty$ instead of $\lambda \to \infty$.

To find the holographic dual of the $U(0|N)$ $\mathcal{N}=4$ theory, we consider the backreaction of $N$ negative D3-branes. This generates the background
\begin{equation}
\begin{aligned}
\de s^2 &= H^{-\half} \de s^2_{3,1}+ H^{\half} \de s^2_6 \,, & e^{\Phi} &= \gs\,, &
\tilde{F}_5 &= \gs^{-1} (1+\star) \de H^{-1} \wedge \Omega_4 \,,
\end{aligned}
\end{equation}
where
\begin{equation}
H = 1 - \frac{\lambda \alpha'^2}{r^{4}}\,.
\end{equation}
As argued in \S\ref{subsec:negbraneresults}, this describes a bubble of the exotic string theory IIB$^{--}_{7,3}$ surrounding the negative D3-brane. The D3-brane horizon at $r=0$ lies within this bubble, and the near horizon region $r\ll L \equiv \lambda^{1/4} \alpha'^{1/2}$ is therefore described by this string theory in the background
\begin{equation} \label{eqn:negD3nearhorizon}
\begin{aligned}
\de s^2 &= \frac{r^2}{L^2} \de s^2_{1,3}+ \frac{L^2}{r^2} \de r^2 + L^2 \de s^2_{S^5} \,, & e^{\Phi} &= \gs\,, &
\tilde{F}_5 &= \frac{4 r^3}{\gs L^4}  (1+\star) \Omega_4 \wedge \de r \,,
\end{aligned}
\end{equation}
where $\de s^2_{S^5}$ is the round metric on the unit five-sphere.

To identify the remaining metric factor in~(\ref{eqn:negD3nearhorizon}), we briefly review maximally symmetric spaces. Besides flat space itself, these can be realized as quadratic hypersurfaces in flat space of one higher dimension,\footnote{For simplicity we ignore the global properties of these hypersurfaces, such as the distinction between them and their universal covers.} either
\begin{equation} \label{eqn:Spq}
\sum_{i=1}^{s+1} U_i^2 - \sum_{i=1}^t V_i^2 = L^2\,,
\end{equation}
for the space which we will notate dS$_{s,t}$ of radius $L$, or
\begin{equation} \label{eqn:bSpq}
\sum_{i=1}^{s} U_i^2 - \sum_{i=1}^{t+1} V_i^2 = -L^2\,,
\end{equation}
for the space which we will notate AdS$_{s,t}$ of radius $L$. In this notation, dS$_{d-1,1}$ and AdS$_{d-1,1}$ are the standard $d$-dimensional de Sitter and anti-de Sitter spaces. Other special cases are dS$_{d,0} \cong S^d$ and AdS$_{d,0} \cong H^d$ (hyperbolic space). In general, dS (AdS) has constant positive (negative) scalar curvature. However, caution is needed, because dS$_{s,t} \cong $ AdS$_{t,s}$ upon exchanging space and time, consistent with the transformation of the Ricci scalar, $\cR \to -\cR$ under $g\to -g$. Thus, for instance dS$_{1,d-1}$ is the space we usually label ``anti-de Sitter space'', but with space and time labels reversed.

\begin{figure}
\begin{center}
\includegraphics[width=0.6\textwidth]{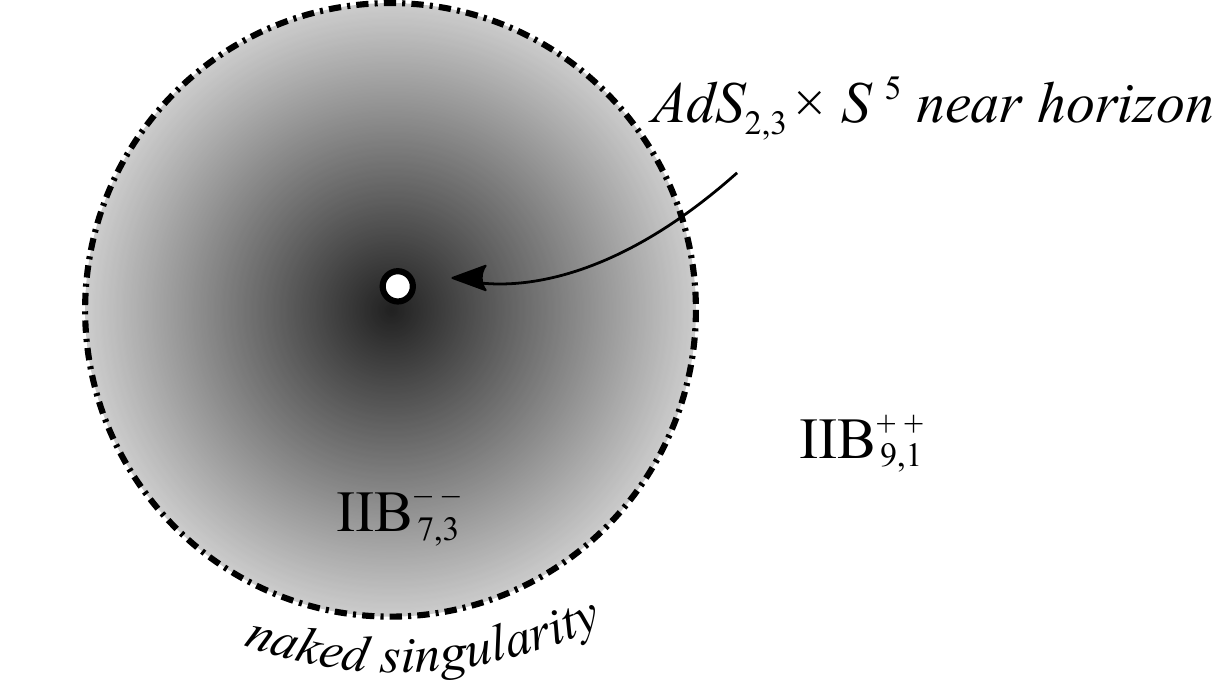}
\caption{The near horizon geometry of negative branes lies within the bubble of exotic string theory surrounding the brane.}
\label{fig:nearhorizon}
\end{center}
\end{figure}

The hypersurface equation~(\ref{eqn:bSpq}) can be rewritten as
\begin{equation} \label{eqn:bSpq2}
\eta_{a b} X^a X^b = X_+ X_- -L^2\,,
\end{equation}
for $p>0$, where $X_\pm = V_{q+1} \pm U_p$ and $X^a = \{U_i, V_j\}$ with the signature $(p-1,q)$ metric $\eta_{a b} = \diag(1,\ldots,-1,\ldots)$. For $X_+ > 0$, we can solve
\begin{align}
X_+ &= r\,, & X_- &= \frac{L^2}{r} + \frac{r}{L^2} \eta_{a b} x^a x^b\,, & X^a &= \frac{r x^a}{L}\,,
\end{align}
which gives the geometry
\begin{equation} \label{eqn:Poincare}
\de s^2 = \frac{r^2}{L^2} \eta_{a b} \de x^a \de x^b + \frac{L^2}{r^2} \de r^2 = \frac{L^2}{u^2} \left(\de u^2 + \eta_{a b} \de x^a \de x^b \right) \,,
\end{equation}
known as the Poincare patch, where $u=L^2/r$. Comparing with~(\ref{eqn:negD3nearhorizon}), we conclude that the near horizon geometry of negative D3-branes is AdS$_{2,3} \times \text{S}^5$, depicted in Figure~\ref{fig:nearhorizon}. 
The bosonic symmetry group $SO(2,4) \times SO(6) \cong SO(4,2) \times SO(6)$ is unchanged, and matches the bosonic part of the $\mathcal{N}=4$ superconformal group as expected. Thus, we are led to conjecture that the $\mathcal{N}=4$ $U(0|N)$ theory is holographically dual to IIB$^{--}_{7,3}$ string theory on AdS$_{2,3} \times \text{S}^5$. 

A mildly annoying feature of this proposed duality is that the boundary of AdS$_{2,3}$ has signature $(1,3)$. We can remedy this moving to the spacetime mirror description,~(\ref{eqn:exchangepairs}),\footnote{We assume for the remainder of this section that the spacetime mirrors are exactly equivalent. All our our subsequent results still follow if this is not the case, but the notation would differ.} for which the dual theory is IIB$^{--}_{3,7}$ compactified on dS$_{3,2}\times \bar{\text{S}}^5$, where $\bar{\text{S}}^5$ denotes a timelike five-sphere. The two descriptions are physically equivalent, but from the CFT perspective the latter is more natural. In this case, the emergent fifth dimension is timelike, and the proposed duality should perhaps be thought of as ``dS/CFT'', albeit unrelated to dS$_{4,1}$.

A classic test of the AdS/CFT correspondence for $U(N)$ is the agreement between the dimensions of chiral primary operators in the gauge theory and the masses of supergravity modes in AdS$_5$~\cite{Aharony:1999ti}, which are related by
\begin{equation}
(\Delta-2)^2 = m^2 L^2 + 4 \,,
\end{equation}
for scalar operators, with similar expressions for other spins. Scalar chiral primary operators are operators of the form $\Tr[\phi^{(i_1} \ldots \phi^{i_n)}]$ with all $SO(6)$ traces removed, where $\phi^i$ denotes one of the six adjoint scalars of $U(N)$ transforming in the vector representation of the $SO(6)$ R-symmetry. These can be matched to the harmonics of the supergravity fields on $S^5$~\cite{Kim:1985ez}.

The same spectrum of chiral primary operators exists in the $U(0|N)$ theory, which should match the mode decomposition of IIB$^{--}_{3,7}$ on dS$_{3,2}\times \bar{\text{S}}^5$ according to our conjecture. In fact, the computation is almost trivially the same, because the two supergravities are related by an analytic continuation of the background by $L^2 \to \pm i L^2$, $g\to \pm i g$, as in~\S\ref{subsec:negbraneresults}, (\ref{eqn:crossing}), whereas the spectrum of $m^2 L^2$ does not depend on $L$. Similar considerations apply to other observables which are independent of $\lambda = L^4/\alpha'^2$.

To obtain nontrivial tests of our conjectural holographic duality we consider $\lambda$-dependent observables in both the $U(0|N)$ gauge theory and in IIB$^{--}_{3,7}$ string theory, where the latter arise from $\alpha'$ corrections to the supergravity action. On the gauge theory side, these observables can be computed by analytically continuing $\lambda \to - \lambda$. By~(\ref{eqn:Poincare}), each factor of the metric $g$ contributes an $L^2 = \alpha' \sqrt{\lambda}$, so that $\lambda \to - \lambda$ corresponds to $g \to \pm i g$, which is the singularity crossing prescription of~\S\ref{subsec:negbraneresults}, (\ref{eqn:crossing})!

So far, we have not shown how to compute $\alpha'$ corrections in exotic string theories such as IIB$^{--}_{3,7}$. As we have just shown, the AdS/CFT correspondence predicts that, at least for IIB$^{--}_{3,7}$ (equivalently IIB$^{--}_{7,3}$), these corrections are related to those for IIB$^{++}_{9,1}$ by the singularity crossing prescription. To test this prediction, we will compute the $\alpha'^3 R^4$ corrections to the IIB$^{--}_{3,7}$ low-energy effective action, first in~\S\ref{sec:curvature} via a chain of S- and T-dualities relating them to those of ordinary M-theory (M$^+_{10,1}$) and then in~\S\ref{sec:worldsheet} by worldsheet methods.

Before proceeding with the calculation we encounter an immediate puzzle, since the prescription $g \to \pm i g$ takes
\begin{equation} \label{eqn:imaginaryR4}
(\cR + \alpha'^3 \cR^4) \to (\cR \pm i \alpha'^3 \cR^4)
\end{equation}
up to an overall phase that cancels against the volume integration measure. Thus (unlike at the two-derivative level) the singularity crossing prescription introduces imaginary terms into the action, which moreover have signs which depend on the choice of branch in continuing around the $H=0$ singularity. This is surprising because the $U(N)$ planar loop expansion is analytic in $\lambda$, hence $\lambda \to - \lambda$ produces a real result for planar observables in the $U(0|N)$ theory.

\begin{figure}
\begin{center}
\includegraphics[width=0.7\textwidth]{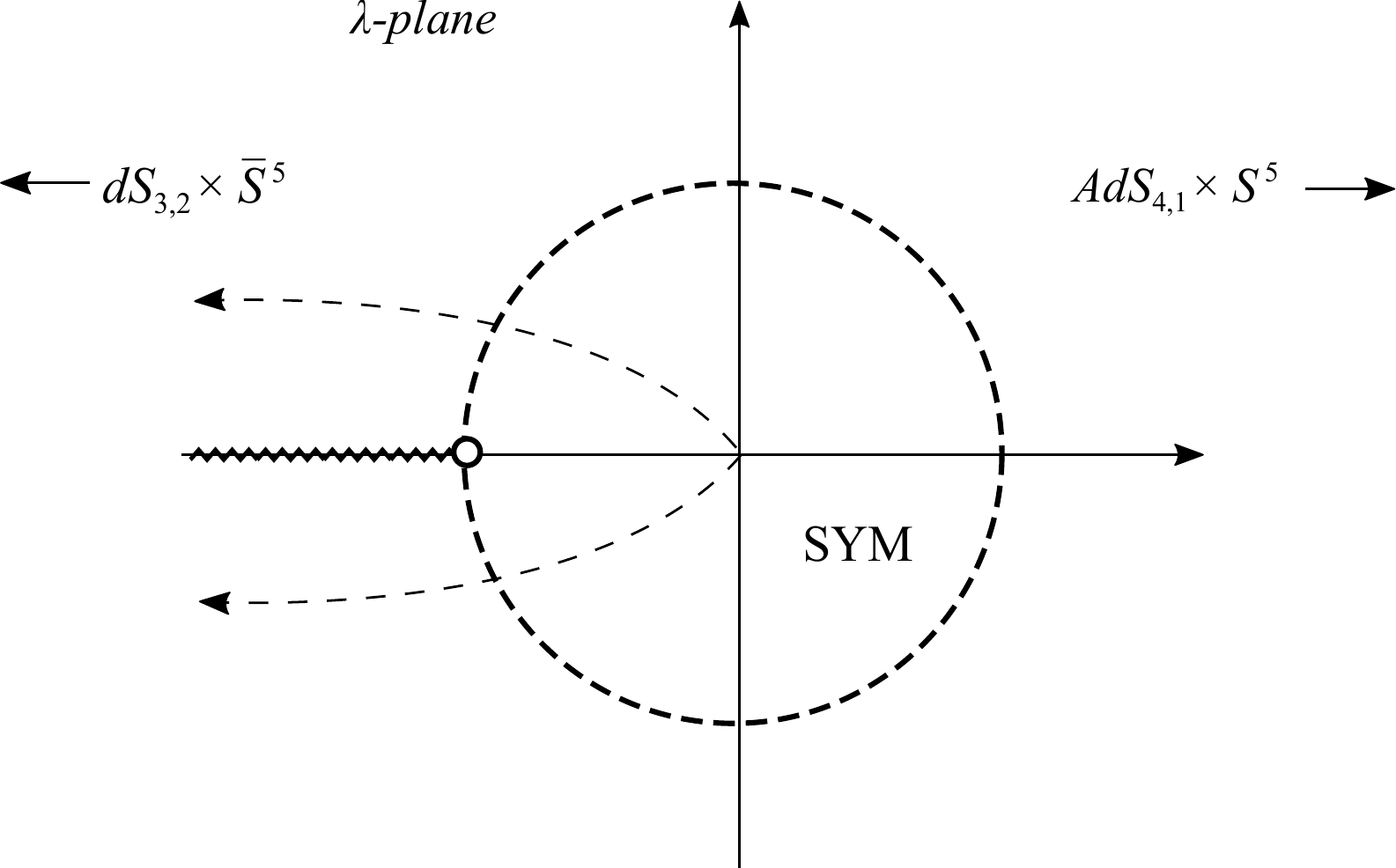}
\caption{The analytic structure of generic planar observables as a function of $\lambda$. The dotted circle $|\lambda| < \lambda_0$ denotes the radius of convergence of the loop expansion in the gauge theory description. Analytically continuing $\lambda \to +\infty$, we obtain tree-level string theory on AdS$_{4,1}\times \text{S}^5$ according to the ordinary AdS/CFT correspondence at the planar level. Analytically continuing $\lambda \to -\infty$ we encounter a branch cut starting at (or outside) the radius of convergence. Depending on which way we go around the branch cut, we obtain an ambiguous answer of the form $\Delta_1 \pm i \Delta_2$, corresponding to the sign ambiguity in $\alpha'$ corrections to IIB$^{--}_{3,7}$ on dS$_{3,2}\times \bar{\text{S}}^5$.}
\label{fig:branchcut}
\end{center}
\end{figure}

This apparent inconsistency is resolved by the fact that the planar loop expansion has a finite radius of convergence beyond which branch cuts and other singularities can appear~\cite{'tHooft:1982tz,Brezin:1977sv}. For instance, the spectrum of string excitations in the plane wave limit of AdS$_5\times \text{S}^5$ describe a set of operators with twists~\cite{Berenstein:2002jq}
\begin{align}
\Delta - J &= \sum_{n=-\infty}^{\infty} N_n \sqrt{1+\lambda \frac{n^2}{J^2}} \,, & \sum_{n=-\infty}^{\infty} n N_n&=0 \,,
\end{align}
for large spin $J\sim O(N^{1/2})$. Expanding $\Delta - J$ in $\lambda \ll 1$, we find a radius of convergence $|\lambda| < \sqrt{J/n_{\rm max}}$, beyond which a branch cut appears on the negative $\lambda$ axis. Analytically continuing $\lambda \to - \infty$, we obtain opposite imaginary results for $\Delta - J$, depending on which way we go around the branch cut, as depicted in Figure~\ref{fig:branchcut}. Similar factors of $i$ inducing a sign-change in the gauge coupling have been anticipated in the context of analytic continuation from AdS to dS spacetimes, for which it has been suggested the corresponding gauge theory is nonunitary; see~\cite{Polyakov:2007mm}. 

More generally, the appearance of half-integer powers of $\lambda$ in the $\lambda\gg 1$ supergravity expansion indicates the presence of such a branch cut in generic $\lambda$-dependent observables, and the ambiguity in the supergravity effective action corresponds to the ambiguity in going around the branch cut on the negative $\lambda$ axis.

\subsection{Other holographic duals}

Besides the conjectural duality between the $\mathcal{N}=4$ $U(0|N)$ negative brane gauge theory and IIB$^{--}_{3,7}$ on AdS$_{2,3}\times {\text{S}}^5$, there are several other holographic dualities predicted by our work, most notably between the worldvolume theory on negative M2-branes and M$^+_{2,9}$ on AdS$_{2,2} \times {\text{S}}^7$ and between the worldvolume theory on negative M5-branes and M$^-_{5,6}$ on AdS$_{2,5} \times {\text{S}}^4$. The latter two examples do not have exactly marginal couplings, hence the only expansion parameter is $N$. In principle, the negative brane theories are related to their ordinary cousins by analytically continuing $N \to -N$, and a similar analysis to that given above may be possible, though it is more difficult in the absence of a weak coupling limit for the dual CFT.\footnote{For negative M2-branes we can introduce a second parameter by placing the M2-branes at a $\mathbb{C}^4/\mathbb{Z}_k$ singularity (breaking some of the supersymmetry), where $k$ corresponds to the Chern-Simons level in the dual ABJM theory~\cite{Aharony:2008ug}, which is perturbative for $k\gg N$. An analysis of this configuration and its dual is beyond the scope of the present work.} We leave any further discussion of these prospective holographic dualities for a future work.

In principle, we can derive holographic duals of the worldvolume theories on D3-branes, M2-branes, or M5-branes in any of the exotic string/M-theories considered above by taking the near horizon limit of their large $N$ backreacted geometries. However, additional difficulties arise in cases other than those considered above, because the ``horizon'' $\text{S}^d$ becomes a non-compact space (A)dS$_{s,t}$, making it unclear whether standard arguments apply or how to interpret the result if they do. This, too, is left for future work.
 
\section{Curvature Corrections} \label{sec:curvature}

In this section, we analyze the $\cR^4$ corrections in exotic string theories by relating them to one loop corrections in the low energy effective supergravity description.
 The results of this section will be used to confirm the predictions of AdS/CFT derived in~\S\ref{sec:AdSCFT}, and will be cross checked against explicit worldsheet calculations in~\S\ref{sec:worldsheet}.

\subsection{$\cR^4$ corrections from KK loops} \label{subsec:R4kk}

There are two $\cR^4$ corrections to the effective action of type IIA string theory, of the schematic form:
\begin{equation} \label{eqn:IIAR4}
  S_{\cR^4} \sim \frac{\alpha'^3}{2 \kappa_{10}^2} \int \de^{10} x \sqrt{| g |}  (e^{- 2
  \Phi} \cR^4 + \cR^4) \,,
\end{equation}
appearing at tree level and one loop in the string loop expansion, respectively. Here we suppress the overall coefficient and index structure for simplicity, as this is not essential to our analysis (see, e.g.,~\cite{Green:1997di} for the details).  Lifting to $M$-theory, we obtain (\ref{eqn:R11}, \ref{eqn:Fg11})
\begin{align}
   \mathcal{R}  &\sim e^{- \frac{2}{3} \Phi} \mathcal{R}_{(11)}\,, &  \sqrt{|g|} &= e^{\frac{8}{3} \Phi}  \sqrt{| g_{(11)} |} \,.
\end{align}
In the M-theory lift, the Einstein-Hilbert term $\sqrt{|g|} e^{-2 \Phi} \cR$ and the one-loop correction $\sqrt{|g|} \cR^4$ are independent of $R_{11} \sim e^{2 \Phi/3}$, but the tree-level correction $\sqrt{|g|} e^{-2 \Phi} \cR^4$ scales as $1/R_{11}^3$ and vanishes as we take the eleven-dimensional limit, $R_{11} \to \infty$. Thus, the M-theory effective action is corrected:
\begin{equation} \label{eqn:MR4}
S_{\cR^4}^{(11)} \sim \frac{\ell_{\rm P}^6}{2 \kappa_{11}^2} \int \de^{11} x \sqrt{| g |}  \cR^4 \,,
\end{equation}
due to the one-loop correction of type IIA string theory, but no trace of the tree-level correction remains in the eleven-dimensional theory.

The reason for this discrepancy is that the ``tree-level'' correction in type IIA string theory is generated by one loop diagrams involving KK modes in the eleven dimensional effective supergravity description~\cite{Green:1997as}, whereas the ``one-loop'' correction corresponds to the tree-level coupling~(\ref{eqn:MR4}) in eleven dimensions.

This means that the string tree-level $\cR^4$ correction of type IIA string theory can be obtained by a one-loop computation in eleven-dimensional supergravity. The string one-loop correction can also be obtained by one-loop supergravity calculation when we consider the T-duality between type IIA and type IIB string theory. In particular, compactifying type IIA and type IIB string theory on a circle leads to the same nine-dimensional theory, whose two-derivative low-energy effective action is (\ref{IIAreduced}). The radion $\sigma$ and dilaton $\Phi$ in nine-dimensions are related to the ten-dimensional dilaton and compactification radius as follows:
\begin{align}
e^\sigma &= R_{\rm (IIA)}/\ls = \ls/R_{\rm (IIB)}\,, &
e^\Phi &= e^{\Phi_{\rm (IIA)}}\,, & e^{\Phi - \sigma} &= e^{\Phi_{\rm (IIB)}}\,.
\end{align}
The $\cR^4$ corrections (\ref{eqn:IIAR4}) reduce to
\begin{equation} \label{eqn:R49A}
  S_{\cR^4}^{(9A)} \sim \frac{\alpha'^3}{2 \kappa_{9}^2} \int \de^{9} x \sqrt{| g |}  (e^{\sigma- 2
  \Phi} \cR^4 + e^{\sigma} \cR^4) \,.
\end{equation}
An analogous set of corrections exist in type IIB string theory. These reduce to
\begin{equation}  \label{eqn:R49B}
  S_{\cR^4}^{(9B)} \sim \frac{\alpha'^3}{2 \kappa_{9}^2} \int \de^{9} x \sqrt{| g |}  (e^{\sigma- 2
  \Phi} \cR^4 + e^{-\sigma} \cR^4) \,.
\end{equation}
To explain the discrepancy between (\ref{eqn:R49A}) and (\ref{eqn:R49B}), note that in the ten-dimensional type IIA description, the $\sqrt{|g|} e^{-\sigma} \cR^4$ correction scales as $1/R_{\rm (IIA)}^2$, and disappears as $R_{\rm (IIA)} \to \infty$. In direct analogy with the M-theory compactification considered above, we conclude that the additional $\sqrt{|g|} e^{-\sigma} \cR^4$ correction in the compact theory is generated a loop of KK modes.\footnote{The relation between the computations can be established by considering M-theory compactified on a torus. Exchanging which cycles we label as the M-theory and IIA circles exchanges the two computations.} Likewise, the $\sqrt{|g|} e^{\sigma} \cR^4$ correction is generated by a loop of KK modes in the ten-dimensional IIB description. The nine-dimensional effective theory has all three corrections
\begin{equation}
  S_{\cR^4}^{(9)} \sim \frac{\alpha'^3}{2 \kappa_{9}^2} \int \de^{9} x \sqrt{| g |}  (e^{\sigma- 2
  \Phi} \cR^4 +e^{\sigma} \cR^4+ e^{-\sigma} \cR^4) \,,
\end{equation}
in addition to an infinite set corrections which are non-perturbative in $g_s$, and which we ignore in the present analysis.\footnote{These additional corrections are also present in type IIB but not in type IIA. In the type IIA description, they are generated by the KK modes of the D0 brane and its marginal bound states. In the eleven-dimensional description, they are generated by the lattice of KK modes on $T^2$.}

Thus, all the $\cR^4$ corrections in string/M-theory can be thought of as effective couplings generated a loop of KK modes in some dual description. We now exploit this fact to compute the $\cR^4$ corrections in the exotic string/M-theories considered previously.
To do so, we first consider how the spacetime signature and the difference between spacelike and timelike compactification affect these corrections.

\subsection{Feynman rules and spacetime mirrors} \label{subsec:quantmirrors}

We now derive the Feynman rules for the effective supergravity theories described in~\S\ref{subsec:SGaction}.
To avoid any possible ambiguities, we do so by relating the quantized effective theories to each other and to the ordinary supergravities IIA$^{++}_{9,1}$, IIB$^{++}_{9,1}$ and M$^+_{10,1}$ via spacelike and timelike compactification in the language of Feynman diagrams.

Since the vertices are local, the physics of compactification is encoded in the propagator. Before compactifying, we have the Feynman propagator\footnote{We follow the conventions of~\cite{Peskin:1995ev}, converting to a mostly-plus metric signature.}
\begin{equation}
G_D(x-y;m^2) = \int \frac{\de^D p}{(2 \pi)^D}\,\frac{-i e^{i p\cdot(x-y)}}{p^2+m^2 - i \epsilon} \,,
\end{equation}
for a scalar field in $D$ dimensions. Compactifying on a spatial circle,  $x^\mu \cong x^\mu + 2 \pi R^\mu$ for spacelike $R^\mu$, the Green's function becomes a sum over images
\begin{equation}
\widehat{G}_D(x;m^2) = \sum_{n=-\infty}^{\infty} G_D(x^\mu+2 \pi n R^\mu; m^2) \,.
\end{equation}
The infinite sum on $n$ can be evaluated using Poisson resummation,
$\sum_{n=-\infty}^{\infty} e^{2 \pi i n x} = \sum_{k=-\infty}^{\infty} \delta(x-k)$.
Fixing $R^\mu = (\ldots,0,R)$ with a Lorentz transformation, we obtain
\begin{equation}  \label{eqn:propagatorLor}
\widehat{G}_D(\hat{x},y;m^2) =  \sum_{k=-\infty}^{\infty} \frac{e^{\frac{i k y}{R}}}{2\pi R} \, G_{d}(\hat{x}; m^2 + k^2/R^2) \,,
\end{equation}
where $x^\mu = (\hat{x}^\alpha, y)$ and $d= D-1$. Thus, for each $D$-dimensional scalar there is an infinite tower of KK modes in $d$ dimensions, labeled by their KK number $k\in\mathbb{Z}$ with masses $M^2 = m^2 + k^2/R^2$. For each vertex, the integral $\int_0^{2 \pi R} d y \prod_i e^{\frac{i k_i y}{R}} = (2\pi R) \delta_{\sum_i k_i}$ enforces conservation of KK number. The Feynman rules of the $d$-dimensional theory are otherwise the same as those of the $D$-dimensional theory.\footnote{The factors of $\frac{1}{2\pi R}$ ($2\pi R$) for each propagator (vertex) appear because we have not canonically normalized the fields and couplings in $d$ dimensions.}

The case where $R^\mu$ is timelike is closely analogous. We fix $R^\mu = (T,0,\ldots)$ with a Lorentz transformation, which gives
\begin{equation}  \label{eqn:propagatorEuc}
\widehat{G}_D(t,\vec{x};m^2) =  \sum_{k=-\infty}^{\infty} \frac{e^{-\frac{i k y}{T}}}{2\pi T} \, G_{d}(\vec{x}; m^2 - k^2/R^2) \,.
\end{equation}
Thus, the Feynman rules are the same as for spacelike compactification, except that the metric signature is $(+\ldots+)$ instead of $(-+\ldots+)$ and the masses of the KK modes are $M^2 = m^2 - k^2/R^2$.

Comparing the various supergravities via S and T-dualities, we conclude that all of them have Feynman rules which follow from a path integral of the standard form
\begin{equation}
Z = \int [D \phi] e^{i S[\phi]} \,,
\end{equation}
with the epsilon prescription $p^2 + m^2 - i \epsilon$ for the propagators, where now $p^2$ involves the signature $(s,t)$ metric. While this is the expected result, it is important to establish definitively, as our subsequent conclusions will follow from it.

As a preliminary exercise, we verify the equivalence of the spacetime mirror theories~(\ref{eqn:exchangepairs}) at the quantum level in effective field theory. Mapping $g\to -g$ (reversing the metric signature) has two effects on the quantum theory:
\begin{enumerate}
\item
The Feynman rules are mapped onto those of the parity-reversed spacetime mirror, with an extra minus sign for each vertex and propagator, see~(\ref{eqn:mirrormatch}) and following discussion.
\item
The epsilon prescription is reversed ($\epsilon \to - \epsilon$).
\end{enumerate}
Combining these two effects, we see that
\begin{equation} \label{eqn:mirrorM}
\overline{\mathcal{M}(p,j,Q;\mathrm{in})}_{g \to -g} = -\mathcal{M}(\tilde{p},\tilde{j},Q;\mathrm{out})^\ast \,,
\end{equation}
where $\mathcal{M}$ denotes any scattering amplitude, $\overline{\mathcal{M}}$ the same scattering amplitude in the spacetime mirror, $p$, $j$, and $Q$ the momenta, spins, and charges of the external particles---considered to be incoming on one side of the equation and outgoing on the other---and $\tilde{p}, \tilde{j}$ the parity reversed momenta and spins. Thus, the scattering amplitudes for the spacetime mirror theories are related by T (up to a physically insignificant overall minus sign),\footnote{For theories with multiple times, we define T as the antilinear operator which exchanges in and out states combined with a spacetime parity flip, where spatial and temporal parity flips are equivalent for $s,t>1$.} and the two effective theories are physically equivalent. Notice that the minus sign in~(\ref{eqn:mirrormatch}) is crucial to the success of~(\ref{eqn:mirrorM}). Without this, the two theories would be inequivalent!

\subsection{KK loop diagrams}

We now consider the effect of one loop diagrams involving KK modes, which will generate effective couplings for the KK zero modes in $d=D-1$ dimensions. Denote the signature of the $D$-dimensional theory as $(s,t)$. The amplitude contains an integral of the form
\begin{equation}
i \mathcal{M} \sim I_R = \int \frac{\de^{d} \ell}{(2\pi)^{d}} \frac{(\ell^2)^m}{\left[\ell^2 + \sum_{i<j} u_i u_j (p_i - p_j)^2 + \sum_i u_i m_i^2 + \frac{k^2}{R^2} - i \epsilon\right]^n} \,,
\end{equation}
in the spacelike case, where we introduce Feynman parameters $u_i\ge 0$ with $\sum_i u_i =1$, and $p_i$ are related to the external momenta. In the low-energy limit $|\vec{p}_i|, m_i \ll 1/R$, the $u_i$-dependent terms can be neglected, and we obtain
\begin{equation}
I_R = i^t \frac{\Gam{n-m-\frac{d}{2}}\Gam{m+\frac{d}{2}}}{(4 \pi)^{\frac{d}{2}} \Gam{n} \Gam{\frac{d}{2}}}\, (R^2/k^2)^{n-m-\frac{d}{2}} \,,
\end{equation}
after Wick rotating $\ell_{\hat{\mu}} \to i \ell_{\hat{\mu}}$ for $\hat{\mu}$ timelike, as dictated by $\Re \epsilon > 0$. By contrast, a tree-level effective coupling contributes $i \mathcal{M} \sim i \lambda_{\rm eff}$, so that
\begin{equation} \label{eqn:Xeff}
\lambda_{\rm eff} \sim i^{t-1} \,,
\end{equation}
up to an overall sign. In particular, for $t=1$ the effective couplings are real, as expected from unitary field theory, but for $t\ne 1$ this no longer holds in general.

Similar reasoning applies to the case of time-like compactification, except that the denominator is of the form $\vec{\ell}^2 - k^2/T^2 - i \epsilon$, so we Wick rotate $\ell_{\hat{\mu}} \to - i \ell_{\hat{\mu}}$ for $\hat{\mu}$ spacelike, giving
\begin{equation}
I_T = (-1)^{m+n} (-i)^{s} \frac{\Gam{n-m-\frac{d}{2}}\Gam{m+\frac{d}{2}}}{(4 \pi)^{\frac{d}{2}} \Gam{n} \Gam{\frac{d}{2}}}\, (T^2/k^2)^{n-m-\frac{d}{2}} \,.
\end{equation}
and the effective coupling takes the form
\begin{equation} \label{eqn:Teff}
\lambda_{\rm eff} \sim (-1)^{m+n} (-i)^{s+1} \,,
\end{equation}
so the result can be real or imaginary, depending on the spacetime signature.

Using~(\ref{eqn:Xeff}), (\ref{eqn:Teff}), 
we can fix the phase of the $\cR^4$ corrections in all of the exotic string/M-theories by relating them to KK loop calculations. We find
\begin{align} \label{eqn:R4allsig}
\begin{split}
  S_{\cR^4}[\mathrm{M}^{\alpha}] &\sim \frac{\ell_{\rm P}^6}{2 \kappa_{11}^2} \int \de^{11} x \sqrt{| g |} \cR^4 \,, \\
  S_{\cR^4}[\mathrm{IIA}^{\alpha \beta}] &\sim \frac{\alpha'^3}{2 \kappa_{10}^2} \int \de^{10} x \sqrt{| g |}  ( i^{\frac{1-\alpha}{2}} e^{- 2
  \Phi} \cR^4 + \cR^4) \,, \\
  S_{\cR^4}[\mathrm{IIB}^{\alpha \beta}] &\sim \frac{\alpha'^3}{2 \kappa_{10}^2} \int \de^{10} x \sqrt{| g |}  ( i^{\frac{1-\alpha}{2}} e^{- 2
  \Phi} \cR^4 + i^{\frac{1-\alpha}{2}} \cR^4) \,,
\end{split}
\end{align}
for $\alpha = \pm 1$, where we have not attempted to compute the overall sign, coefficient, or index structure for each term. For each correction, there are multiple ways to relate it to a KK loop along the lines of~\S\ref{subsec:R4kk}, all of which give the same phase. Notice in particular that spacetime mirrors have $\cR^4$ corrections with the same phases, and that these phases agree with the singularity crossing prescription of~\S\ref{subsec:negbraneresults}, (\ref{eqn:crossing}), and with the AdS/CFT analysis of~\S\ref{sec:AdSCFT}.

In type IIB theories there are additional corrections which are non-perturbative in the string loop expansion. These are discussed briefly in~\S\ref{subsec:lowenergyconsistency}.
 
\section{Worldsheet Theories} \label{sec:worldsheet}

In the previous section, we quantized the low energy effective theories described in~\S\ref{subsec:SGaction} and used them to learn something about $\cR^4$ corrections. In this section, we will try to understand the full UV behavior of these theories. We will work under the assumption that the exotic IIA/B theories have a worldsheet description, much like ordinary type IIA/B string theory.

A novel feature which occurs in some of these theories is that the fundamental strings have Euclidean signature. To better understand the consequences of this, we will study bosonic string theory with a Euclidean worldsheet. This should share some features with the putative worldsheet theories for the exotic IIA/B$^{-\pm}$ theories. We leave a construction of the full superstring in these exotic theories for future work.\footnote{Note that there is no guarantee that the exotic string theories admit a worldsheet description---even if they exist as mathematically consistent spacetime theories with stringlike excitations. However, the results of this section suggest that a worldsheet description is possible.}

\subsection{The classical bosonic Euclidean string}

The Polyakov action generalized to arbitrary worldsheet signature is
			\begin{align}
				\label{eqn:freeaction} S &= \frac{\varepsilon}{4\pi \alpha^\prime} \int_\Sigma \mathrm{d}^2 \sigma \sqrt{\varepsilon \det \gamma} \gamma^{ab} \eta_{\mu \nu} \partial_a X^\mu \partial_b X^\nu\,,
			\end{align}
where the worldsheet metric $\gamma_{a b}$ has signature $(\varepsilon, 1)$ with $\varepsilon=+1$ ($\varepsilon=-1$) for a Euclidean (Lorentzian) worldsheet, and the overall sign is chosen so that $(\partial_\tau X)^2$ has a positive coefficient.

We proceed to solve the classical theory for the closed string in the usual fashion, keeping the parameter $\varepsilon$ generic throughout the calculation. Gauge-fixing to a flat worldsheet metric $\gamma_{\tau \tau} = \varepsilon$, $\gamma_{\sigma \sigma} =1$, $\gamma_{\tau \sigma} = 0$ and introducing left and right-moving coordinates
				$\sigma^{\pm{}} = \tau \pm{} \sqrt{-\varepsilon} \sigma$,
		 the $X^\mu$ equations of motion have the general solution
			\begin{align}
				X^\mu = X_L^\mu(\sigma^+) + X_R^\mu(\sigma^-). 
			\end{align}
		The mode expansion takes the form
			\begin{align}
			\begin{split}
			\label{eqn:Fourierexpansion}
				X^\mu(\sigma^+, \sigma^-) &= x^\mu + {\alpha^\prime} p^\mu \tau - \sqrt{\frac{\alpha^\prime}{2 \varepsilon}} \sum_{n \ne 0 } \left[\frac{\alpha_n^\mu}{n} e^{- \sqrt{\varepsilon} n \sigma^+} + \frac{\tilde \alpha_n^\mu}{n} e^{- \sqrt{\varepsilon} n \sigma^-}\right] \,,
			\end{split}
			\end{align}
		where $x^\mu, p^\mu$ are real and the oscillator modes satisfy 
				\begin{align}
				\label{eqn:strangereality}
					\alpha_n^{\mu *} ,\tilde  \alpha_n^{\mu*} = \begin{cases} \alpha_{-n}^{\mu }, \tilde \alpha_{-n}^\mu &\varepsilon = -1 \,,\\
					\tilde \alpha_n^\mu , \alpha_n^\mu & \varepsilon = +1 \,.
				\end{cases}
				\end{align}
			Notice that the reality conditions for the Euclidean worldsheet theory differ from those for the Wick rotated Lorentzian worldsheet theory. Wick rotation does not alter the reality conditions because the Wick rotated time coordinate is imaginary $\tau_E^* = (i \tau)^* = - \tau_E$. By contrast, the time coordinate in the Euclidean worldsheet theory is real, implying that complex conjugation exchanges left and right movers. 
						
Defining $\alpha_0^\mu =\tilde \alpha_0^\mu = \sqrt{\alpha^{\prime}/2}\, p^\mu$, the Virasoro constraints $(\partial_+ X)^2 = (\partial_- X)^2 = 0$ take the form
$\sum_n \alpha_n\cdot \alpha_{m-n} = \sum_n \tilde{\alpha}_n\cdot \tilde{\alpha}_{m-n} = 0\,,$
for every $m$. In particular, the $m=0$ constraint determines the spectrum
\begin{equation} \label{eqn:massshell}
\frac{\alpha^\prime p^2}{4} + \sum_{n> 0} \alpha_{-n} \cdot \alpha_n =\frac{ \alpha^\prime p^2}{4} + \sum_{n> 0} \tilde \alpha_{-n} \cdot \tilde{\alpha}_n =0.
\end{equation}

		\subsection{The quantized Euclidean string} \label{subsec:QuantString}
			We now compute the spectrum of the quantized Euclidean string. Surprisingly, all the massive states will turn out to have imaginary $m^2$. We later show that this novel feature is in fact required for the consistency of the theory when toroidally compactified. 
			
			The quantum dynamics associated to the Euclidean worldsheet theory is formally encoded in the path integral over Euclidean worldsheets,
			\begin{align} \label{eqn:ZEuclWS}
				Z=\int [D \gamma] [D X] \, e^{i S} \,.
			\end{align}
			The Lorentzian worldsheet theory can also be written as a Euclidean path integral, but now of the form:
			\begin{align} \label{eqn:ZLorWS}
				Z=\int [D \gamma] [D X] \, e^{-S_E} \,.
			\end{align}
			The crucial difference between~(\ref{eqn:ZEuclWS}) and~(\ref{eqn:ZLorWS}) is that the former is oscillatory, whereas the latter is damped. Thus, although both variants admit a Euclidean path integral description, they are physically distinct. The effect of these differences will soon become apparent.
			
	To canonically quantize the theory, we introduce the equal-time commutator
			\begin{align}
			\label{eqn:bracket}
				[X^\mu(\sigma,\tau), \Pi^\nu(\sigma^\prime, \tau)] &= i \eta^{\mu \nu} \delta(\sigma - \sigma^\prime)\,, & \Pi^\mu &= \frac{\delta S}{\delta \partial_\tau X_\mu} = \frac{1}{2\pi \alpha'} \partial_\tau X^\mu.
			\end{align}
		 Using~(\ref{eqn:Fourierexpansion}), we obtain the mode algebra
			\begin{align}
			\label{eqn:modebracket}
				[x^\mu, p^\nu] &= i \hbar \eta^{\mu \nu}\,, & [\alpha^\mu_{m}, \alpha^\nu_{n} ] & =[\tilde \alpha^\mu_{m}, \tilde \alpha^\nu_{ n} ] = \frac{m}{\sqrt{-\varepsilon}} \delta_{m, -n} \eta^{\mu \nu}.
			\end{align}
		Observe that for $\varepsilon =+1$ the commutator (\ref{eqn:modebracket}) is imaginary. Define the number operators
		\begin{align}
			N &= \sqrt{-\varepsilon} \sum_{n>0} \alpha_{-n} \cdot \alpha_n \,, & \tilde N &= \sqrt{-\varepsilon} \sum_{n> 0} \tilde \alpha_{-n} \cdot \tilde \alpha_n\,,
		\end{align}
		whose normalization is chosen to ensure real eigenvalues.
		The mass-shell condition~(\ref{eqn:massshell}) is then
			\begin{align}
			- p^2 = \frac{4}{\alpha^\prime \sqrt{-\varepsilon}}(N + A) = \frac{4}{\alpha^\prime \sqrt{-\varepsilon}} (\tilde N + \tilde{A})
			\end{align}
			where the normal ordering constants are $A = \tilde{A} = -1$ by a standard computation. 
			Thus, the quantized Euclidean string has an imaginary spectrum!\footnote{In the above, we have not been careful about the subtleties of gauge fixing and physical states. This can be addressed systematically either through BRST quantization or the use of light-cone gauge. Both approaches proceed in straightforward analogy with the usual, Lorentzian string, and we omit any further discussion of these issues.}

		\subsection{T-duality}
		
		We now consider the spectrum of the Euclidean worldsheet theory compactified on a circle of signature $\eta^{DD} = \pm 1$.
	The mode expansion becomes
	\begin{align}
	\begin{split}
	\label{eqn:compactmodeexpansion}
		X_L &= x_L + \frac{\alpha^\prime}{2} p_L \sigma^+ - \sqrt{\frac{\alpha^\prime}{2 \varepsilon} } \sum_{n \ne 0} \frac{\alpha_n}{n} e^{- \sqrt{\varepsilon} n \sigma^+} \,,\\
		X_R &= x_R + \frac{\alpha^\prime}{2} p_R \sigma^- - \sqrt{\frac{\alpha^\prime}{2 \varepsilon}} \sum_{n \ne 0} \frac{\tilde \alpha_n}{n} e^{- \sqrt{\varepsilon} n \sigma^-}\,,
	\end{split}
	\end{align}
	where
	\begin{align} \label{eqn:pLR}
		p_L^D &= \frac{n}{R} + \frac{1}{\sqrt{-\varepsilon}} \frac{w R}{\alpha^\prime} \,, & p_R^D &=  \frac{n}{R} - \frac{1}{\sqrt{-\varepsilon}} \frac{w R}{\alpha^\prime} \,,
	\end{align}
	for a periodic coordinate $X^D \cong X^D + 2\pi R$. Here $w$ is the winding number, $X^D(\tau,\sigma+2 \pi) = X^D(\tau,\sigma) + 2\pi R w$, and the compact momentum $(p_L^D + p_R^D)/2$ is quantized in units of $1/R$. The mass-shell condition is
	\begin{equation} \label{eqn:compactspectrum0}
	-k^2 = \eta^{DD} \biggl(\frac{n}{R}+\frac{1}{\sqrt{-\varepsilon}} \frac{w R}{ \alpha'} \biggr)^2 + \frac{4}{\alpha' \sqrt{-\varepsilon}} (N-1) = \eta^{DD} \biggl(\frac{n}{R}-\frac{1}{\sqrt{-\varepsilon}} \frac{w R}{ \alpha'} \biggr)^2 + \frac{4}{\alpha' \sqrt{-\varepsilon}} (\tilde{N}-1) \,,
	\end{equation}
	where $k^m$ is the momentum in the non-compact directions. This can be rewritten as
	\begin{align}
		\label{eqn:compactspectrum}
			-k^2 &= \eta^{DD} \frac{n^2}{R^2} -\varepsilon \eta^{DD} \frac{R^2 w^2}{\alpha^{\prime 2}} + \frac{2}{\alpha' \sqrt{-\varepsilon}} (N + \tilde N -2)\,, &
			0&=\eta^{DD} nw + N - \tilde{N} \,.
		\end{align}
	Consider the limit $R\to 0$. The low-energy modes have $n=0$, with the mass-shell condition
	$0 = k^2+(-\varepsilon \eta^{DD}) \frac{w^2}{(\alpha'/R)^2} + \frac{4}{\alpha' \sqrt{-\varepsilon}} (N - 1)$ and $\tilde{N} = N$. The winding contribution can be interpreted as the quantized momentum on a T-dual circle of radius $R' = \alpha'/R$ and signature $(\eta^{DD})'=-\varepsilon\eta^{DD}$. In particular, for a Euclidean worldsheet $\varepsilon = +1$, the T-dual circle has opposite signature! This matches the D-brane based results of~\S\ref{subsec:dualityweb} and the effective supergravity analysis of Appendix~\ref{app:supergravity}.

T-duality extends to the complete theory as follows
\begin{align}
R &\overset{\text{T}}{\longrightarrow} \frac{\alpha'}{R}\,, & \eta^{DD} &\overset{\text{T}}{\longrightarrow} - \varepsilon \eta^{DD}\,, & X^D_L &\overset{\text{T}}{\longrightarrow} \sqrt{-\varepsilon} X^D_L\,, & X^D_R &\overset{\text{T}}{\longrightarrow} -\sqrt{-\varepsilon} X^D_R\,,
\end{align}
where $p_L^D \longrightarrow \sqrt{-\varepsilon} p_L^D$ and $\alpha^D_n \longrightarrow \sqrt{-\varepsilon} \alpha^D_n$ pick up the same overall phase as $X_L^D$, and likewise for the right-movers
The effect on the spectrum is
\begin{align}
n &\overset{\text{T}}{\longrightarrow} w\,, & w &\overset{\text{T}}{\longrightarrow} -\varepsilon n\,, & N &\overset{\text{T}}{\longrightarrow} N\,, & \tilde{N} &\overset{\text{T}}{\longrightarrow} \tilde{N}\,,
\end{align}
where the phases picked up by the oscillators $\alpha_n^D$, $\tilde{\alpha}_n^D$ and $\eta^{DD}$ cancel, leaving $N$, $\tilde{N}$ invariant.

Note that~(\ref{eqn:compactspectrum0}), (\ref{eqn:compactspectrum}) contain an interesting interplay between the zero modes and oscillators. The change of signature during T-duality for $\varepsilon = +1$ arises because of the relative $\pm i$ between the momentum and winding contributions to $p_{L,R}$, but this creates an imaginary level mismatch which must be cancelled by the oscillators. Thus, the imaginary oscillator spectrum is a necessary consequence of the change of spacetime signature during T-duality.

The spectrum~(\ref{eqn:compactspectrum}) admits extra massless states with nonzero compact momentum and winding for the radii $R=\sqrt{2 \alpha'}$ and its T-dual $R = \sqrt{\alpha'/2}$. Consider the former case without loss of generality. Massless states appear for $n=\pm 2$, $w=\pm1$, and either $(N,\tilde{N}) = (2,0)$ or $(0,2)$ depending on the relative sign of $n$ and $w$. The spectrum for $(N,\tilde{N})=(2,0)$ consists of two scalars, a vector, and a symmetric tensor from $\alpha_{-1}^M \alpha_{-1}^N |0\rangle$ and a further scalar and vector from $\alpha_{-2}^M |0\rangle$. Due to the appearance of additional massless symmetric tensors (gravitons) charged under momentum and winding, we conclude that the physics at this radius must be quite exotic!

\subsection{Modular invariance}

A thorough treatment of string interactions in the Euclidean worldsheet theory is beyond the scope of this paper. It is relatively straightforward, however, to verify the modular invariance of the one loop zero-point amplitude.\footnote{Our treatment closely follows~\cite{Polchinski:1998rq}.} The torus partition function is
\begin{equation} \label{eqn:torusZ}
Z = \Tr\biggl[ q^{\sqrt{-\varepsilon} \bigl(L_0 - \frac{c}{24}\bigr)} \bar{q}^{\sqrt{-\varepsilon} \bigl(\tilde{L}_0 - \frac{\tilde{c}}{24}\bigr)} \biggr] \,,
\end{equation}
where $q=e^{2\pi i\tau}$, $\tau = \tau_1+i \tau_2$ is the complex structure of the torus, and
\begin{align}
L_0 &= \frac{\alpha^\prime}{4} p^2_L + \sum_{n> 0} \alpha_{-n} \cdot \alpha_n\,, & \tilde{L}_0 &=\frac{ \alpha^\prime }{4} p_R^2 + \sum_{n> 0} \tilde \alpha_{-n} \cdot \tilde{\alpha}_n \,.
\end{align}
The factor of $\sqrt{-\varepsilon}$ appears in~(\ref{eqn:torusZ}) because the generators of $\tau$ and $\sigma$ translations are
\begin{align}
H &= \int_0^{2 \pi} T_{\tau \tau} d \sigma = \frac{1}{2\pi \alpha'} \int_0^{2 \pi} [ (\partial_+ X)^2 + (\partial_- X)^2 ] \,, \\ P &= \int_0^{2 \pi} T_{\tau \sigma} d \sigma = \frac{\sqrt{-\varepsilon}}{2\pi \alpha'} \int_0^{2 \pi} [ (\partial_+ X)^2 - (\partial_- X)^2 ] \,,
\end{align}
whereas the time translation is imaginary (real) for the Lorentzian (Euclidean) worldsheet theory, due to the Wick rotation involved in the former case. Thus, the oscillators contribute $q^N \bar{q}^{\tilde{N}}$, independent of the signature, and the partition function for a non-compact boson $X^\mu$ is virtually unchanged:
\begin{align} \label{eqn:ZX}
Z_{X}(\tau,\bar{\tau}) \equiv Z/V = (-\varepsilon)^{-1/4} (4 \pi^2 \alpha' \tau_2)^{-1/2} |\eta(\tau)|^{-2} \,,
\end{align}
where the phase for $\varepsilon=+1$ comes from the oscillatory integral over zero modes and $V$ is the target space volume. Modular invariance of the one-loop amplitude follows in close analogy with the Lorentzian case.

The compact free boson $X^\mu \cong X^\mu + 2 \pi R$ is relevant to our study of T-duality above. In this case, the integral over zero modes becomes a sum,\footnote{For $\varepsilon = +1$, the sum is oscillatory, and can be made convergent by taking $R$ to have a small positive imaginary part.}
\begin{equation}
Z = \frac{1}{|\eta(\tau)|^2}\sum_{n,w=-\infty}^{\infty} \exp\biggl[-\pi \tau_2 \biggl( \sqrt{-\varepsilon}\, \frac{\alpha' n^2}{R^2} + \frac{1}{\sqrt{-\varepsilon}}\, \frac{w^2 R^2}{\alpha'} \biggr) + 2 \pi i \tau_1 n w \biggr]\,,
\end{equation}
for spacelike $X^\mu$. Modular invariance can be established using Poisson resummation on $n$:
\begin{equation}
Z = (2 \pi R) Z_X(\tau,\bar{\tau}) \sum_{m,w=-\infty}^{\infty} \exp\biggl[-\frac{\pi R^2}{\sqrt{-\varepsilon} \alpha'}  \frac{|m-\tau w|^2}{\tau_2} \biggr] \,,
\end{equation}
where $Z_X(\tau,\bar{\tau})$ is the non-compact result,~(\ref{eqn:ZX}). This is manifestly invariant under the modular transformations
\begin{align}
\tau &\longrightarrow \frac{a \tau+b}{c \tau + d}\,, & \begin{pmatrix} m \\ w \end{pmatrix} &\longrightarrow \begin{pmatrix} a & b \\ c & d \end{pmatrix}\!\!  \begin{pmatrix} m \\ w \end{pmatrix} \,.
\end{align}

		\subsection{The effective action}
		
The non-linear sigma model for the bosonic string generalizes to the Euclidean worldsheet as follows 
	\begin{equation}
	\label{eqn:sigmamodel}
		 S = \frac{1}{4\pi \alpha^\prime} \int \mathrm{d}^2\sigma \biggl[ \Bigl( \varepsilon |\gamma|^{1/2}  \gamma^{ab} g_{\mu \nu}(X) - \epsilon^{a b} B_{\mu \nu}(X)\Bigr) \partial_a X^\mu \partial_b X^\nu 
		 -\beta_{\varepsilon} \alpha^\prime |\gamma|^{1/2} \mathcal{R} \Phi(X) \biggr] \,,
	\end{equation}
	where $\mathcal{R}$ is the worldsheet Ricci scalar and $\epsilon^{12} = - \epsilon^{21} = +1$. This follows from the Polyakov action~(\ref{eqn:freeaction}) by replacing $\eta_{\mu \nu} \to g_{\mu \nu}(X)$ and adding a B-field and dilaton in the usual way. For the dilaton term we including an extra phase $\beta_\varepsilon$ (with $\beta_{-1} = 1$), which will turn out to be necessary.\footnote{The phase of the B-field is fixed (up to a conventional overall sign) by the requirement that the path integral weight $e^{i S}$ is invariant under large gauge transformations $\int B \to \int B + (2\pi)^2 \alpha' n$.}

	 In principle, by requiring the beta functions of this sigma model to vanish we can compute the low energy effective action in a derivative expansion. Instead, we relate
	  the Euclidean ($\varepsilon = +1$) and Lorentzian ($\varepsilon = -1$) sigma models to each other and use known results for the latter. In the Lorentzian case, we first Wick rotate to obtain the action
	\begin{equation}
	\label{eqn:sigmamodelLor}
		 i S_E = \frac{i}{4\pi \alpha^\prime} \int \mathrm{d}^2\sigma \biggl[ \Bigl( |\gamma|^{1/2}  \gamma^{ab} g_{\mu \nu}(X) + i \epsilon^{a b} B_{\mu \nu}(X)\Bigr) \partial_a X^\mu \partial_b X^\nu 
		 + \alpha^\prime |\gamma|^{1/2} \mathcal{R} \Phi(X) \biggr] \,.
	\end{equation}
	The Euclidean sigma models~(\ref{eqn:sigmamodelLor}) and~(\ref{eqn:sigmamodel}) (with $\varepsilon = +1$) are related by an analytic continuation of the background fields
	\begin{align} \label{eqn:sigmaModelCrossing0}
	g_{\mu \nu} &\longrightarrow -i g_{\mu \nu}\,, & B_{\mu \nu} &\longrightarrow B_{\mu \nu}\,, & \Phi &\longrightarrow i \beta_{1} \Phi\,,
	\end{align}
	where we equate the path integral weights $e^{i S} = e^{i (i S_E)} = e^{-S_E}$. Since $g_{\mu \nu}(X)$, $B_{\mu \nu}(X)$, and $\Phi(X)$ are essentially the couplings of the sigma model, this can be thought of as an analytic continuation of the couplings.

To fix the phase $\beta_{\varepsilon}$, we expand
\begin{align}
g_{\mu \nu}(X) &= \eta_{\mu \nu} + s_{\mu \nu} e^{i k\cdot X} + \ldots, & B_{\mu \nu}(X) &= a_{\mu \nu} e^{i k\cdot X} + \ldots, & \Phi(X) = \phi e^{i k\cdot X} + \ldots,
\end{align}
in a plane wave background. The coefficients $s_{\mu \nu}$, $a_{\mu \nu}$, and $\phi$ correspond to vertex operators on the worldsheet. By a calculation in~\cite{Polchinski:1998rq}, for the ordinary bosonic string $\phi \propto \eta^{\mu \nu} s_{\mu \nu}$ in light-cone gauge, i.e., the dilaton arises from the trace part of the graviton vertex operator. Making the replacements $\eta_{\mu \nu} \to - i \eta_{\mu \nu}$, $s_{\mu \nu} \to - i s_{\mu \nu}$, $\phi \to i \beta_{1} \phi$ according to~(\ref{eqn:sigmaModelCrossing0}), we obtain $i \beta_{1} \phi \propto \eta^{\mu \nu} s_{\mu \nu}$ for the Euclidean worldsheet theory. Since the dilaton remains the trace part of the graviton, we must have $i \beta_{1} = 1$ up to a conventional sign, so that $\beta_{\varepsilon} = 1/\sqrt{-\varepsilon}$.

Using the above results, we can fix the two-derivative effective action of the bosonic string:
\begin{equation}
S=\frac{1}{2 \kappa_{26}^2} \int \de^{26} x \sqrt{|g|} e^{-2 \Phi} \Bigl[\mathcal{R} + \frac{\varepsilon}{2} |H|^2 + 4 (\nabla \Phi)^2 \Bigr] \,,
\end{equation}
for $\varepsilon = \pm 1$. Here we apply the analytic continuation~(\ref{eqn:sigmaModelCrossing0}) to the known result for $\varepsilon = -1$. Apart from the different critical dimension of the bosonic string, this agrees precisely with the NSNS part of the superstring result~(\ref{eqn:sugraaction}), which we derived by duality arguments. Moreover, the analytic continuation~(\ref{eqn:sigmaModelCrossing0}) is essentially the singularity crossing prescription~(\ref{eqn:crossing}). The only difference is the signature-dependent phase acquired by $e^{-2 \Phi}$ in~(\ref{eqn:crossing}), but in the absence of RR fields this only affects loop calculations, which we have not addressed.
 
In particular,~(\ref{eqn:sigmaModelCrossing0}) is sufficient to reproduce the $i$ prefactor of the $e^{-2 \Phi} \mathcal{R}^4$ curvature correction for Euclidean worldsheets that we found in~(\ref{eqn:R4allsig}). Heuristically, since derivative corrections introduce new propagating degrees of freedom, the imaginary $\mathcal{R}^4$ correction can be thought of as a consequence of the imaginary spectrum of massive string modes for the Euclidean theory.

\medskip

A more detailed development of the worldsheet description of the exotic string theories considered in our work would likely provide further insight into their properties, but we leave this for a future work.

\section{Non-Perturbative Dynamics of Supergroup Gauge Theories} \label{sec:seibergwitten}

For much of this paper, we have focused on isolated negative branes. In this section, we consider supergroup gauge theories, arising from positive and negative branes together. We present evidence that these gauge theories exist non-perturbatively.

\subsection{The difference between $U(N | M)$ and $U(N - M)$}

We begin by commenting on a subtle issue associated with supergroup gauge theories. A long-standing observation about these theories is that $U(N | M)$ closely mimics $U(N - M)$~\cite{Yost:1991ht, AlvarezGaume:1991zc, Okuda:2006fb}. Consider for example gauge-invariant correlation functions in the $\mathcal{N}=4$ theory discussed above. Using the 't Hooft double line notation for Feynman diagrams, $N$ and $M$ only appear in the combination $\Str 1 = N - M$, so that $U(N | M)$ is perturbatively equivalent to $U(N - M)$!

The partition function of the $\mathcal{N}=4$ $U(N)$ theory on $S^4$ is given by that of a Gaussian matrix model~\cite{Pestun:2007rz}. By a supersymmetric localization argument the same should apply to supergroup gauge theories, and we can learn something about their non-perturbative physics by studying Gaussian supermatrix models. Recently it was observed that, despite earlier claims to the contrary~\cite{AlvarezGaume:1991zc}, the $U(N | M)$ supermatrix model is not equivalent to the $U(N - M)$ matrix model~\cite{Vafa:2014iua}, as there are gauge invariant operators whose expectation
value is zero in the latter but non-zero in the former.  It turns out that even the exact partition
functions of the two theories are not the same:  The differences appear to arise from subtleties associated to fermionic gauge symmetries~\cite{supermatrix}, which are best resolved by gauge-fixing.

This suggests that $U(N | M)$ and $U(N - M)$ are inequivalent theories, distinguished at least by their behavior on $S^4$, and likely in other ways as well. If so, this has interesting consequences for negative branes in string theory. For instance, it implies that an equal number of superposed positive and negative branes is not equivalent to the string vacuum, contrary to~\cite{Okuda:2006fb}. Nonetheless, this configuration looks identical to the vacuum from the perspective of closed strings in perturbation theory, and it is not immediately clear what distinguishes it non-perturbatively. One possibility is that the supermatrix model only sees a difference because it describes a brane wrapping a compact cycle, for which the Coulomb branch is integrated over, as opposed to a brane with a non-compact worldvolume, for which the Coulomb branch parameterizes different superselection sectors.  However, this cannot be the full
answer, because the set of operators in the  $U(N | M)$ theory is strictly bigger than the
set of operators in the $U(N - M)$  theory.

\subsection{Negative D-brane intersections and negative matter} \label{subsec:negmat}

As is well known, the boundary of a brane can live on the worldvolume of another brane. Given the above discussion, we anticipate that this kind of configuration can also be realized with negative branes.
For example, consider $N$ D4 branes ending on a pair of parallel NS5-branes---this should give
rise to ${\cal N}=2$ $SU(N)$ Yang-Mills in four dimensions. Now instead, imagine a collection of $N$ ordinary D4-branes and $M$ negative D4-branes. Since ${\cal N}=2$
SYM makes sense, this construction should be able to accommodate negative branes. In particular, suspending $N$ ordinary D4 and $M$ negative D4-branes between two NS5-branes should produce a $SU(N|M)$ gauge theory with
${\cal N}=2$ supersymmetry. As far as the mechanism of branes ending on branes is concerned, the only difference between a positive brane
and a negative brane is that the respective charges they induce on the boundary are opposite; the amount of supersymmetry
they preserve is the same.

The above example also illustrates the type of matter these theories can have.  Consider $N$ D4-branes suspended between a pair of parallel NS5-branes, and add one D4-brane ending on the other side of one of the NS5-branes.  This gives rise to a theory with one fundamental $SU(N)$ matter. In terms of geometric engineering, this
configuration can be viewed as the result of breaking a supersymmetric $SU(N+1)$ theory to an $SU(N)\times U(1)$ theory with some
of the matter of the former theory appearing as matter content in the latter theory.  Now we might ask what happens if we replace the D4-brane
on the other side of the NS5-brane with a negative D4 brane? Similarly, this kind of configuration can be viewed as the result of breaking of a $SU(N|1)$ theory
to a $SU(N)\times U(0|1)$ theory, where the extra matter is now in the fundamental representation of $SU(N)$ with opposite statistics. We call this ``fermionic'' or ``negative'' fundamental matter.

\subsection{$\mathcal{N}=2$ supergroup theories}

As discussed in~\S\ref{sec:supergroup}, in order to provide evidence for the non-perturbative existence
of gauge theories with supergroup symmetry, it would be useful to perform some exact non-perturbative computations
in these theories. It turns out to be possible to do computations similar to those done in usual case of ordinary Lie groups.  In this section we explain how such computations can be carried for ${\cal N}=2$ supersymmetric theories with
gauge supergroup $SU(N|M)$. 

First of all we have to note that the negative of the beta function for an ${\cal N}=2$ $SU(N)$ theory is proportional to $N$, which implies
that in the supergroup case the negative of the beta function is proportional to $N-M$. If $N>M$ we have an asymptotically free theory; if $N=M$
we should get a conformal theory; and if $N<M$ the theory is not UV complete. We will study the Seiberg-Witten (SW) curve
for this theory, which will depend on $N+M-1$ parameters. In fact, as we will explain shortly, we learn that the SW curve in this case is
the same as that of an ordinary $SU(N)$ gauge theory with $2M$ fundamental matters, where the masses of the $2M$ flavors are pairwise
equal. Such a theory has $(N-1)$ Coulomb and $M$ mass parameters, also leading to a total of $N+M-1$ parameters.  
In this context, we find that the Seiberg-Witten curve has genus $N-1$,
suggesting that the coupling constants for the $SU(M)$ part do not get corrected.  Moreover, the Coulomb branch parameters
of the $SU(M)$ part behave as if they are the mass parameters of the $SU(N)$ theory.

Recall that for an ordinary $SU(N)$ theory, the Seibeg-Witten curve is given by
\begin{equation}
z+\text{det}(x-\Phi)+{1\over z}=0,
\end{equation}
with $\Phi$ representing the scalar $SU(N)$ matrix of the Coulomb branch parameters. The
SW differential is given by 
\begin{equation}
\lambda=x{dz\over z}.
\end{equation}
The SW curve we find for the $SU(N|M)$ gauge theory is given again by the same formula except that the determinant
is replaced with the super-determinant:
\begin{equation}
z+\text{sdet}(x-\Phi)+\frac{1}{z}=0.
\end{equation}
In particular, if we diagonalize, $\Phi= {\rm diag}(a_1,...a_N;b_1,...,b_M)$, we get the SW curve
\begin{align}
\label{eqn:SWcurve1}
z+{\prod_{i=1}^N (x-a_i)\over \prod_{j=1}^M (x-b_j)}+{1\over z}=0,
\end{align}
which can be expressed alternatively as
\begin{align}
\label{eqn:SWcurve2}
z\prod_{j=1}^M (x-b_j) +\prod_{i=1}^N (x-a_i)+{ 1\over z} \prod_{j=1}^M (x-b_j) =0
\end{align}
In this form we recognize this as the SW curve for $SU(N)$ theory with $2M$ fundamental flavors with two fundamentals
of masses $b_j$.

We will now derive the above results using negative branes in string theory in two different ways, as well as check them
against the instanton calculus of Nekrasov.

\subsection{The brane perspective}
As discussed in~\S\ref{subsec:negmat}, following the construction in \cite{Witten:1997sc}, we can construct ${\cal N}=2$
supersymmetric $SU(N|M)$ gauge theories by suspending $N$ ordinary D4-branes and $M$ negative D4-branes
between two parallel (ordinary) NS5-branes. We now show how we can use this picture
to obtain the SW curve for this theory. 

\begin{figure}
\begin{center}
\includegraphics[width=\textwidth]{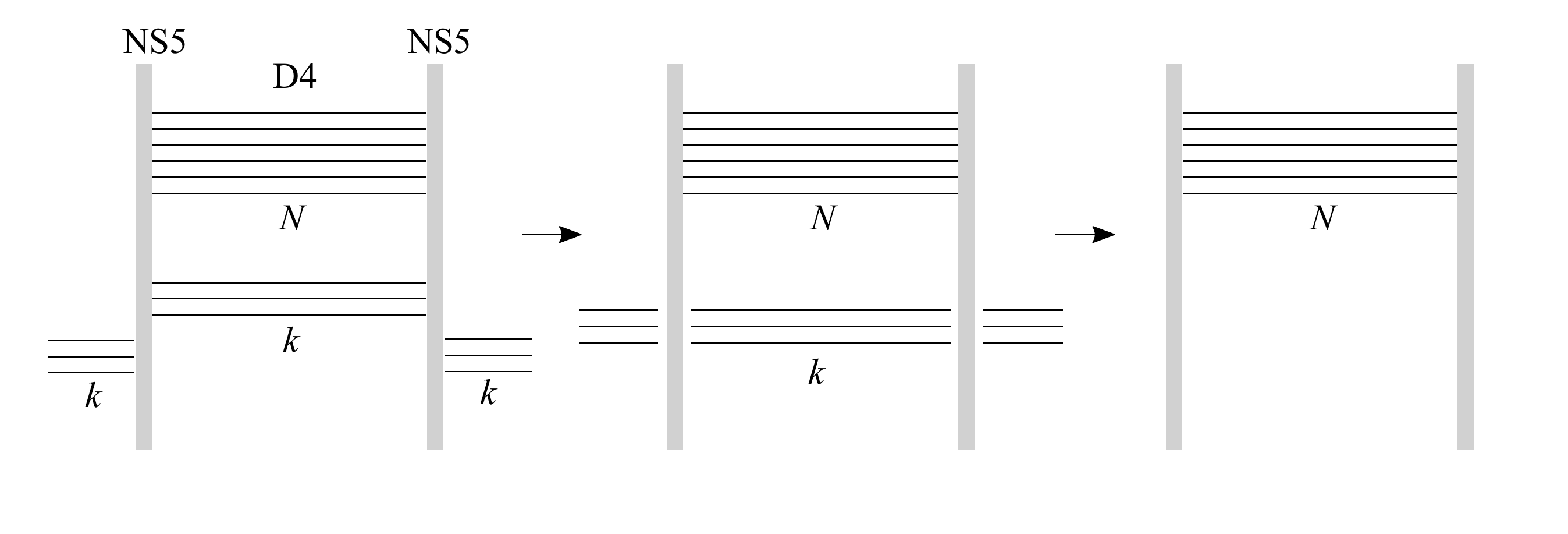}
\caption{A configuration of D4 branes suspended between two NS5 branes that reproduces a $SU(N+k)$ gauge theory with $2k$ flavors. It can be Higgsed to a pure $SU(N)$ theory.}
\label{fig:suspendedbranes}
\end{center}
\end{figure}

As a warmup exercise, let us consider an $SU(N+k)$ theory with $2k$ flavors,
where the Coulomb branch and flavor masses are such that the $SU(N+k)$ theory
can be Higgsed to an $SU(N)$ theory as depicted in Figure~\ref{fig:suspendedbranes}. In other words
we are now in a configuration where we can remove $k$ D4-branes that pass through both NS5-branes.
 This subspace of mass and Coulomb
branch parameters of the $SU(N+k)$ theory with $2k$ flavors will have the same SW geometry
as a pure $SU(N)$ theory because the hypermultiplet and Coulomb branch moduli
spaces are decoupled.  We could have described this setup in an ``inverse'' manner:  adding extra
D4 branes that pass through both NS5-branes of the pure $SU(N)$ theory does not
change the geometry of the Coulomb branch. We now use this idea to solve for the SW curve
of the $SU(N|M)$ theory.    

\begin{figure}
\begin{center}
\includegraphics[width=\textwidth]{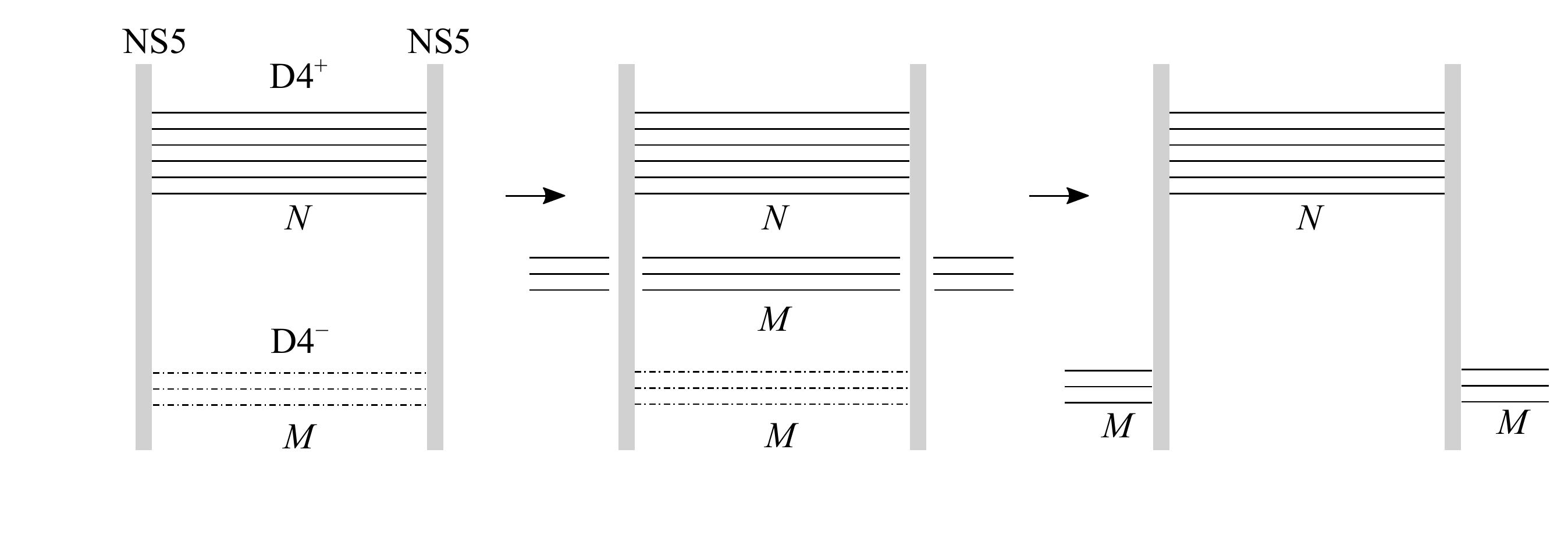}
\caption{A configuration of positive and negative D4 branes suspended between two NS5 branes that reproduces a $SU(N|M)$ gauge theory. It can be deformed to a  $SU(N)$ theory with $2M$ flavors. }
\label{fig:posnegbranes}
\end{center}
\end{figure}

The idea is simple: Consider now the brane realization of a $\mathcal N = 2 $ $SU(N|M)$ gauge theory and introduce $M$ additional ordinary D4-branes which
pass through both NS5-branes. As already discussed, this will not affect the SW geometry of the Coulomb
branch.  Next, move these additional $M$ D4-branes so that they coincide with the $M$ negative D4 branes
in the region between the two NS5-branes.  In this way, the $M$ D4-branes and $M$ negative D4-branes
cancel each other out in the region between the parallel NS5-branes and we are left with only positive branes!
More precisely, the resulting configuration consists of $N$ ordinary D4-branes suspended between the NS5-branes, $M$ semi-infinite ordinary D4-branes attached to the `outside' of one NS5-brane and $M$ semi-infinite D4-branes attached the `outside' of the \emph{other} NS5-brane, see Figure~\ref{fig:suspendedbranes}.  This configuration is exactly the same
as the brane construction for a $SU(N)$ theory with $2M$ flavors where the masses of the flavors are
pairwise equal to the Coulomb branch parameters for the $U(M)\subset SU(N|M)$ part of the
theory and hence leads to the SW curve given in (\ref{eqn:SWcurve2}).
 
Incidentally, the construction we have discussed here also explains some of the results in~\cite{Mikhaylov:2014aoa} where
it was shown that $N$ and $M$ D3-branes ending on opposite sides of an NS5-brane
engineer a $U(N|M)$ Chern-Simons theory on the three dimensional boundary.  This configuration can be derived
from the present set up by considering $N$ ordinary D3 branes and $M$ negative D3 branes
on the same side, which manifestly realizes a ${\cal N}=4$ $U(N|M)$ theory and naturally
leads to a $U(N|M)$ Chern-Simons theory on the boundary in the same fashion.  Adding
$M$ ordinary D3 branes passing through the NS5-branes and coinciding with the negative D3 branes reproduces the two-sided configuration (involving only positive D3 branes) realizing a boundary $U(N|M)$ Chern-Simons theory as described in~\cite{Mikhaylov:2014aoa}.

\subsection{Geometric engineering and mirror symmetry} \label{subsec:geomeng}
Next we use geometric engineering to identify the SW geometry.  Let us recall
  how this is done
in the $SU(N)$ case~\cite{Klemm:1996bj, Katz:1997eq} and then generalize the construction to the supergroup case.

We consider an $A_{N-1}$ singularity in ordinary type IIA string theory, 
\begin{equation}
x^N = uv,
\end{equation}
which gives rise to a $SU(N)$ gauge theory in $d=6$.
We then compactify this theory on a ${\mathbb P}^1$ which gives rise to an ${\cal N}=2$, $SU(N)$
gauge theory in $d=4$. Next, we apply mirror symmetry to this geometry to obtain a type IIB geometry,
which leads to the SW geometry. The mirror geometry can be assembled in pieces:
The mirror of ${\mathbb P}^1$ leads to the $z+1/z$ monomial
and the mirror of $SU(N)$ geometry leads to $\text{det} (\Phi -x)=uv$, the total being 
the local Calabi-Yau 3-fold geometry
\begin{equation}
z+\text{det}(\Phi -x)+{1\over z}=uv.
\end{equation}
As usual, the SW curve is the surface where we set the right hand side of the above equation equal to zero.
Now we apply the same idea to a $SU(N|M)$ theory with ${\cal N}=2$.  Here, to engineer a
 geometry corresponding to the $SU(N|M)$ theory we require a multi-center Taub-NUT geometry, specifically with $N$ positive charge and $M$ negative charge centers,
as discussed in~\S\ref{sec:diffsigs}.  

Moreover, since mirror symmetry in two complex
dimensions is simply a hyperK\"ahler rotation, to get the type IIB mirror we only need
to write the complex geometry associated with this TN geometry. 
Recall that the standard Taub-NUT space on $S^1 \times {\mathbb R}^3$ is given by
\begin{equation}
ds^2 = {1\over V}(d\theta + A)^2 + V dy^2,
\end{equation}
with
\begin{equation}
V=1 + \sum_i \frac{N_i}{2 |\vec{y} -\vec{y}_i|},
\end{equation}
where the $N_i$ are positive integers and $A$ is the vector potential dual to $V$, $\partial_i V = \epsilon^{i j k} \partial_j A_k$. It can be given a complex structure by defining complex coordinates $x = y^1+ i y^2$ and 
\begin{align}
\log u &= y^3 +i\theta + \sum_i {N_i \over 2} \log\bigl(|\vec{y} - \vec{y}_i| +( y^3 - y_i^3)\bigr),\\
\log v &= -y^3 - i\theta' + \sum_i {N_i \over 2} \log\bigl(|\vec{y} - \vec{y}_i| - (y^3 - y_i^3)\bigr),
\end{align}
where $\theta$ and $\theta'$ are the local coordinates on $S^1$ associated to vector potentials with Dirac strings located at $x = x_i$ and $y^3 < y_i^3$ or $y^3 > y_i^3$, respectively. The transition function is
\begin{equation}
\theta - \theta' = \sum_i N_i \arg(x - x_i) \,,
\end{equation}
therefore $x, u, v$ obey the equation
\begin{equation} \label{eqn:defANsing}
u v = \prod_i (x-x_i)^{N_i} \,,
\end{equation}
with the holomorphic two-form $\Omega = \frac{d u}{u} \wedge d x=-\frac{d v}{v} \wedge d x$. This describes a collection of $A_{N_i - 1}$ singularities at the points $u=v=0$, $x=x_i$, corresponding to the Taub-NUT charges located at $\vec{y} = \vec{y}_i$.

The above analysis can be extended to the case with both positive and negative Taub-NUT charges $N_i$, with the result that the right-hand side of~(\ref{eqn:defANsing}) will contain poles as well as zeros. This can be cast in terms of a superdeterminant with eigenvalues $a_i$, $b_j$:
\begin{align} \label{eqn:defsupersing}
\text{sdet}(\Phi -x) = \frac{\prod_i (x-a_i)}{\prod_j (x - b_j)} =uv.
\end{align}
Putting these two pieces together we obtain the mirror geometry
\begin{align}
z+\text{sdet}(\Phi -x) +{1\over z}=0\,,
\end{align}
as was to be shown.

There is an interesting subtlety in this derivation. By construction, every point in the Taub-NUT geometry is mapped to a point on the complex hypersurface~(\ref{eqn:defsupersing}), but the mapping is not one-to-one in the presence of poles (negatively charged). Since $x = y^1 + i y^2$ and $\arg u = \theta$ are shared between the two descriptions, the injectivity of the map depends on
\begin{equation}
\frac{\partial \log |u|}{\partial y^3} = V \,.
\end{equation}
So long as $V>0$ everywhere (or $V<0$ everywhere), the map is one-to-one, but precisely when $N_i < 0$ for some $i$, there is a $V=0$ surface surrounding each negative charge, and the map is not one-to-one in this vicinity.\footnote{It is interesting to note that this $V=0$ surface is also a signature-changing singularity.}

In other words, the Taub-NUT geometry is a multiple cover of~(\ref{eqn:defsupersing}), with a complicated and unusual map between the two near the negative charges. Since the Seiberg-Witten curve depends only on the complex structure, it is possible that this subtlety does not affect its calculation. However, this point deserves further study, which might provide additional insight into the dynamics of the corresponding $\mathcal{N}=2$ supergroup gauge theory.

\subsection{Instanton Calculus}
We can also use Nekrasov's instanton calculus.  It is easiest to first study the $SU(N|N)$ case
and then generalize to $SU(N|M)$ for $N>M$ case by taking $N-M$ of the Coulomb branch parameters for the
$SU(0|M)$ part to be large.
 One way to solve for the SW curve of the $SU(N|N)$ theory, as was explained to us by
Nekrasov, is to relate this problem to the following ordinary ${\cal N}=2$ theory:  $SU(N)\times SU(N)$
with two bifundamentals where the coupling constantes of the two $SU(N)$s have opposite
signs $\tau_1=-\tau_2$, or in the exponentiated form, $q_1q_2=1$.  The reason this comes
about is that if we first break the $SU(N|M)$ theory to a $SU(N)\times SU(M)$ theory, the
off-diagonal blocks, which are fermionic, have ghosts associated with them that behave
as if they are ordinary matter. Therefore, the ${\cal N}=2$ instanton calculus
and localization computation maps the supergroup case to the ordinary ${\cal N}=2$
case noted above, with one restriction:  Since there is only one coupling $\tau$ in the $SU(N|N)$ theory
and the coupling of the $SU(N|0)$ and $SU(0|N)$ theories differ by a sign, because of the supertrace
we need to impose $\tau_1=-\tau_2=\tau$.  Of course this is different from the physical
region of the $SU(N)\times SU(N)$ theory where $\text{Im }\tau_1$ and $\text{Im }\tau_2$ are both positive.
Nevertheless we can analytically continue the answer for the $SU(N)\times SU(N)$ theory to the
case of interest for the supergroup case.

The SW curve for the $SU(N)\times SU(N)$ with two bi-fundamental matter fields has been
worked out in~\cite{Nekrasov:2012xe} (see (7.81)) and was found to be:
\begin{align}
 \left({q_2\over q_1} \right)^{1/4}{{\vartheta}_{2}( z^2; q^2) \over {\vartheta}_{3} (z^2; q^2)} = {P_{2}(x)\over P_{1}(x)}\,,
 \end{align}
where $P_1(x)$ and $P_2(x)$ are the polynomials controlling the Coulomb branches
of the two $SU(N)$s and $q=q_1q_2$.  We are instructed to take $q\rightarrow 1$.  Using the product formulas 
\begin{align}
\vartheta_3(z^2; q^2) &= \prod_{m=1}^\infty 
\left( 1 - q^{2m}\right)
\left( 1 + q^{2m-1}z^2\right)
\left( 1 + q^{2m-1}/z^2\right)\,, \\
\vartheta_{2}(z^2;q^2) &= 2 q^{1/4}(z+z^{-1})\prod_{m=1}^\infty 
\left( 1 - q^{2m}\right)
\left( 1 + q^{2m}z^2\right)
\left( 1 + q^{2m}/z^2\right)\,,
\end{align}
we see that 
\begin{equation}
{\vartheta_3\over \vartheta_2}\overset{q\rightarrow 1}{\longrightarrow} 2(z+z^{-1}).
\end{equation}
Identifying
\begin{equation}
{P_{2}(x)\over P_{1}(x)}=\text{sdet}(\Phi-x) 
\end{equation}
 leads to
\begin{equation}
2q_1^{-1/2}(z+z^{-1})-\text{sdet}(\Phi-x)=0\,,
\end{equation}
which, up to rescaling of $z$ and a choice of the scale in the theory leads to the curve (\ref{eqn:SWcurve1}) obtained using the other string theoretic methods.

\section{Potential Issues and Concluding Remarks} \label{sec:problems}

In this section, we discuss some potential issues with the ideas discussed in this
paper and their possible resolution. We hope to address some of these issues in future work.

\subsection{The Cauchy problem} \label{subsec:Cauchy}

Field theories with multiple times generically do not have a well posed initial value problem. While the spacetime can be foliated by choosing a fiducial time direction, the initial value surface now has mixed signature, implying that signals can propagate \emph{along} the surface, rather than just forwards in ``time''. This means that for generic initial data there is no solution to the field equations. A solution can be guaranteed by specifying initial data on a spacelike hypersurface, but these have codimension greater than one, so the resulting solution is highly non-unique.

While this naively seems to rule out classical determinism in a theory with multiple times, the issue can be solved---at least in the free field case~\cite{craig-weinstein,Weinstein:2008aj}---by imposing constraints on the initial data sufficient to guarantee a solution. As analyzed in~\cite{craig-weinstein}, a sufficient condition is to restrict the support of the initial data to spacelike/null momenta along the initial value surface. This is an interesting, non-local constraint on the physics which also removes unstable exponentially growing and decaying modes from the theory.

While it remains unproven whether a similar condition can be applied to an interacting theory, this suggests that issues with determinism in multiple time theories may be circumventable. A more thorough development of this topic would be of great physical and mathematical interest.

 \subsection{Stability}

If negative branes exist then ordinary string theory is potentially unstable. For instance, in type IIA string theory negative D0 branes have negative squared mass, and appear to be tachyons. More generally, nothing seems to prevent the pair creation of an arbitrary number of negative branes, a process which releases energy due to their negative tension.

We suggest two possible resolutions. Firstly, one could speculate that this problem is analogous to that encountered by Dirac in quantizing the electron field. Perhaps what is required is an appropriate analog of ``filling the negative brane sea.'' Unfortunately, branes do not obey the Pauli exclusion principle, so it is unclear how ``filling the sea'' will prevent the instability, but some yet unknown variation of this idea may succeed.

More concretely, we have seen that negative branes induce a dynamic change of spacetime signature. This suggests that it is perhaps too naive to treat them as negative energy probes. In particular, while we have focused exclusively on BPS configurations of branes, brane pair creation is inherently non-BPS, and it may be that the spacetime associated to negative brane pair creation actually has \emph{positive} energy, due to the need to create a bubble connecting to one of the exotic string theory vacua. Indeed, large contributions to the energy may arise from the divergent background fields near the singular signature-changing domain wall.

Unfortunately, while this idea is plausible, it is nearly impossible to check: the only reason we have been able to work with singular spacetimes with any confidence is the high degree of supersymmetry involved. Calculating the tension of the signature-changing domain wall without supersymmetry is presently out of reach. One approach would be to study negative D0 branes exclusively and seek a smooth solution describing their pair creation in the M-theory lift. Even this is technically quite involved, and we will not attempt it in the present work.

Another process of interest is the mixing between vacua with no branes and those with mutually BPS coincident pairs of positive and negative branes. Since the brane tensions and charges cancel, these vacua are nearly indistinguishable from each other, and there may be instantons connecting them. If so, the true vacuum of the theory will involve a superposition, and may realize the vague notion of ``filling the sea'' discussed above.

Addressing such questions of stability is imperative for the consistency of negative branes and the associated exotic string theories. We leave this as an important open problem for future work.

 \subsection{Complex actions and holography}

Another confusing aspect of many of the exotic string and field theories we have described are the complex effective actions that naturally appear, as we saw in the context of $\mathcal{R}^4$ corrections in~\S\ref{sec:AdSCFT}--\ref{sec:worldsheet}. However, in some situations this appears to be a feature, not a bug. Besides the natural connection between imaginary curvature corrections and branch cuts in the $\lambda$-plane that we discovered in~\S\ref{sec:AdSCFT}, holography with emergent time such as the dS/CFT correspondence \cite{Strominger:2001pn} appears to be another such example.
  Conventional logic implies that CFTs dual to de Sitter space are pure Euclidean. However, a holographic representation of quantum mechanics will necessarily lead to complex amplitudes, reproducing the familiar quantum phase factors $e^{iS/\hbar}$. This suggests that the corresponding Euclidean CFT should have a complex, or even purely imaginary action.
  
  Note that the path-integral for a Euclidean supergroup gauge theory
\begin{equation}
\int [DA] ~e^{i\int \Str F^2} \,,
\end{equation}
is less problematic then the (Wick-rotated) Lorentzian case, since it is merely rapidly oscillating instead of divergent, and may admit a convergent regulator.  
Of course, the issue with complex actions in the exotic string theories we consider is somewhat more severe
because, unlike the above example, these actions do not have a definite phase. Nonetheless, while puzzling in its implications for semiclassical dynamics, a complex effective action is not obviously inconsistent.

\subsection{Consistency of the low energy limit} \label{subsec:lowenergyconsistency}

In section~\S\ref{sec:curvature} we have attempted to extract as much information as possible about the exotic string theories from their low energy effective actions. This is advantageous, as these actions are perhaps simpler to work with than any worldsheet description, are available in M-theory as well as string theory, and are derived from dualities and known results (unlike the consistent but conjectural Euclidean worldsheet description developed in~\S\ref{sec:worldsheet}).

Nonetheless, there could be problems with taking the low energy limit in a theory without a positive-definite notion of energy. Because of the existence of negative energy excitations, in principle such a theory can generate high energy excitations from minimal input energy. If this occurs, the low-energy effective description would likely be invalidated. However, the solution to the stability problem mentioned above may prevent this from happening. So long as negative energy excitations only appear off shell and not in the final state, high energy modes cannot be produced on shell either, and an effective field theory approach may still be justified.

To illustrate these issues, we consider the exotic string theories IIB$^{+-}_{9,1}$ and IIB$^{-+}_{9,1}$. These can be described as M-theory on a mixed signature torus $T^{1,1}$ in the small volume limit, by analogy with the usual M-theory construction of F-theory. Similar to the discussion in~\S\ref{subsec:Cauchy} above, null rays can propagate along the $T^{1,1}$ due to its mixed signature. As a result, some KK modes will be massless for most choices of metric on $T^{1,1}$. In particular, the KK mode masses are
\begin{equation} \label{eqn:T11spectr}
m^2_{\rm KK} = - \frac{1}{R^2}\cdot \frac{2 (m \tau_+ - n)(m \tau_- - n)}{\tau_+-\tau_-} \,,
\end{equation}
where the metric on $T^{1,1}$ takes the form\footnote{To allow for all possible non-degenerate metrics on $T^{1,1}$, we take $\tau_{\pm} \in \mathbb{R} \cup \{\infty\}$ with $\tau_+ \ne \tau_-$.}
\begin{equation}
ds^2 = \frac{1}{\tau_2} (d x+\tau_1 dy)^2 - \tau_2 dy^2 = \frac{2}{\tau_+ - \tau_-} (d x+\tau_+ dy)(dx+\tau_- dy)\,,
\end{equation}
for $\tau_{\pm} = \tau_1 \pm \tau_2$, and $\tau_1 = C_0$, $\tau_2 = e^{-\Phi}$ ($\tau_2 = - e^{-\Phi}$) in the dual type IIB$^{+-}$ (IIB$^{-+}$) description. Here $x \cong x+2\pi R$ and $y \cong y+2\pi R$ parameterize the fundamental cycles of $T^{1,1}$, and $\tau_+ \ne \tau_-$ describe its shape. Modular transformations on the torus map $\tau_\pm \to \frac{a \tau_\pm + b}{c \tau_\pm + d}$, for $a,b,c,d\in\mathbb{Z}$, $a d -b c=1$.

When either $\tau_+$ or $\tau_-$ is rational, there are infinitely many non-trivial solutions to $m_{\rm KK}^2 = 0$ in~(\ref{eqn:T11spectr})! Even when $\tau_+$ and $\tau_-$ are both irrational, there are an infinite number of KK modes in the range
\begin{equation} \label{eqn:KKmodedensity}
-\frac{2}{\sqrt{5}}-\varepsilon < m^2_{\rm KK} R^2 < \frac{2}{\sqrt{5}}+\varepsilon \,,
\end{equation}
for any $\varepsilon >0$,\footnote{This can be shown using Hurwitz's theorem: for irrational $\tau_-$ there are infinitely many $m$, $n$ with $\gcd(m,n)=1$ such that $|\tau_- - \frac{n}{m}| < \frac{1}{\sqrt{5} m^2}$. Putting this into the mass formula and noting that $\frac{\tau_+ - n/m}{\tau_+ - \tau_-} = 1 + O(1/m^2)$ for $m \gg 1$, we obtain the desired result.} which is almost as bad. The failure of KK modes to decouple suggests that naive dimensional reduction on $T^{1,1}$ may be inconsistent.

\paragraph{$\mathcal{R}^4$ corrections at strong coupling}

To further illustrate the problem, we reconsider the calculation of the $\mathcal{R}^4$ corrections in these theories. Our discussion in~\S\ref{sec:curvature} ignored non-perturbative corrections in $\gs$, which are present in type IIB string theory, and are generated by integrating out the KK modes on $T^2$ in the M-theory description, or equivalently by D-instantons in the type IIB description. In ordinary IIB$_{9,1}^{++}$, these corrections take the form~\cite{Green:1997di}
\begin{equation}
S_{\mathcal{R}^4} \sim \int \de^{10} x \sqrt{|g|} f(\tau,\bar{\tau})  \mathcal{R}^4\,,
\end{equation}
in Einstein frame, where $\tau = C_0 + i e^{-\Phi} = \tau_1 + i \tau_2$ and
\begin{equation} \label{eqn:Eisenstein}
f(\tau,\bar{\tau}) = \zeta(3) E_{\frac{3}{2}}(\tau,\bar{\tau}) = \sum_{(m,n)\ne (0,0)} \frac{\tau_2^{3/2}}{|m \tau - n |^{3}} 
= 2 \zeta(3) \tau_2^{3/2} + \frac{2 \pi^2}{3} \tau_2^{-1/2} +O(e^{-2 \pi \tau_2}) \,.
\end{equation}
Here the summand is related to the KK mode masses, $m_{\rm KK}^2 R^2 = \frac{|m \tau - n|^2}{\tau_2}$, and modular invariance is manifest because the various KK modes are interchanged under $\tau \to \frac{a \tau+ b}{c\tau+d}$, $ad-bc =1$.
The first two terms are the string tree-level and one-loop corrections discussed in~\S\ref{sec:curvature}, and the remaining corrections are suppressed by factors of $e^{-2\pi/g_s}$.

Since the additional corrections are small, ignoring them is justifiable in this case. However, for IIB$^{+-}_{9,1}$, the sum over KK modes becomes
\begin{equation} \label{eqn:splitEisenstein0}
\sum_{(m,n)\ne (0,0)} \biggl(\frac{\tau_+-\tau_-}{2 (m \tau_+ - n)(m \tau_- - n)} \biggr)^{3/2} \,,
\end{equation}
which diverges regardless of $\tau_\pm$ due to the infinite number of KK modes satisfying~(\ref{eqn:KKmodedensity})!

We can attempt to regulate~(\ref{eqn:splitEisenstein0}), of course, but most regulators will break modular invariance. There are two simple possibilities that respect modular invariance:
\begin{enumerate}
\item
Assume that $\Im\tau_\pm \ne 0$. The sum converges, and is invariant under $\tau_\pm \to \frac{a \tau_\pm + b}{c \tau_\pm+d}$, but we cannot restrict $|\Im\tau_\pm| \ll 1$ in a modular invariant fashion, so the underlying physics is complexified.
\item
Assume that $\tau_\pm \in \mathbb{Q} \cup \{\infty\}$ and remove the terms from the sum where $n/m = \tau_+$ or $n/m = \tau_-$. The result converges and is modular invariant, but is not a smooth function of $\tau_\pm$.
\end{enumerate}
Both approaches have their disadvantages.\footnote{Another possibility would be to form new modular invariant combinations, similar to what is done for theta functions for indefinite lattices.  We leave the study of such a possibility to future work.} Ideally, we might like a smooth modular invariant function of real $\tau_\pm$, but this can be shown to be impossible.\footnote{Suppose that $f(\tau_+,\tau_-)$ is smooth for generic $\tau_\pm$ and modular invariant. Define $f_+(\tau_+) = f(\tau_+,\infty)$, $f_-(\tau_-) = f(\infty, -\tau_-)$. Because the modular group maps $\infty$ to every rational point, $f_\pm$ must also be smooth at a generic point by assumption.
Modular invariance requires $f_\pm(x+1)=f_\pm(x)$ and $f_+(a/c) = f_-(d/c)$ for any $a,d,c$ such that $a d \equiv 1 \pmod c$. Using $a = p^2 k$, $d = q^2 k$, $c=p q k+1$ for any $p,q,k$, we conclude that $f_+(p^2 x) - f_-(q^2 x)$ has zeroes $x_k=\frac{k}{p q k +1} = \frac{1}{p q}\bigl(1-\frac{1}{pq k+1}\bigr)$ which accumulate near $x = 1/pq$, implying a singularity there unless the difference vanishes everywhere. Therefore, if $f_\pm(x)$ both have smooth regions, then $f_+(p^2 x) - f_-(q^2 x)$ for more than one value of $p,q\ne0$, but then $f_\pm(x+1)=f_\pm(x)$ implies $f_\pm(x)$ is constant, therefore $f(\tau_+,\tau_-)$ is also constant.}

We will not attempt a definitive resolution this puzzle in the present paper. The appearance of complex actions discussed above suggests that the first option may be the right one.
 We note, however, that the second option has an intriguing parallel to~\S\ref{subsec:Cauchy}, in that we remove by hand certain Fourier modes along $T^{1,1}$, in this case those with null momenta. In type IIB language, this corresponds to removing certain tensionless $(p,q)$ string bound states,\footnote{Another possibility is that the correct low-energy description is a higher spin theory generated by these tensionless strings.} whereas in M-theory language we remove the massless non-zero modes. This cures the problems with the KK spectrum~(\ref{eqn:T11spectr}), in that the number of KK modes with mass below any fixed threshold is finite.

Moreover, the series~(\ref{eqn:splitEisenstein0}) now has a finite sum. Consider the generalized Eisenstein-like series~\cite{rsk2005}
\begin{equation} \label{eqn:splitEisenstein}
\tilde{E}_s^{\pm}(\tau_+,\tau_-) \equiv \frac{1}{\zeta(2s)} \sum_{\substack{(m,n)\ne (0,0) \\ n/m\ne\tau_\pm}} \biggl(\frac{\tau_+-\tau_-}{2 (m \tau_+ - n)(m \tau_- - n)} \pm i \epsilon \biggr)^s \,,
\end{equation}
for $\tau_{\pm} \in \mathbb{Q} \cup \{\infty\}$, $\tau_+ \ne \tau_-$, where $\epsilon>0$ indicates the choice of branch cut. This can be summed explicitly 
in terms of the Hurwitz zeta function $\zeta(s,x) \equiv \sum_{n=0}^\infty (x+n)^{-s}$. Fixing $\tau_+ = \infty$ by a modular transformation, the result can be expressed in terms of $\tau_- = q/p$, $p>0$, $\gcd(p,q)=1$:
\begin{equation}
\tilde{E}_s^{\pm}(\tau_+,\tau_-) = \frac{2}{(2p)^s \zeta(2s)} \sum_{k=1}^p \zeta\Bigl(s,\frac{k}{p}\Bigr) \biggl(\zeta\Bigl(s,\Bigl[\frac{k q}{p}\Bigr] \Bigr) \pm i \zeta\Bigl(s,\Bigl[-\frac{k q}{p}\Bigr] \Bigr)\biggr)\,,
\end{equation}
where we define the symbol $[x] = x + 1 - \ceil{x}$ for convenience. The modular transformations which fix $\tau_+ = \infty$ map $q \to q + n p$ ($\tau_- \to \tau_- + n$), so the result is indeed modular invariant.

The regulated $\mathcal{R}^4$ correction is then proportional to $f(\tau_+,\tau_-) = \zeta(3) \tilde{E}_{\frac{3}{2}}^{\pm}(\tau_+,\tau_-)$ for some choice of branch cut we do not attempt to fix. This result is interesting and finite, but it is unclear how to interpret the $\tau_\pm$ dependence as a coupling to the background fields $C_0$ and $\Phi$, since the latter are continuous dynamical fields.

The above example demonstrates that low energy effective description we have used in parts of our paper may be subject to important corrections or even may break down entirely in some cases. This bears further consideration.

\subsection{Concluding Thoughts}

In this paper we argued that two unusual ideas about gauge theory and string theory are connected. Previous work has explored, on the one hand, the possibility of supergroup gauge theories and their connection to negative branes in string theory~\cite{Vafa:2001qf,Okuda:2006fb,Vafa:2014iua}, and on the other hand, the possibility of exotic spacetime signatures and timelike compactifications in string theory~\cite{Moore:1993zc,Moore:1993qe,Hull:1998ym,Hull:1998vg}. We have shown that these two ideas are directly related. In particular, negative D0 branes arise from a smooth M-theory background with an $S^1$ that becomes timelike near the branes, giving a dynamic change of spacetime signature in the type IIA description. Using duality arguments, all other types of negative branes can be related to dynamic signature changes and the exotic string theories originally explored by Hull.

It remains possible that neither of these ideas is consistent, in which case we have added nothing to the general knowledge of string theory and gauge theory. However, the deep connections we have found between these two seemingly disparate areas are intriguing, and provide circumstantial evidence that both may actually be realized in string theory.

In our paper we have presented both consistency checks and puzzles. The most promising approach to definitively establish the existence of these exotic string theories would be to non-perturbatively construct supergroup gauge theories in four dimensions, e.g., on a lattice, which would in principle prove the existence of the exotic string theory IIB$^{--}_{3,7}$ via the AdS/CFT correspondence. While this remains a difficult (though perhaps not impossible) problem, understanding supermatrix models is a small step in the right direction~\cite{Vafa:2014iua,supermatrix}.

\acknowledgments{

We would like to thank N. Berkovits, G. Horowitz, S. Katz, J. Maldacena, N. Nekrasov, M. Reece, J. Scholtz, S. Shakirov, and S. Shao for illuminating discussions.  We would also like to thank the SCGP for hospitality during the 2015 summer workshop.

The research of CV is supported in part by NSF grant  PHY-1067976. BH is supported by the Fundamental Laws Initiative of the Center for the Fundamental Laws of Nature.}

\appendix

\section{Duality and effective actions in various signatures} \label{app:supergravity}

To determine the supergravity actions for the exotic string theories obtained by timelike compactification and T-duality, all we need to do is keep track of the two-derivative effective action during compactification. Our approach is to consider a general compactification of the supergravity actions posited in~\S\ref{subsec:SGaction} and to show that these actions are related to each other under KK reduction and T-duality as claimed in the text. After matching the actions to known results for the standard string theories, this shows that the exotic string theories have the low energy effective actions given in~\S\ref{subsec:SGaction}.

In~\S\ref{subsec:MtoIIA} we consider the dimensional reduction from eleven-dimensional supergravity to type IIA supergravity, in~\S\ref{subsec:IIAcompact}--\ref{subsec:buscher} we discuss T-duality between type IIA and type IIB supergravities, and in~\S\ref{subsec:Sduality} we discuss S-duality in type IIB supergravities. A computation of the Ricci scalar for KK reduction in arbitrary spacetime signature is included in~\S\ref{app:Riemann} for completeness.

\subsection{Reduction of eleven-dimensional supergravity to IIA} \label{subsec:MtoIIA}

We consider the theory $M^{\beta}$:
\begin{equation}
  S = \frac{1}{2 \kappa_{11}^2}  \int \de^{11} x \sqrt{| g |}  \left[ \cR -
  \frac{\beta}{2}  | F_4 |^2 \right] - \frac{1}{12 \kappa_{11}^2} \int C_3
  \wedge F_4 \wedge F_4 \,,
\end{equation}
where $\beta = \pm 1$ and $\beta = + 1$ gives the standard M-theory action in
signature $(10, 1)$.

We reduce this theory on a circle with the ansatz:
\begin{align}
  \de s^2_{11} &= e^{- \frac{2}{3} \Phi} \de s_{10}^2 + \eta_{y y}
  e^{\frac{4}{3} \Phi}  (\de y + C_1)^2 \,,\\
  F_4^{(11)} &= F_4 + H_3 \wedge \de y = \tilde{F}_4 + H_3 \wedge (\de
  y + C_1)  \,, &  C_3^{(11)} &= C_3 + B_2 \wedge \de y \,,
\end{align}
where $F_p \equiv \de C_{p - 1}$, $H_3 \equiv \de B_2$, and $\tilde{F}_4
\equiv F_4 - H_3 \wedge C_1$. Here $\eta_{y y} = \pm 1$ encodes the signature
of the compact dimension, and $y \cong y + 2 \pi R$. Using~(\ref{eqn:RicciReduced}), we find:
\begin{equation} \label{eqn:R11}
  e^{- \frac{2}{3} \Phi} \mathcal{R}^{(11)} = \mathcal{R} - \frac{\eta_{y y}}{2} e^{2 \Phi}  | F_2
  |^2 - \frac{16}{3}  (\nabla \Phi)^2 + \frac{14}{3} \nabla^2 \Phi \,,
\end{equation}
where $F_2 = \de C_1$. We also have
\begin{align} \label{eqn:Fg11}
  | F_4^{(11)} |^2 &= e^{\frac{8}{3} \Phi}  | \tilde{F}_4 |^2 + \eta_{y y}
  e^{\frac{2}{3} \Phi}  | H_3 |^2 \,, &   \sqrt{| g_{11} |} &= e^{- \frac{8}{3} \Phi} \sqrt{| g_{10} |}  \,.
\end{align}
Putting this into the action and integrating by parts, we obtain:
\begin{multline} \label{eqn:MtoIIAred}
  S = \frac{1}{2 \kappa_{10}^2} \int \de^{10} x \sqrt{| g |}  \left( e^{- 2
  \Phi}  \left[ \mathcal{R} - \frac{\alpha}{2}  | H_3 |^2 + 4 (\nabla \Phi)^2 \right] -
  \frac{\alpha \beta}{2}  | F_2 |^2 -\frac{\beta}{2} | \tilde{F}_4 |^2 \right)\\
  - \frac{1}{4 \kappa_{10}^2}  \int B_2 \wedge F_4 \wedge F_4 \,,
\end{multline}
for $\alpha \equiv \beta \eta_{y y} = \pm 1$ and $\kappa_{10}^2 \equiv \kappa_{11}^2/(2\pi R)$, which is the action for
IIA$^{\alpha \beta}$.

\subsection{IIA compactified on a circle} \label{subsec:IIAcompact}

To determine the T-dual of IIA$^{\alpha \beta}$, we start with the action~(\ref{eqn:MtoIIAred}),\footnote{For simplicity, we set
the Romans mass to zero.}
and compactify on a circle with the ansatz:
\begin{equation}
\begin{aligned}
  \de s_{10}^2 &= \de s_9^2 + \gamma e^{2 \sigma}  (\de y + A_1)^2 \,,\\
  \tilde{F}_4^{(10)} &= \tilde{F}_4 + \tilde{F}_3 \wedge (\de y + A_1)\,, & 
  C_3^{(10)} &= C_3 + C_2 \wedge (\de y + A_1)\,, \\
  \tilde{F}_2^{(10)} &= \tilde{F}_2 + F_1 \wedge (\de y + A_1)\,, & 
  C_1^{(10)} &= C_1 + C_0 \wedge (\de y + A_1) \,,\\
  H_3^{(10)} &= \tilde{H}_3 + H_2 \wedge (\de y + A_1)\,, & 
  B_2^{(10)} &= B_2 + B_1 \wedge (\de y + A_1)\,,
\end{aligned}
\end{equation}
where $\gamma = \pm 1$ specifies the signature of the compact dimension,
\begin{equation}
\begin{aligned}
  \tilde{F}_4 &\equiv F_4 + G_2 \wedge C_2 - \tilde{H}_3 \wedge C_1 \,, &
  \tilde{F}_2 &\equiv F_2 + G_2 \wedge C_0 \,, \\  
  \tilde{F}_3 &\equiv F_3 + H_2 \wedge C_1 - \tilde{H}_3 \wedge C_0 \,, &
  \tilde{H}_3 &\equiv H_3 - G_2 \wedge B_1 \,, \label{IIAdefs}
\end{aligned}
\end{equation}
and $G_2 = \de A_1$, $H_p = \de B_{p - 1}$, and $F_p = \de C_{p - 1}$. These satisfy the modified Bianchi identities:
\begin{equation}
\begin{aligned}
  \de \tilde{F}_4 &= G_2 \wedge \tilde{F}_3 + \tilde{H}_3 \wedge  \tilde{F}_2\,, &
  \de \tilde{F}_2 &= G_2 \wedge F_1 \,, \\
  \de \tilde{F}_3 &= H_2 \wedge \tilde{F}_2 + \tilde{H}_3 \wedge F_1\,, &
    \de \tilde{H}_3 &= - G_2 \wedge H_2\,. 
\end{aligned}
\end{equation}
Using
\begin{equation}
\begin{aligned}
  R^{(10)} &= R^{(9)} - \frac{1}{2} \gamma e^{2 \sigma}  | G_2 |^2 - 2 e^{-
  \sigma} \nabla^2 e^{\sigma}\,, &
  \sqrt{| g_{(10)} |} &= e^{\sigma}  \sqrt{| g_{(9)} |} \,, \\
  | \tilde{F}_4^{(10)} |^2 &= | \tilde{F}_4 |^2 + \gamma e^{- 2 \sigma}  |
  \tilde{F}_3 |^2 \,, &
  | \tilde{F}_2^{(10)} |^2 &= | \tilde{F}_2 |^2 + \gamma e^{- 2 \sigma}  |
  \tilde{F}_1 |^2 \,, \\
  | H_3^{(10)} |^2 &= | \tilde{H}_3 |^2 + \gamma e^{- 2 \sigma}  | H_2 |^2 \,,
\end{aligned}
\end{equation}
we obtain the dimensionally reduced action:
\begin{multline}
  S =  \frac{1}{2 \kappa_9^2}\! \int \!\de^9 x \sqrt{| g |} {e^{\sigma - 2
  \Phi}}  {\left[ \cR + 4 \nabla \Phi\! \cdot\! \nabla (\Phi - \sigma) -
  \frac{\alpha}{2}  | \tilde{H}_3 |^2 - \frac{\alpha \gamma}{2} e^{- 2 \sigma}
  | H_2 |^2 - \frac{\gamma}{2} e^{2 \sigma}  | G_2 |^2 \right]}\\
   - \frac{1}{4 \kappa_9^2}  \int \de^9 x \sqrt{| g |}  \left[\alpha \beta
  \gamma e^{- \sigma} | F_1 |^2 + \alpha \beta e^{\sigma}  | \tilde{F}_2 |^2 +
  \beta \gamma e^{- \sigma}  | \tilde{F}_3 |^2 + \beta e^{\sigma}  |
  \tilde{F}_4 |^2\right]\\
   - \frac{1}{4 \kappa_9^2} \int [B_1 \wedge \hat{F}_4 \wedge \hat{F}_4 + 2
  B_2 \wedge F_3 \wedge \hat{F}_4]\,,  \label{IIAreduced}
\end{multline}
where $\kappa_9^2 = \kappa_{10}^2 / (2 \pi R)$ and $\hat{F}_4 \equiv F_4 + G_2
\wedge C_2$. Here, gauge invariance of the Chern-Simons term can be proven by
taking the formal exterior derivative of the integrand, which is
\begin{equation}
  \de [B_1 \wedge \hat{F}_4 \wedge \hat{F}_4 + 2 B_2 \wedge F_3 \wedge
  \hat{F}_4] = H_2 \wedge \tilde{F}_4 \wedge \tilde{F}_4 + 2 \tilde{H}_3
  \wedge \tilde{F}_3 \wedge \tilde{F}_4 \,,
\end{equation}
equivalent to the dimensional reduction of $\de (B_2 \wedge F_4 \wedge F_4) = H_3 \wedge \tilde{F}_4 \wedge \tilde{F}_4$.

\subsection{IIB compactified on a circle} \label{subsec:IIBcompact}

We compare this result with the compactification of IIB$^{\alpha \beta}$, which has the pseudo action
\begin{multline}
  S = \frac{1}{2 \kappa_{10}^2} \int \de^{10} x \sqrt{| g |} e^{- 2
  \Phi}  \left[ \cR - \frac{\alpha}{2}  | H_3 |^2 + 4 (\nabla \Phi)^2 \right] \\
 - \frac{1}{4 \kappa_{10}^2}\! \int\! \de^{10} x \sqrt{| g |} \left[ \alpha \beta | F_1 |^2 + \beta | \tilde{F}_3 |^2 +
  \frac{\alpha \beta}{2} | \tilde{F}_5 |^2 \right] - \frac{1}{4 \kappa_{10}^2}\!  \int \! B_2 \wedge F_3 \wedge F_5 \,,
\end{multline}
where $\tilde{F}_3 = F_3 - H_3 \wedge C_0$, $\tilde{F}_5 = F_5 - H_3 \wedge C_2$, and the equations of motion need to be supplemented with the
self-duality constraint, $\tilde{F}_5 = \alpha \beta \star \tilde{F}_5$.
We compactify on a circle with the ansatz:\footnote{We do not consider the
case where $F_1$ has a leg along the compact circle, which is T-dual to a
non-zero Romans mass.}
\begin{equation}
\begin{aligned}
  \de s_{10}^2 &= \de s_9^2 + \gamma e^{2 \sigma}  (\de y + A_1)^2 \,, \\
  \tilde{F}_5^{(10)} &= \tilde{F}_5 + \tilde{F}_4 \wedge (\de y + A_1) \,, &
  C_4^{(10)} &= C_4 + C_3 \wedge (\de y + A_1) \,, \\
  \tilde{F}_3^{(10)} &= \tilde{F}_3 + \tilde{F}_2 \wedge (\de y + A_1) \,, &
  C_2^{(10)} &= C_2 + C_1 \wedge (\de y + A_1) \,, \\
  H_3^{(10)} &= \tilde{H}_3 + H_2 \wedge (\de y + A_1) \,, &
  B_2^{(10)} &= B_2 + B_1 \wedge (\de y + A_1) \,,
\end{aligned}
\end{equation}
where $\gamma = \pm 1$ specifies the signature of the compact dimension,
\begin{equation}
\begin{aligned}
  \tilde{F}_5 &\equiv F_5 - G_2 \wedge C_3 - \tilde{H}_3 \wedge C_2 \,, \\
  \tilde{F}_4 &\equiv F_4 - H_2 \wedge C_2 - \tilde{H}_3 \wedge C_1 \,, &
  \tilde{F}_2 &\equiv F_2 - H_2 \wedge C_0 \,, \\
  \tilde{F}_3 &\equiv F_3 - G_2 \wedge C_1 - \tilde{H}_3 \wedge C_0 \,, & 
  \tilde{H}_3 &\equiv H_3 - G_2 \wedge B_1 \,, \label{IIBdefs}
\end{aligned}
\end{equation}
and $G_2
= \de A_1$, $H_p = \de B_{p - 1}$, and $F_p = \de C_{p - 1}$. These satisfy the modified Bianchi identities:
\begin{equation}
\begin{aligned}
  \de \tilde{F}_5 &= - G_2 \wedge \tilde{F}_4 + \tilde{H}_3 \wedge \tilde{F}_3 \,, \\
  \de \tilde{F}_4 &= - H_2 \wedge \tilde{F}_3 + \tilde{H}_3 \wedge \tilde{F}_2 \,, &
  \de \tilde{F}_2 &= - H_2 \wedge F_1 \,, \\
  \de \tilde{F}_3 &=  - G_2 \wedge \tilde{F}_2 + \tilde{H}_3 \wedge F_1 \,, &
  \de \tilde{H}_3 &= - G_2 \wedge H_2 \,.
\end{aligned}
\end{equation}

Using
\begin{equation}
\begin{aligned}
  \cR^{(10)} &= \cR^{(9)} - \frac{1}{2} \gamma e^{2 \sigma}  | G_2 |^2 - 2 e^{-\sigma} \nabla^2 e^{\sigma} \,, &
  \sqrt{| g_{(10)} |} &= e^{\sigma}  \sqrt{| g_{(9)} |} \,, \\
  |\tilde{F}_5^{(10)}|^2 &= |\tilde{F}_5|^2 + \gamma e^{- 2 \sigma}  |\tilde{F}_4|^2 \,, &
  |\tilde{F}_3^{(10)}|^2 &= |\tilde{F}_3|^2 + \gamma e^{- 2 \sigma}  |\tilde{F}_2|^2 \,, \\
  | H_3^{(10)} |^2 &= | \tilde{H}_3 |^2 + \gamma e^{- 2 \sigma}  | H_2 |^2 \,, 
\end{aligned}
\end{equation}
we obtain the dimensionally-reduced pseudo-action:
\begin{multline}
  S = \frac{1}{2 \kappa_9^2} \!\int\! \de^9 x \sqrt{|g|} {e^{\sigma - 2 \Phi}}  {\left[\cR + 4 \nabla \Phi \!\cdot\! \nabla (\Phi - \sigma) 
  - \frac{\alpha}{2} |\tilde{H}_3|^2 - \frac{\alpha \gamma}{2} e^{- 2 \sigma}  |H_2|^2 - \frac{\gamma}{2} e^{2 \sigma}  |G_2|^2 \right]}\\
   - \frac{\beta}{4 \kappa_9^2} \!\int\! \de^9 x \sqrt{| g |}  \left[ \alpha
   e^{\sigma} | F_1 |^2 +  e^{\sigma}  | \tilde{F}_3 |^2 + 
  \gamma e^{- \sigma}  | \tilde{F}_2 |^2 + \frac{\alpha}{2} e^{\sigma} 
  | \tilde{F}_5 |^2 + \frac{\alpha \gamma}{2} e^{- \sigma}  |
  \tilde{F}_4 |^2 \right]\\
   + \frac{1}{4 \kappa_9^2} \int [\tilde{F}_4 \wedge \tilde{F}_5 + A_1
  \wedge \hat{F}_4 \wedge \hat{F}_4 - 2 \hat{B}_2 \wedge F_3 \wedge \hat{F}_4]\,,
\end{multline}
where $\kappa_9^2 = \kappa_{10}^2 / (2 \pi R)$, $\hat{F}_4 \equiv F_4 - H_2
\wedge C_2$, and $\hat{B}_2 \equiv B_2 + B_1 \wedge A_1$. The
manipulations leading to the Chern-Simons term given above can be quite complicated. To
simplify them, we write the ten-dimensional Chern-Simons term in a formal,
manifestly gauge-invariant way:
\begin{equation}
  S_{\mathrm{CS}} = - \frac{1}{4 \kappa_{10}^2}  \int_{X_{11}} H_3^{(10)} \wedge
  \tilde{F}_3^{(10)} \wedge \tilde{F}_5 \,,
\end{equation}
where $X_{11}$ is an eleven-dimensional manifold whose boundary is the
ten-dimensional space of interest. Reducing on a circle, we obtain
\begin{equation}
  S_{\mathrm{CS}} = \frac{1}{4 \kappa_9^2}  \int_{X_{10}} [- H_2 \wedge
  \tilde{F}_3 \wedge \tilde{F}_5 + \tilde{H}_3 \wedge \tilde{F}_2 \wedge
  \tilde{F}_5 - \tilde{H}_3 \wedge \tilde{F}_3 \wedge \tilde{F}_4] \,.
\end{equation}
Using
\begin{equation}
\begin{gathered}
  \de [\tilde{F}_4 \wedge \tilde{F}_5] = - H_2 \wedge \tilde{F}_3 \wedge
  \tilde{F}_5 + \tilde{H}_3 \wedge \tilde{F}_2 \wedge \tilde{F}_5 - G_2 \wedge
  \tilde{F}_4 \wedge \tilde{F}_4 + \tilde{H}_3 \wedge \tilde{F}_3 \wedge
  \tilde{F}_4 \,, \\
  \de [A_1 \wedge \hat{F}_4 \wedge \hat{F}_4 - 2 \hat{B}_2 \wedge F_3
  \wedge \hat{F}_4] = G_2 \wedge \tilde{F}_4 \wedge \tilde{F}_4 - 2
  \tilde{H}_3 \wedge \tilde{F}_3 \wedge \tilde{F}_4  \,,
\end{gathered}
\end{equation}
we recover the above result.

The self-duality constraint is now
\begin{equation}
  \tilde{F}_5 = - e^{- \sigma} \alpha \beta \gamma \star \tilde{F}_4 \,,
\end{equation}
hence
the potentials $C_3$ and $C_4$ are electromagnetic duals, and not
independent. 
The terms of the pseudo-action which depend on $C_4$
are
\begin{equation}
  \hat{S} = - \frac{1}{4 \kappa_9^2}  \int  \left[ \frac{\alpha \beta}{2}
  e^{\sigma}  \tilde{F}_5 \wedge \star \tilde{F}_5 - \tilde{F}_4 \wedge
  \tilde{F}_5 \right] \,,
\end{equation}
where $C_4$ does not appear explicitly
and the variation of the action with respect to $F_5$ is
\begin{equation}
  \hat{S} = - \frac{1}{4 \kappa_9^2}  \int \delta F_5 \wedge (e^{\sigma}
  \alpha \beta \star \tilde{F}_5 - \tilde{F}_4) \,,
\end{equation}
which is proportional to the self-duality constraint. Thus, if we replace $F_5 = \de C_4$
with an auxilliary field $\Lambda_5$, then the $\Lambda_5$ equation of motion
enforces the constraint and $\Lambda_5$ can be integrated out to give:
\begin{equation}
  \hat{S} = - \frac{1}{4 \kappa_9^2}  \int  \frac{\alpha \beta \gamma}{2} e^{-
  \sigma}  \tilde{F}_4 \wedge \star \tilde{F}_4 \,.
\end{equation}
Having solved the constraint, the pseudo-action becomes the bona fide
action
\begin{multline}
  S = \frac{1}{2 \kappa_9^2} \!\int\! \de^9 x \sqrt{|g|} {e^{\sigma - 2 \Phi}} {\left[\cR + 4 \nabla \Phi \!\cdot\! \nabla (\Phi - \sigma)
   - \frac{\alpha}{2}  | \tilde{H}_3 |^2 - \frac{\alpha \gamma}{2} e^{- 2 \sigma}  |H_2|^2 - \frac{\gamma}{2} e^{2 \sigma}  | G_2 |^2 \right]}\\
   - \frac{1}{4 \kappa_9^2}  \int \de^9 x \sqrt{| g |}  \left[\alpha \beta e^{\sigma} | F_1 |^2 + \beta \gamma e^{- \sigma}  | \tilde{F}_2 |^2
    + \beta e^{\sigma}  | \tilde{F}_3 |^2 + \alpha \beta \gamma e^{- \sigma}  | \tilde{F}_4 |^2\right]\\
   - \frac{1}{4 \kappa_9^2} \int [- A_1 \wedge \hat{F}_4 \wedge \hat{F}_4 + 2 \hat{B}_2 \wedge F_3 \wedge \hat{F}_4] \,. \label{IIBreduced}
\end{multline}

\subsection{Buscher rules}  \label{subsec:buscher}

To read off the Buscher rules, we compare the actions (\ref{IIAreduced}, \ref{IIBreduced}), bearing in mind
the differences between (\ref{IIAdefs}) and (\ref{IIBdefs}). We find that the
actions are mapped to each other by the field redefinitions
\begin{equation}
\begin{aligned}
  \sigma &\rightarrow - \sigma \,, &
  \Phi &\rightarrow \Phi - \sigma \,, \\
  A_1 &\rightarrow - B_1 \,, &
  B_1 &\rightarrow - A_1 \,, &
  B_2 &\rightarrow B_2 + B_1 \wedge A_1 \,,  
\end{aligned} \label{buscher}
\end{equation}
in the NS-NS sector, with $C_p$ and $g_{{\mu} \nu}^{}$ invariant, where the signs $\alpha, \beta, \gamma$ are related by
\begin{equation}
  \mathrm{IIA}^{\alpha, \beta}_{\gamma} \cong \mathrm{IIB}^{\alpha, \beta
  \gamma}_{\alpha \gamma} \,,
\end{equation}
and the subscript denotes the signature of the compact dimension. These are the T-duality relations derived from brane considerations in~\S\ref{subsec:dualityweb}. Written in components, the rules~(\ref{buscher}) become
\begin{align}
\label{eqn:Buscher}
  g'_{y y} &= \alpha \frac{1}{g_{y y}} \,, &
  g'_{y m} &= \alpha \frac{B_{y m}}{g_{y y}} \,, &
    g'_{m n} &= g_{m n} + \frac{\alpha B_{y m} B_{y n} - g_{y m} g_{y n}}{g_{y y}} \,, \\
     \Phi' &= \Phi - \frac{1}{2} \log | g_{y y} | \,, &
  B_{y m}' &= \frac{g_{y m}}{g_{y y}} \,, &
  B_{m n}' &= B_{m n} + \frac{g_{y m} B_{y n} - g_{y n} B_{y m}}{g_{y y}} \,,
\end{align}
which reproduce the usual Buscher rules for $\alpha = +1$.

\subsection{Einstein frame and S-duality} \label{subsec:Sduality}

The above discussion covers all possible T-dualities between type IIA and type IIB theories, as well as S-dualities between type IIA theories and M theories, at the two derivative level. This is sufficient to connect any two theories by dualities, hence to derive the two-derivative effective action for all of the exotic string theories considered by Hull. As a further consistency check, we verify the S-dualities between type IIB theories shown in Figures~\ref{fig:dualityweb1}, \ref{fig:dualityweb2} at the two derivative level.

The type IIB$^{\alpha \beta}$ pseudo-action becomes
\begin{multline}
  S = \frac{1}{2 \kappa_{10}^2} \int \de^{10} x \sqrt{| g |}   \left[ \cR - \frac{1}{2} (\nabla \Phi)^2 - \frac{\alpha \beta}{2} e^{2 \Phi} |F_1|^2  -  \frac{\alpha \beta}{4} |\tilde{F}_5|^2 \right] \\
 - \frac{1}{4 \kappa_{10}^2}\! \int\! \de^{10} x \sqrt{| g |} \left[ \beta e^{\Phi} | \tilde{F}_3 |^2 + \alpha e^{- \Phi} | H_3 |^2 \right] - \frac{1}{4 \kappa_{10}^2}\!  \int \! C_4 \wedge H_3 \wedge F_3 \,,
\end{multline}
in Einstein frame, where
\begin{align}
g_{\mu \nu} &= e^{-\Phi/2} g_{\mu \nu}^{({\rm str})}\,, & C_4 &= C_4^{({\rm str})} - \frac{1}{2} B_2 \wedge C_2 \,,
\end{align}
so that
$
\tilde{F}_5 = F_5 - \frac{1}{2} C_2 \wedge H_3 + \frac{1}{2} B_2 \wedge F_3 \,,
$
with $\tilde{F}_5 = \alpha \beta \star \tilde{F}_5$ as before. To make the $SL(2,\mathbb{Z})$ invariance explicit, we form the hypercomplex combinations
\begin{align}
\tau &\equiv C_0 + j e^{-\Phi}\,, & G_3 &\equiv F_3 - \tau H_3 \,,
\end{align}
where $j$ is a hypercomplex number satisfying $j^2 = -\alpha \beta$ and $j^\ast = -j$. Written in terms of $\tau$ and $G_3$, the pseudo-action becomes
\begin{multline}
  S = \frac{1}{2 \kappa_{10}^2} \int \de^{10} x \sqrt{| g |}   \left[ \cR - \frac{\alpha \beta}{2 (\im \tau)^2} |\de \tau|^2  - \frac{\beta}{2 \im \tau} | G_3 |^2-  \frac{\alpha \beta}{4} |\tilde{F}_5|^2 \right] \\
 - \frac{1}{4 \kappa_{10}^2}\!  \int \! C_4 \wedge H_3 \wedge F_3 \,,
\end{multline}
where we define $| G_p |^2 \equiv \frac{1}{p!} G^{{\mu}_1 \ldots {\mu}_p} G_{{\mu}_1 \ldots {\mu}_p}^\ast$ for hypercomplex forms. This action is invariant under the $SL(2,\mathbb{R})$ transformation
\begin{align}
\tau' &= \frac{a \tau+b}{c \tau+d}\,, & \begin{pmatrix} F_3' \\ H_3' \end{pmatrix} &= \begin{pmatrix}a&b\\c&d\end{pmatrix} \begin{pmatrix} F_3 \\ H_3 \end{pmatrix} \,, & a d-b c&=1\,,
\end{align}
where
\begin{align}
\de \tau' &= \frac{\de \tau}{(c \tau+d)^2}\,, & G_3' &= \frac{G_3}{c \tau+d}\,, & \im \tau' &=  \frac{\im \tau}{|c \tau+d|^2} \,.
\end{align}
The spectrum of branes is only invariant for $a,b,c,d\in \mathbb{Z}$, so that the symmetry of the full theory is $SL(2,\mathbb{Z})$.

Note that for $\alpha \beta = +1$, we can choose $j = i$ so that $\tau$ is a complex number, whereas for $\alpha \beta = -1$, $j$ generates the split complex numbers. In the former case, $\tau \to \frac{a \tau+b}{c \tau+d}$ maps the upper half plane $\im \tau > 0$ to itself, whereas in the latter it does not. In cases where $\im \tau' < 0$ we can redefine $\tau \to \tau^\ast$, $G_3 \to G_3^\ast$ to fix $\im \tau > 0$, but this maps $\beta \to - \beta$ with $\alpha \beta = -1$ fixed, exchanging IIB$^{+-}$ with IIB$^{-+}$. This implies that for $C_0=0$ IIB$^{+-}$ and IIB$^{-+}$ are S-dual (related by $\tau \to -1/\tau$). More generally, it is convenient to think of IIB$^{+-}$ and IIB$^{-+}$ as occupying the upper and lower half of the split-complex plane, with $\tau \to \frac{a \tau+b}{c \tau+d}$ mapping some points to the same half plane and some to the opposite. By contrast, for IIB$^{++}$ and IIB$^{--}$ $\tau$ is confined to the upper half of the complex plane, and the theory is self-dual for any $\tau$.

\subsection{Riemann tensor computation} \label{app:Riemann}

We consider the metric ansatz:
\begin{equation} \label{eqn:RiemannAnastz}
  \de \hat{s}_{d + 1}^2 = \de s_d^2 + \eta_{y y} e^{2 \lambda}  (\de y +
  A)^2 = \eta_{i j} \theta^i \theta^j + \eta_{y y} \theta^2 \,,
\end{equation}
where $\eta_{i j}$ and $\eta_{y y}$ are arbitrary constants (allowing any signature), and $\theta = e^{\lambda}  (\de y + A)$.
We compute:
\begin{align}
  \de \theta &= \de \lambda \wedge \theta + e^{\lambda} F \,, &
   \de \theta^i &= - \omega^i_{\; j} \wedge \theta^j \,, 
\end{align}
where $\omega^i_{\; j}$ is the spin connection for $\de s_d^2$ and $F\equiv \de A$.
The solution to Cartan's structure equations is:
\begin{align}
  \hat{\omega}^i_{\; j} &= \omega^i_{\; j} - \eta_{y y} e^{\lambda}  \frac{1}{2}
  F^i_{\; j} \theta\,, &
  \hat{\omega}^y_{\; i} &= (\nabla_i \lambda) \theta + e^{\lambda}  \frac{1}{2}
  F_{i j} \theta^j \,, & \hat{\omega}^i_{\; y} = - \eta_{y y} \eta^{i j} \hat{\omega}^y_{\; j}\,,
\end{align}
where $\hat{\omega}^i_{\; j}$ is the spin connection for $\de \hat{s}_{d+1}^2$.

To compute the curvature two-form, we note that all factors of $\omega^i_{\; j}$ must eventually cancel from the result by general covariance,
except for derivatives of $\omega^i_{\; j}$, such as
$\de \omega^i_{\; j} = \mathcal{R}^i_{\; j} - \omega^i_{\; j} \wedge
\omega^j_{\; k}$, so we can set $\omega^i_{\; j} = 0$ without loss of generality where it appears undifferentiated. We then have:
\begin{equation}
  \de F_{i j} = (\nabla_k F_{i j}) \theta^k \qquad \left(
 \omega^i_{\; j} = 0 \right)\,,
\end{equation}
and so on for other tensors. We find:
\begin{align}
  \hat{\mathcal{R}}^i_{\; j} &= \mathcal{R}^i_{\; j} - \frac{1}{4} \eta_{y y} e^{2 \lambda}
  F^i_{\; k} F_{j l} \theta^k \wedge \theta^l - \frac{1}{2} \eta_{y y} e^{2
  \lambda} F^i_{\; j} F - \frac{1}{2} \eta_{y y} e^{\lambda}  \left( \nabla_k
  F^i_{\; j} \right) \theta^k \wedge \theta \nonumber\\
  &\noeq - \frac{1}{2} \eta_{y y} e^{\lambda}  (\nabla^i \lambda) F_{j l} \theta
  \wedge \theta^l - \frac{1}{2} \eta_{y y} e^{\lambda}  (\nabla_j \lambda)
  F^i_{\; k} \theta^k \wedge \theta - \eta_{y y} e^{\lambda} F^i_{\; j} \de
  \lambda \wedge \theta \,,\\
  \hat{\mathcal{R}}^y_{\; i} &= (\nabla_i \lambda)  [\de \lambda \wedge \theta +
  e^{\lambda} F] + \frac{1}{2} e^{\lambda} F_{i j} \de \lambda \wedge
  \theta^j + (\nabla_i \nabla_j \lambda) \theta^j \wedge \theta \nonumber \\
  &\noeq + \frac{1}{2}
  e^{\lambda}  (\nabla_k F_{i j}) \theta^k \wedge \theta^j
   - \frac{1}{4} \eta_{y y} e^{2 \lambda} F_{i j} F^j_{\; k} \theta \wedge \theta^k \,, \\
   \hat{\mathcal{R}}^i_{\; y} &= - \eta_{y y} \eta^{i j} \hat{\mathcal{R}}^y_{\; j} \,,
\end{align}
where $\mathcal{R}^i_{\; j}$ denotes the curvature two-form for $\de s_d^2$.
From this, we compute the Ricci one-form $\mathcal{R}_B =i_{E_A} \mathcal{R}^A_{\; B}$:
\begin{align}
  \hat{\mathcal{R}}_j &= \mathcal{R}_j - \frac{1}{2} \eta_{y y} e^{2 \lambda} F^i_{\; j} F_{i
  k} \theta^k  - \frac{1}{2} \eta_{y y} e^{-2 \lambda} \nabla_i (e^{3 \lambda}
  F^i_{\; j} ) \theta  - e^{-\lambda} (\nabla_i \nabla_j e^{\lambda}) \theta^i\,,\\
  \hat{\mathcal{R}}_y &= \frac{1}{4} \eta_{y y}^2 e^{2 \lambda} F^{i j} F_{i j} \theta - \frac{1}{2} \eta_{y y}
  e^{-2\lambda} \nabla_i (e^{3 \lambda} F^i_{\; j} ) \theta^j  - \eta_{y y} e^{-\lambda} (\nabla^2 e^{\lambda}) \theta \,,
\end{align}
and the Ricci scalar $\mathcal{R} = \eta^{A B} i_{E_A} \mathcal{R}_B$:
\begin{equation}
  \hat{\mathcal{R} }= \mathcal{R} - \frac{1}{4} \eta_{y y} e^{2 \lambda} F^{i j} F_{i j} - 2
  e^{- \lambda} \nabla^2 e^{\lambda} \,.
\end{equation}
This result can be combined with the well-known conformal transformation of the
Ricci scalar to generalize the ansatz. For $\de \hat{s}^2_d = e^{2 \omega} \de s_d^2$ we have
\begin{align}
  e^{2 \omega} \hat{\mathcal{R}} &= \mathcal{R} - (d - 1) (d - 2)  (\nabla \omega)^2 - 2 (d - 1)
  \nabla^2 \omega \,,\\
  e^{2 \omega} \hat{\nabla}^2 \phi &= \nabla^2 \phi + (d - 2)  \nabla
  \omega \cdot \nabla \phi\,.
\end{align}
Thus, for the more general ansatz
\begin{equation}
  \de \hat{s}_{d + 1}^2 = e^{2 \omega} \de s_d^2 + \eta_{y y} e^{2 \lambda} 
  (\de y + A)^2 \,,
\end{equation}
we obtain
\begin{multline} \label{eqn:RicciReduced}
  e^{2 \omega} \hat{\mathcal{R}} = \mathcal{R} - \frac{1}{4} \eta_{y y} e^{2 (\lambda - \omega)}
  F^{i j} F_{i j}
  - (d - 1) (d - 2)  (\nabla \omega)^2 \\ - 2 (d - 1) \nabla^2
  \omega - 2 (\nabla \lambda)^2 - 2 \nabla^2 \lambda - 2 (d - 2) \nabla \omega
  \cdot \nabla \lambda \,.
\end{multline}

\bibliographystyle{JHEP}
\bibliography{refs}

\providecommand{\href}[2]{#2}\begingroup\raggedright\begin{thebibliography}{10}

\bibitem{Bars:2000qm}
I.~Bars, {\it {Survey of two time physics}},  {\em Class. Quant. Grav.} {\bf
  18} (2001) 3113--3130, [\href{http://arxiv.org/abs/hep-th/0008164}{{\tt
  hep-th/0008164}}].

\bibitem{Hull:1998ym}
C.~Hull, {\it {Duality and the signature of space-time}},  {\em JHEP} {\bf
  9811} (1998) 017, [\href{http://arxiv.org/abs/9807127}{{\tt 9807127}}].

\bibitem{Vafa:2014iua}
C.~Vafa, {\it {Non-Unitary Holography}},
  \href{http://arxiv.org/abs/1409.1603}{{\tt arXiv:1409.1603}}.

\bibitem{Vafa:2001qf}
C.~Vafa, {\it {Brane / anti-brane systems and $U(N|M)$ supergroup}},
  \href{http://arxiv.org/abs/hep-th/0101218}{{\tt hep-th/0101218}}.

\bibitem{Okuda:2006fb}
T.~Okuda and T.~Takayanagi, {\it {Ghost D-branes}},  {\em JHEP} {\bf 03} (2006)
  062, [\href{http://arxiv.org/abs/hep-th/0601024}{{\tt hep-th/0601024}}].

\bibitem{Hull:1998fh}
C.~M. Hull and R.~R. Khuri, {\it {Branes, times and dualities}},  {\em Nucl.
  Phys.} {\bf B536} (1998) 219--244,
  [\href{http://arxiv.org/abs/hep-th/9808069}{{\tt hep-th/9808069}}].

\bibitem{Hull:1999mt}
C.~M. Hull and R.~R. Khuri, {\it {World volume theories, holography, duality
  and time}},  {\em Nucl. Phys.} {\bf B575} (2000) 231--254,
  [\href{http://arxiv.org/abs/hep-th/9911082}{{\tt hep-th/9911082}}].

\bibitem{Blencowe:1988sk}
M.~P. Blencowe and M.~J. Duff, {\it {Supermembranes and the Signature of
  Space-time}},  {\em Nucl. Phys.} {\bf B310} (1988) 387.

\bibitem{Parkhomenko:2008dt}
S.~E. Parkhomenko, {\it {Free Field Construction of D-Branes in Rational Models
  of CFT and Gepner Models}},  {\em SIGMA} {\bf 4} (2008) 025,
  [\href{http://arxiv.org/abs/0802.3445}{{\tt arXiv:0802.3445}}].

\bibitem{supermatrix}
R.~Dijkgraaf, B.~Heidenreich, and C.~Vafa, ``Supermatrix models revisited.''
\newblock To appear.

\bibitem{Cardy:1985yy}
J.~L. Cardy, {\it {Conformal Invariance and the Yang-lee Edge Singularity in
  Two-dimensions}},  {\em Phys. Rev. Lett.} {\bf 54} (1985) 1354--1356.

\bibitem{Belavin:1984vu}
A.~A. Belavin, A.~M. Polyakov, and A.~B. Zamolodchikov, {\it {Infinite
  Conformal Symmetry in Two-Dimensional Quantum Field Theory}},  {\em Nucl.
  Phys.} {\bf B241} (1984) 333--380.

\bibitem{Fisher:1978pf}
M.~E. Fisher, {\it {Yang-Lee Edge Singularity and $\varphi^3$ Field Theory}},
  {\em Phys. Rev. Lett.} {\bf 40} (1978) 1610--1613.

\bibitem{Quella:2007hr}
T.~Quella and V.~Schomerus, {\it {Free fermion resolution of supergroup WZNW
  models}},  {\em JHEP} {\bf 09} (2007) 085,
  [\href{http://arxiv.org/abs/0706.0744}{{\tt arXiv:0706.0744}}].

\bibitem{Fei:2014yja}
L.~Fei, S.~Giombi, and I.~R. Klebanov, {\it {Critical $O(N)$ models in
  $6-\epsilon$ dimensions}},  {\em Phys. Rev.} {\bf D90} (2014), no.~2 025018,
  [\href{http://arxiv.org/abs/1404.1094}{{\tt arXiv:1404.1094}}].

\bibitem{Mati:2016wjn}
P.~Mati, {\it {Critical scaling in the large-$N$ $O(N)$ model in higher
  dimensions and its possible connection to quantum gravity}},
  \href{http://arxiv.org/abs/1601.00450}{{\tt arXiv:1601.00450}}.

\bibitem{Arnone:2000qd}
S.~Arnone, {\relax Yu}.~A. Kubyshin, T.~R. Morris, and J.~F. Tighe, {\it {A
  Gauge invariant regulator for the ERG}},  {\em Int. J. Mod. Phys.} {\bf A16}
  (2001) 1989, [\href{http://arxiv.org/abs/hep-th/0102054}{{\tt
  hep-th/0102054}}].

\bibitem{Metsaev:1998it}
R.~R. Metsaev and A.~A. Tseytlin, {\it {Type IIB superstring action in
  AdS$_5\times$S$^5$ background}},  {\em Nucl. Phys.} {\bf B533} (1998)
  109--126, [\href{http://arxiv.org/abs/hep-th/9805028}{{\tt hep-th/9805028}}].

\bibitem{Kallosh:1998qs}
R.~Kallosh and A.~Rajaraman, {\it {Vacua of M theory and string theory}},  {\em
  Phys. Rev.} {\bf D58} (1998) 125003,
  [\href{http://arxiv.org/abs/hep-th/9805041}{{\tt hep-th/9805041}}].

\bibitem{Bars:2011aa}
I.~Bars, S.-H. Chen, P.~J. Steinhardt, and N.~Turok, {\it {Antigravity and the
  Big Crunch/Big Bang Transition}},  {\em Phys. Lett.} {\bf B715} (2012)
  278--281, [\href{http://arxiv.org/abs/1112.2470}{{\tt arXiv:1112.2470}}].

\bibitem{Araya:2015fva}
I.~J. Araya, I.~Bars, and A.~James, {\it {Journey Beyond the Schwarzschild
  Black Hole Singularity}},  \href{http://arxiv.org/abs/1510.03396}{{\tt
  arXiv:1510.03396}}.

\bibitem{Moore:1993zc}
G.~W. Moore, {\it {Finite in all directions}},
  \href{http://arxiv.org/abs/hep-th/9305139}{{\tt hep-th/9305139}}.

\bibitem{Moore:1993qe}
G.~W. Moore, {\it {Symmetries and symmetry breaking in string theory}},  in
  {\em {International Workshop on Supersymmetry and Unification of Fundamental
  Interactions (SUSY 93) Boston, Massachusetts, March 29-April 1, 1993}}, 1993.
\newblock \href{http://arxiv.org/abs/hep-th/9308052}{{\tt hep-th/9308052}}.

\bibitem{Hull:1998vg}
C.~M. Hull, {\it {Timelike T duality, de Sitter space, large N gauge theories
  and topological field theory}},  {\em JHEP} {\bf 07} (1998) 021,
  [\href{http://arxiv.org/abs/hep-th/9806146}{{\tt hep-th/9806146}}].

\bibitem{Gutperle:2002ai}
M.~Gutperle and A.~Strominger, {\it {Space - like branes}},  {\em JHEP} {\bf
  04} (2002) 018, [\href{http://arxiv.org/abs/hep-th/0202210}{{\tt
  hep-th/0202210}}].

\bibitem{Aharony:2013hda}
O.~Aharony, N.~Seiberg, and Y.~Tachikawa, {\it {Reading between the lines of
  four-dimensional gauge theories}},  {\em JHEP} {\bf 08} (2013) 115,
  [\href{http://arxiv.org/abs/1305.0318}{{\tt arXiv:1305.0318}}].

\bibitem{Sethi:2013hra}
S.~Sethi, {\it {A New String in Ten Dimensions?}},  {\em JHEP} {\bf 09} (2013)
  149, [\href{http://arxiv.org/abs/1304.1551}{{\tt arXiv:1304.1551}}].

\bibitem{Duff:2006iy}
M.~J. Duff and J.~Kalkkinen, {\it {Metric and coupling reversal in string
  theory}},  {\em Nucl. Phys.} {\bf B760} (2007) 64--88,
  [\href{http://arxiv.org/abs/hep-th/0605274}{{\tt hep-th/0605274}}].

\bibitem{Duff:2006ix}
M.~J. Duff and J.~Kalkkinen, {\it {Signature reversal invariance}},  {\em Nucl.
  Phys.} {\bf B758} (2006) 161--184,
  [\href{http://arxiv.org/abs/hep-th/0605273}{{\tt hep-th/0605273}}].

\bibitem{Cornalba:2003kd}
L.~Cornalba and M.~S. Costa, {\it {Time dependent orbifolds and string
  cosmology}},  {\em Fortsch. Phys.} {\bf 52} (2004) 145--199,
  [\href{http://arxiv.org/abs/hep-th/0310099}{{\tt hep-th/0310099}}].

\bibitem{Liu:2002ft}
H.~Liu, G.~W. Moore, and N.~Seiberg, {\it {Strings in a time dependent
  orbifold}},  {\em JHEP} {\bf 06} (2002) 045,
  [\href{http://arxiv.org/abs/hep-th/0204168}{{\tt hep-th/0204168}}].

\bibitem{Aharony:1999ti}
O.~Aharony, S.~S. Gubser, J.~M. Maldacena, H.~Ooguri, and Y.~Oz, {\it {Large N
  field theories, string theory and gravity}},  {\em Phys. Rept.} {\bf 323}
  (2000) 183--386, [\href{http://arxiv.org/abs/hep-th/9905111}{{\tt
  hep-th/9905111}}].

\bibitem{Kim:1985ez}
H.~J. Kim, L.~J. Romans, and P.~van Nieuwenhuizen, {\it {Mass spectrum of
  chiral ten-dimensional $\mathcal{N}=2$ supergravity on $S^5$}},  {\em Phys.
  Rev.} {\bf D32} (1985) 389.

\bibitem{'tHooft:1982tz}
G.~'t~Hooft, {\it {On the Convergence of Planar Diagram Expansions}},  {\em
  Commun. Math. Phys.} {\bf 86} (1982) 449.

\bibitem{Brezin:1977sv}
E.~Brezin, C.~Itzykson, G.~Parisi, and J.~B. Zuber, {\it {Planar Diagrams}},
  {\em Commun. Math. Phys.} {\bf 59} (1978) 35.

\bibitem{Berenstein:2002jq}
D.~E. Berenstein, J.~M. Maldacena, and H.~S. Nastase, {\it {Strings in flat
  space and pp waves from N=4 superYang-Mills}},  {\em JHEP} {\bf 04} (2002)
  013, [\href{http://arxiv.org/abs/hep-th/0202021}{{\tt hep-th/0202021}}].

\bibitem{Polyakov:2007mm}
A.~M. Polyakov, {\it {De Sitter space and eternity}},  {\em Nucl. Phys.} {\bf
  B797} (2008) 199--217, [\href{http://arxiv.org/abs/0709.2899}{{\tt
  arXiv:0709.2899}}].

\bibitem{Aharony:2008ug}
O.~Aharony, O.~Bergman, D.~L. Jafferis, and J.~Maldacena, {\it {N=6
  superconformal Chern-Simons-matter theories, M2-branes and their gravity
  duals}},  {\em JHEP} {\bf 10} (2008) 091,
  [\href{http://arxiv.org/abs/0806.1218}{{\tt arXiv:0806.1218}}].

\bibitem{Green:1997di}
M.~B. Green and P.~Vanhove, {\it {D instantons, strings and M theory}},  {\em
  Phys. Lett.} {\bf B408} (1997) 122--134,
  [\href{http://arxiv.org/abs/hep-th/9704145}{{\tt hep-th/9704145}}].

\bibitem{Green:1997as}
M.~B. Green, M.~Gutperle, and P.~Vanhove, {\it {One loop in
  eleven-dimensions}},  {\em Phys. Lett.} {\bf B409} (1997) 177--184,
  [\href{http://arxiv.org/abs/hep-th/9706175}{{\tt hep-th/9706175}}].

\bibitem{Peskin:1995ev}
M.~E. Peskin and D.~V. Schroeder, {\em {An Introduction to quantum field
  theory}}.
\newblock 1995.

\bibitem{Polchinski:1998rq}
J.~Polchinski, {\em {String theory. Vol. 1: An introduction to the bosonic
  string}}.
\newblock Cambridge University Press, 2007.

\bibitem{Yost:1991ht}
S.~A. Yost, {\it {Supermatrix models}},  {\em Int. J. Mod. Phys.} {\bf A7}
  (1992) 6105--6120, [\href{http://arxiv.org/abs/hep-th/9111033}{{\tt
  hep-th/9111033}}].

\bibitem{AlvarezGaume:1991zc}
L.~Alvarez-Gaume and J.~L. Manes, {\it {Supermatrix models}},  {\em Mod. Phys.
  Lett.} {\bf A6} (1991) 2039--2050.

\bibitem{Pestun:2007rz}
V.~Pestun, {\it {Localization of gauge theory on a four-sphere and
  supersymmetric Wilson loops}},  {\em Commun. Math. Phys.} {\bf 313} (2012)
  71--129, [\href{http://arxiv.org/abs/0712.2824}{{\tt arXiv:0712.2824}}].

\bibitem{Witten:1997sc}
E.~Witten, {\it {Solutions of four-dimensional field theories via M theory}},
  {\em Nucl. Phys.} {\bf B500} (1997) 3--42,
  [\href{http://arxiv.org/abs/hep-th/9703166}{{\tt hep-th/9703166}}].

\bibitem{Mikhaylov:2014aoa}
V.~Mikhaylov and E.~Witten, {\it {Branes And Supergroups}},
  \href{http://arxiv.org/abs/1410.1175}{{\tt arXiv:1410.1175}}.

\bibitem{Klemm:1996bj}
A.~Klemm, W.~Lerche, P.~Mayr, C.~Vafa, and N.~P. Warner, {\it {Selfdual strings
  and N=2 supersymmetric field theory}},  {\em Nucl.Phys.} {\bf B477} (1996)
  746--766, [\href{http://arxiv.org/abs/hep-th/9604034}{{\tt hep-th/9604034}}].

\bibitem{Katz:1997eq}
S.~Katz, P.~Mayr, and C.~Vafa, {\it {Mirror symmetry and exact solution of 4-D
  N=2 gauge theories: 1.}},  {\em Adv.Theor.Math.Phys.} {\bf 1} (1998) 53--114,
  [\href{http://arxiv.org/abs/hep-th/9706110}{{\tt hep-th/9706110}}].

\bibitem{Nekrasov:2012xe}
N.~Nekrasov and V.~Pestun, {\it {Seiberg-Witten geometry of four dimensional
  N=2 quiver gauge theories}},  \href{http://arxiv.org/abs/1211.2240}{{\tt
  arXiv:1211.2240}}.

\bibitem{craig-weinstein}
W.~{Craig} and S.~{Weinstein}, {\it {On determinism and well-posedness in
  multiple time dimensions}},  {\em Proc. Roy. Soc. A} {\bf 465} (2009) 3023,
  [\href{http://arxiv.org/abs/0812.0210}{{\tt arXiv:0812.0210}}].

\bibitem{Weinstein:2008aj}
S.~Weinstein, {\it {Multiple Time Dimensions}},
  \href{http://arxiv.org/abs/0812.3869}{{\tt arXiv:0812.3869}}.

\bibitem{Strominger:2001pn}
A.~Strominger, {\it {The dS / CFT correspondence}},  {\em JHEP} {\bf 10} (2001)
  034, [\href{http://arxiv.org/abs/hep-th/0106113}{{\tt hep-th/0106113}}].

\bibitem{rsk2005}
R.~S. Krau§har, {\it Generalized analytic automorphic forms for some arithmetic
  congruence subgroups of the vahlen group on the $n$-dimensional hyperbolic
  space},  {\em Bull. Belg. Math. Soc. Simon Stevin} {\bf 11} (03, 2005)
  759--774.

\end{thebibliography}\endgroup

\end{document}